\documentclass[]{aa}
\usepackage[utf8]{inputenc}
\usepackage[english]{babel}
\usepackage{graphicx}
\usepackage{xcolor}
\usepackage{booktabs} 
\usepackage{amssymb,amsmath}
\usepackage{array}
\usepackage{siunitx}
\usepackage{textgreek}
\usepackage{natbib}
\usepackage[varg]{txfonts}
\usepackage{url}
\usepackage[unicode=true]{hyperref}
\bibpunct{(}{)}{;}{a}{}{,} 
\hypersetup{
    colorlinks=true,
    linkcolor={blue!50!black}, 
    citecolor={blue!50!black}, 
    urlcolor={blue!80!black}   
}
\usepackage{etoolbox}
\makeatletter
\patchcmd\@combinedblfloats{\box\@outputbox}{\unvbox\@outputbox}{}{\errmessage{\noexpand patch failed}}
\makeatother
\begin{document}

\title{Skymaps of observables of three-dimensional MHD astrosphere models}
\author{L. R. Baalmann\inst{\ref{tp4}} \and K. Scherer\inst{\ref{tp4},\ref{pci}} \and H. Fichtner\inst{\ref{tp4},\ref{pci}} \and J. Kleimann\inst{\ref{tp4}} \and D. J. Bomans\inst{\ref{pci},\ref{astro}} \and K. Weis\inst{\ref{astro}}}
\institute{Ruhr-Universität Bochum, Fakultät für Physik und Astronomie, Institut für Theoretische Physik IV, Germany\label{tp4}\\
e-mail: \texttt{lb@tp4.rub.de} \and
Ruhr-Universität Bochum, Research Department, Plasmas with Complex Interactions, Germany\label{pci} \and
Ruhr-Universität Bochum, Fakultät für Physik und Astronomie, Astronomisches Institut, Germany\label{astro}}
\date{Received \textcolor{red}{\bf date} / Accepted \textcolor{red}{\bf date}}
\abstract{Three-dimensional models of astrospheres have recently become of interest. However, comparisons between these models and observations are non-trivial because of the two-dimensional nature of observations.}{By projecting selected physical values of three-dimensional models of astrospheres onto the surface of a sphere that is centred on a virtual all-sky observer, these models can be compared to observational data in different observables: the column density, bremsstrahlung flux, rotation measure, H\textalpha\ flux, and synchrotron or cyclotron flux.}{Projections were calculated by rotating and moving the astrosphere model to the desired position and orientation and by then computing the value of a given patch on the sphere by a modified line-of-sight integration. Contributions to the selected observable made by all model cells that are connected to the patch by the line of sight in question were taken into account.}{When the model produces a bow shock, a distinct parabolic structure produced by the outer astrosheath can be seen in every observable of the projection, the exact shape depending on the orientations of the line of sight and the stellar motion. Of all four examined astrosphere models, only that of \textlambda~Cephei shows fluxes that are higher than current observational thresholds. This is due to the strong stellar wind and interstellar inflow of the \textlambda~Cephei model.}{}
\maketitle

\section{Introduction}\label{sec:intro}
The interactions of interstellar and stellar plasmas result in distinct structures around the host stars, as first analytically modelled by \citet{weaver77}. They have been simulated in one-dimensional (1D) hydrodynamics (HD) \citep[e.g.][]{arthur12}, 2D HD \citep[e.g.][and references therein]{scherer16,delvalle18},  2D magnetohydrodynamics (MHD) \citep[e.g.][]{meyer17}, and more recently also in 2.5D MHD \citep[e.g.][]{vanmarle14}, 3D HD \citep[e.g.][]{mohamed12}, and 3D MHD \citep[e.g.][]{gvaramadze18,scherer18}. The resulting models have been successfully compared to observational images and data in various observables \citep[e.g.][and references therein]{kobulnicky17,ayaso2018}. The shapes of these structures strongly depend on the physical parameters of the star and on the surrounding interstellar medium (ISM) \citep[see e.g.][]{gvaramadze12}, as well as on the orientation of the stellar motion relative to the observer's line of sight (LOS) \citep[see e.g.][]{tarangoyong18}. Because Earth is located inside the heliosphere, numerical models of the latter can be compared to observational data with much higher spatial resolution than for other astrospheres. Recently, both numerical 3D MHD models and observational studies have been used to explore various aspects of the heliosphere in detail \citep[see e.g.][]{opher2016,dialynas2017,kornbleuth2018,guo2019}. Because observational images are 2D projections of a 3D structure, 3D models must be treated similarly to allow comparisons between simulation and observation. Such a 3D model can be rotated at will before projecting, simulating different orientations of the modelled structure to the observer's LOS. 

We here use 3D single-fluid MHD simulations of astrospheres as in \citet{scherer16b} as a basis for projections in various observables. The available physical parameters of the simulations are used to calculate further physical values in each model cell that can then be projected through a modified LOS integration to yield observables such as the radiation fluxes of bremsstrahlung, cyclotron radiation and H\textalpha\ emission, the column density, or the rotation measure. The simulation volume can be rotated about all three axes and shifted to a chosen position in Galactic coordinates before integration. 

The astrospheres of two notable stars, \textlambda~Cephei and the Sun, have been examined for example by \citet{scherer16b}. The Sun is interesting because it is our host star; Earth is embedded in the solar astrosphere, the heliosphere \citep{zank1999,opher16,pogorelov17}. \textlambda~Cephei is the brightest runaway O star in the sky \citep{scherer16}, and therefore is the object most likely to have the most visible astrosphere. The astrosphere of \textlambda~Cephei has been proposed as a sink for cosmic rays \mbox{\citep{scherer15}}, which has been examined further by \citet{scherer16c}. 

The shock structures of the models are assumed to be comparable to the shock structure described by \citet{scherer16b}: a termination shock (TS) close to the star, where the stellar wind jumps from supersonic to subsonic speeds, an astropause (AP) separating the stellar and interstellar fluids, and a bow shock (BS) in the direction of the incoming ISM in front of the star, where the interstellar fluid jumps from supersonic to subsonic speeds. In the downwind direction, that is, the direction opposite to the interstellar inflow, the stellar wind jumps from supersonic to subsonic speeds at the Mach disc (MD). A BS may not exist in all models, depending on the relative speed between the star and the surrounding ISM \citep[see e.g.][]{mccomas12,zank13}. The region between the TS and the AP is called the inner astrosheath, and the region between the AP and the BS is the outer astrosheath. The outer astrosheath is divided into a hot outer astrosheath (HOA) close to the BS and a cold outer astrosheath (COA) close to the AP. 

Details of the simulations, including the parameters of the stars, are presented in Sec.~\ref{sec:mods}. An overview of the available physical parameters is given in Sec.~\ref{sec:phys}, and the projection methods are outlined in Sec.~\ref{sec:proj}. The results are presented in Sec.~\ref{sec:res} and compared with observational data in Sec.~\ref{sec:obs}. Conclusions are drawn in Sec.~\ref{sec:conc}.

\section{Astrosphere models}\label{sec:mods}

The astrospheres were modelled in 3D single-fluid MHD with the \textsc{\large cronos} code \citep{kissmann18}, a semi-discrete finite-volume code that advances Eqs.~(\ref{eq:mhdcont}-\ref{eq:mhdindu}) in time, opting for an HLL Riemann solver and a second-order Runge-Kutta scheme,
\begin{align}
 \frac{\partial n}{\partial t} + \nabla \cdot \left( n \vec{u}\right) &= 0 &,\label{eq:mhdcont}\\
 \frac{\partial}{\partial t} \left( mn \vec{u} \right) + \nabla \cdot \left( m n \vec{u} \otimes \vec{u}\right) + \nabla p + \frac{1}{\mu_0} \vec{B} \times \left( \nabla \times \vec{B} \right) &= \vec{0} &,\label{eq:mhdimp}\\
 \frac{\partial e}{\partial t} + \nabla \cdot \left[ \left( e + p + \frac{1}{2\mu_0} \left| \vec{B} \right|^2 \right) \vec{u} - \frac{1}{\mu_0}\left( \vec{u} \cdot \vec{B} \right) \vec{B}\right] &= 0 &,\label{eq:mhdenergy}\\
 \frac{\partial \vec{B}}{\partial t} = - \nabla \times \vec{E} = \nabla \times \left( \vec{u} \times \vec{B} \right) & &,\label{eq:mhdindu}
\end{align}
where $n, \vec{u}, p, \vec{B}, e, \vec{E}$ are the number density, fluid velocity, thermal pressure, magnetic induction, total energy density, and electric field. $m\text{ and } \mu_0$ are the proton mass and the vacuum permeability, respectively; $\otimes$ represents the dyadic product. The system is closed by the equations
\begin{align}
 e &= \frac{p}{\gamma -1} + \frac{1}{2}mn \left| \vec{u} \right|^2 + \frac{1}{2\mu_0}\left| \vec{B} \right|^2 &,\label{eq:mhdgas}\\
 \nabla \cdot \vec{B} &= 0 &,
\end{align}
where $\gamma$ is the polytropic index. 
Instead of the thermal pressure, we usually considered the temperature $T=3 p / (4n k_{\mathrm{B}})$, where $k_{\mathrm{B}}$ is the Boltzmann constant. Factor $1/2$ stems from the assumption of thermodynamic equilibrium between electrons and protons ($p=p_{\mathrm{e}}+p_{\mathrm{p}}=2p_{\mathrm{p}}$).

Additional terms for heating follow \citet{reynolds1999} and were treated as in \citet{kosinski2006}, and cooling following \citet{schure09} was added at the right-hand side of Eq.~(\ref{eq:mhdenergy}), allowing photoelectric heating by dust, dissipation of interstellar turbulence, and Coulomb collisions with cosmic rays, and in particular, radiative cooling in the temperature range $T\in[10^4, 10^7]\,\si{K}$. The polytropic index was set to $\gamma=1.66667\approx 5/3$. 

The model was set up as the interior of a near-spherical polyhedron with cells in spherical coordinates $(r, \vartheta, \varphi)$ that were equidistant in the radial direction $\vec{e}_{\mathrm{r}}$ and equiangular in directions $\vec{e}_{\vartheta, \varphi}$ perpendicular to $\vec{e}_{\mathrm{r}}$. The entire angular ranges of $\vartheta\in[-90\si{\degree}, 90\si{\degree}]$ and $\varphi\in[-180\si{\degree}, 180\si{\degree}]$ are covered, while the radial range extends over the interval $r\in[r_{\mathrm{SW}}, r_{\mathrm{ISM}}]$, $r_{\mathrm{SW}} \neq 0$. The increments and therefore the radial and angular resolutions depend on the model.

Boundary values for the stellar wind (SW) and the ISM were set for the inner- and outermost cells (at $r_{\mathrm{SW, ISM}}$): both $\vec{v}$ and $\vec{B}$ of the ISM are homogeneous with angles $\vartheta_{v,B}, \varphi_{v,B}$, while $\vec{v}$ of the SW is spherically symmetric with only a radial component. $\vec{B}$ of the SW was set up as a spiral field following \citet{parker1958}. Likewise, $n$ and $T$ were set and kept fixed at $r_{\mathrm{SW,ISM}}$ for the SW and ISM. 

By adjusting the boundary values to observed or derived parameters of stars and their surroundings, the astrospheres of these stars were simulated. While any star may be examined this way, the two objects of particular interest, the Sun and \textlambda~Cephei (see Sec.~\ref{sec:intro}), have been closely investigated. We briefly examine the astropheres of two other stars, those of Proxima Centauri and V374 Pegasi. The stellar winds of these stars have been modelled by \citet{garraffo2016} and \citet{vidotto2011}, respectively. The boundary values of the corresponding models are given in Tab.~\ref{tab:modvals}; for \textlambda~Cephei, see \citet{scherer16,scherer18}, and for the Sun, see \citet{pogorelov17}. The values for the ISM of Proxima Centauri were adapted from the Sun; the stellar magnetic field was taken from \citet{reiners2008}, and the density and wind speed were selected to produce a large shock structure. For V374 Pegasi, the ISM was assumed to be a warm ionized medium \citep[WIM, see e.g.][]{mckee1977} with corresponding parameters, while values for the SW were taken from \citet{vidotto2011}. The numbers of simulated cells also differ between the models (see Tab.~\ref{tab:boxvals}). A more detailed description of the simulation can be found in \citet{scherer16}.

\begin{table}
\caption{\label{tab:boxvals}Simulation sizes and resolutions of the various models.}
\centering
\begin{tabular}{lllll}
\toprule\toprule
Param. & \textlambda\ Cep & Helio & Prox. Cen & V374 Peg \\ \midrule
$r$ & 990 & 280 & 1000 & 1000 \\
$\vartheta$ & 30 & 60 & 60 & 90 \\
$\varphi$ & 60 & 120 & 120 & 180 \\ \midrule
$\delta r$ & $0.005\,\si{pc}$ & $3\,\si{AU}$ & $0.5\,\si{AU}$ & $496\,\si{AU}$ \\
$\delta(\vartheta,\varphi)$ & $6\si{\degree}$ & $3\si{\degree}$ & $3\si{\degree}$ & $2\si{\degree}$\\
\bottomrule
\end{tabular}
\tablefoot{Number of cells in $r$, $\vartheta$, and $\varphi$, as well as radial ($\delta r$) and angular ($\delta(\vartheta,\varphi)$) resolutions of the various models.}
\end{table}

\begin{table*}
\caption{\label{tab:modvals}Boundary values of the various models.}
\centering
\begin{tabular}{llllllllll}
\toprule\toprule
\multicolumn{2}{l}{Variable} & \multicolumn{2}{c}{\textlambda~Cephei} & \multicolumn{2}{c}{Heliosphere} & \multicolumn{2}{c}{Proxima Centauri} & \multicolumn{2}{c}{V374 Pegasi} \\ \cmidrule(l{0.2em}){3-4}\cmidrule(l{0.2em}){5-6}\cmidrule(l{0.2em}){7-8}\cmidrule(l{0.2em}){9-10}
 && SW & ISM & SW & ISM & SW & ISM & SW & ISM \\  \midrule
$R$ & & $0.05\,\si{pc}$ & $5\,\si{pc}$ & $60\,\si{AU}$ & $900\,\si{AU}$ & $1\,\si{AU}$ & $500\,\si{AU}$ & $4000\,\si{AU}$ & $5\times 10^5\,\si{AU}$ \\
$n$&$\left[\si{cm^{-3}}\right]$ & 3.4 & 11 & 7 & 0.06 & 0.3 & 0.06 & $7\times 10^4$& 0.01\\
$v$&$\left[\si{km/s}\right]$ & 2500 & 80 & 375 & 26.4 & 1000 & 25 & 1500 & 26.4 \\
$T$&$\left[10^3\,\si{K}\right]$ & 1 & 9 & 73.64 & 6.53 & 73 & 6.5 & 20 & 1000 \\
$B$&$\left[\si{nT}\right]$ & $3\times 10^{-6}$ & 1 & 4 & 0.3 & 300 & 0.3 & 4000 & 3\\ 
$\vartheta_v$&$\left[\si{\degree}\right]$ & & 90 & & 90 & & 90 & & 90 \\
$\varphi_v$&$\left[\si{\degree}\right]$ & & 180 & & 180 & & 180 & & 180 \\
$\vartheta_B$&$\left[\si{\degree}\right]$ & & 30 & & 30 & & 30 & & 30 \\
$\varphi_B$&$\left[\si{\degree}\right]$ & & 150 & & 150 & & 150 & & 150\\
\bottomrule
\end{tabular}
\tablefoot{Boundary values of the SW for \textlambda~Cephei are set at $R_{\mathrm{SW}}$, all other boundary values of the SW are set at a distance of $1\,\si{AU}$ to the model origin.}
\end{table*}

\section{Physical parameters}\label{sec:phys}
Of the dynamic variables, only the number density is directly related to an observable: to the column density. To investigate the projections in other observables, their unprojected physical values must be calculated first. In the cases of the total fluxes of bremsstrahlung (Sec.~\ref{sec:brs}) and synchrotron or cyclotron radiation (Sec.~\ref{sec:cyc}), these values are the total energy loss rates, while the H\textalpha\ flux (Sec.~\ref{sec:rec}) corresponds to the number of recombinations per time and volume, that is, the recombination rate. The rotation measure (Sec.~\ref{sec:frm}) requires a different approach (see Sec.~\ref{sec:proj}).

\subsection{Bremsstrahlung}\label{sec:brs}
When we assume that the electron speed follows a Maxwellian distribution, the total energy loss rate for thermal bremsstrahlung is 
\begin{equation}\label{eq:brs}
 -\left(\frac{\mathrm{d} E}{\mathrm{d} t}\right)_{\mathrm{brs}} \,\left[\si{\frac{erg}{cm^3 \, s}}\right]= 1.435 \times 10^{-27} Z^2 \sqrt{T}\, g_{\mathrm{ff}}(T)\, n n_{\mathrm{e}}
,\end{equation}
with $Z$ the mass number (here $Z=1$), $T$ the temperature in $\si{K}$, $n, n_{\mathrm{e}}$ the number densities of the ions and electrons in $\si{cm^{-3}}$ (here $n=n_{\mathrm{e}}$), and $g_{\mathrm{ff}}(T)$ the temperature-dependent Gaunt factor for free-free transitions \citep{longair92}. A common approximation for $g_{\mathrm{ff}}(T)$ is its temperature-averaged value of $\bar{g}=1.2$. Gaunt factors for any desired non-relativistic temperature were calculated here following \citet{vanhoof14} and result in values $g_{\mathrm{ff}}(T)\in[1.0, 1.5)$ for the temperature range of the examined astrosphere models. 

The emission spectrum of bremsstrahlung was assumed to be flat up to a critical frequency
\begin{equation}
 \nu_{\max}=\frac{4\pi m_{\mathrm{e}} v^2}{h}
,\end{equation}
with $m_{\mathrm{e}}$ the electron mass, $v$ the (thermal) electron speed, and $h$ the Planck constant \citep{longair92}. Because the total spectrum of the complete model is the weighted sum of the spectra of all cells, the total spectrum will be flat up to the lowest critical frequency. This is useful for calculating the spectral flux $F_{\nu}$ by dividing the total flux $F$ by the frequency range, that is, the critical frequency. This is only a preliminary result, however, because self-absorption and similar processes will heavily modify a flat emission spectrum. These processes are currently not considered.

\subsection{Synchrotron and cyclotron radiation}\label{sec:cyc}
The total energy loss rate of a single electron by synchrotron radiation is 
\begin{equation}\label{eq:totsyn}
 -\left(\frac{\mathrm{d} E}{\mathrm{d} t}\right)_{\mathrm{syn}} = \sigma_{\mathrm{T}} c \frac{B^2}{4\pi} \gamma^2 \left( \frac{v}{c} \right)^2 \sin^2 \alpha \ ,
\end{equation}
with $\sigma_{\mathrm{T}}$ the Thomson cross-section, $\gamma$ the Lorentz factor, $c$ the speed of light, $B$, $v$ the absolute values of the magnetic induction and the electron speed, and $\alpha$ the pitch angle between $\vec{v}$ and $\vec{B}$ \citep{blumenthal70}. With $\gamma\approx 1,$ Eq.~(\ref{eq:totsyn}) becomes the total energy loss rate of cyclotron radiation for non-relativistic electrons. The electron speed contains both the absolute value of the fluid velocity and the thermal speed calculated through
\begin{equation}
v_{\mathrm{therm}} = \sqrt{\frac{8 k_{\mathrm{B}} T}{\pi m_e}}
\end{equation} 
\citep{condon58}, so that the angle between the electron motion and the magnetic field is no longer equal to $\alpha$. The factor $\sin^2 \alpha$ must be substituted by its average value of $2/3$ in Eq.~(\ref{eq:totsyn}). 

The emitted power of the ultra-relativistic synchrotron spectrum is of the form
\begin{equation}\label{eq:synspec}
 P_{\mathrm{em}}(\nu) = \frac{\sqrt{3} q_{\mathrm{e}}^3 B}{m_{\mathrm{e}} c^2} \frac{\nu}{\nu_{\mathrm{c}}} \int\limits_{\nu/\nu_{\mathrm{c}}}^{\infty} K_{5/3} (s)\, \mathrm{d} s
,\end{equation}
with $\nu$ the frequency, $q_{\mathrm{e}}$ the elementary charge, $\nu_{\mathrm{c}}$ the critical frequency, and $K_{5/3}$ the modified Bessel function of the second kind \citep{blumenthal70}. The critical frequency is
\begin{equation}\label{eq:synfreq}
\nu_{\mathrm{c}}=\frac{3}{2}\nu_{\mathrm{g}}\gamma^2 \sin\alpha \ ,
\end{equation}
with $\nu_{\mathrm{g}}=eB/(2\pi m_{\mathrm{e}})$ the gyration frequency. Again, the factor $\sin\alpha$ must be substituted by $\pi/4$ because $\alpha$ is no longer the angle between $\vec{v}$ and $\vec{B}$. Integrating over the entire spectral range $\nu\in[0, \infty)$ yields Eq.~(\ref{eq:totsyn}). If the electron speed is non-relativistic, the total power is emitted at $\nu_{\mathrm{g}}$. At higher speeds, radiation is emitted at the gyrofrequency harmonics $\nu_l=l\nu_{\mathrm{g}}, l\in\mathbb{N}$; the power decreases by a factor of $(v/c)^2$ for every harmonic. The harmonics were ignored in the calculations because the electron speed remains lower than $0.1c$ in all examined astrospheres. By multiplying the total power with the number density, we calculated the power density.

\subsection{H\textalpha\ emission}\label{sec:rec}
Following \citet{reynolds84}, the photon flux of H\textalpha\ is connected to the recombination rate by
\begin{equation}\label{eq:reynolds}
 I_{\text{H\textalpha}}\,\left[\frac{\text{photons}}{\si{cm^2 \, s \, sr}}\right]=\frac{\varepsilon\, \tilde{r}_{\mathrm{rr}}}{4\pi} \ ,
\end{equation}
with $\varepsilon$ the average number of H\textalpha\ photons produced per recombination, and $\tilde{r}_{\mathrm{rr}}$ the recombination rate along the LOS in $\si{cm}^{-2}$. The radiation flux was calculated by multiplying with the H\textalpha\ photon energy $h\nu_{\text{H\textalpha}}$, where $\nu_{\text{H\textalpha}}=456.81\,\si{THz}$ is the H\textalpha\ frequency. Because $\tilde{r}_{\mathrm{rr}}$ is the LOS-integrated case B recombination rate,
\begin{equation}
 \tilde{r}_{\mathrm{rr}} = \int_s r_{\mathrm{B}} \mathrm{d} s \ , 
\end{equation}
the emitted power density of a selected volume is
\begin{equation}
 P_{\text{H\textalpha}} = h\nu_{\text{H\textalpha}} \varepsilon r_{\mathrm{B}} \ .
\end{equation}
The case B recombination rate $r_{\mathrm{B}}$ describes recombinations to all atomic states $n$ except for the ground state $n=1$; recombination to the latter would result in subsequent ionization and therefore in no net recombination. The recombination rate can be calculated by the electron and ion number densities $n_{\mathrm{e,H^+}}$ and the case B recombination rate coefficient $\alpha_{\mathrm{B}}$ , which is equivalent to the sum over all recombination rate coefficients $\alpha_n$ except for that of the ground state:
\begin{equation}\label{eq:harb}
 r_{\mathrm{B}} = n_{\mathrm{e}} n_{\mathrm{H^+}} \alpha_{\mathrm{B}} = n_{\mathrm{e}} n_{\mathrm{H^+}} \sum\limits_{n=2}^{\infty} \alpha_n \ .
\end{equation}
The temperature-dependent coefficients $\alpha_n (T)$ were taken from \citet{mao16} and are valid for the entire observed temperature range $T\in[10^1, 10^8]\,\si{K}$ for states $n\in[1,16]$. The coefficients were not only resolved by the principal quantum number $n,$ but also by the azimuthal quantum number $l$ ; the $\alpha_n$ were calculated by summation over all applicable $l$. 

Because no estimation for $\varepsilon$ is given by \citet{reynolds84}, a different approach was chosen to calculate $\varepsilon\alpha_{\mathrm{B}}$. The emission coefficient for the atomic transition $i\rightarrow k$ is given by \citet{draine} as
\begin{equation}\label{eq:jik}
 j_{i\rightarrow k}=\frac{n_e n_{\mathrm{H}^+}}{4\pi} \frac{A_{i\rightarrow k} h \nu_{i\rightarrow k}}{\sum_l A_{i\rightarrow l}} \cdot \left[ \alpha_{i}+ \sum\limits_{l,l>i} \alpha_{l} P_{\mathrm{B},l,i} \right] \ ,
\end{equation}
where $A_{i\rightarrow k}$ is the transition probability for spontaneous emission of transition $i\rightarrow k$, taken from \citet{wiese2009}, and $P_{\mathrm{B},i,k}$ is the branching ratio, that is, the probability that an atom in state $i$ decays through state $k$, 
\begin{align}
 P_{\mathrm{B},i,k} &= \left( A_{i\rightarrow k} + \sum\limits_{j<n<i} A_{i\rightarrow n}P_{\mathrm{B},n,k}\right)\left/ \sum\limits_{n<i}A_{i\rightarrow n}\right. \\ 
 P_{\mathrm{B},i,i+1} &= A_{i\rightarrow i+1} \left/ \sum\limits_{n<i}A_{i\rightarrow n} \right. \ .
\end{align}
An effective recombination rate coefficient $\alpha_{\mathrm{eff},i\rightarrow k}$ was introduced to shorten Eq.~(\ref{eq:jik}) to
\begin{equation}
 j_{i\rightarrow k} = \frac{n_e n_{\mathrm{H}^+}}{4\pi} \alpha_{\mathrm{eff},i\rightarrow k} h \nu_{i\rightarrow k} \ .
\end{equation}
It directly ensues from the above equations that $\alpha_{\mathrm{eff},\text{H\textalpha}}=\varepsilon \alpha_{\mathrm{B}}$. Because H\textalpha\ corresponds to the transition $3\rightarrow 2$ and the $\alpha_n$ quickly decreases for higher $n$, it follows that
\begin{equation}
 \alpha_{\mathrm{eff},3\rightarrow 2}\approx\alpha_3 + \sum_{n=4}^{16} \alpha_n P_{\mathrm{B},n,3} \ .
\end{equation}
Comparison of calculated values of $\alpha_{\mathrm{eff},\text{H\textalpha}}$ and $\alpha_{\mathrm{B}}$ yields $\varepsilon\approx 0.2643^{+0.022}_{-0.019}$ in the temperature range $T\in[10^1, 10^8]\,\si{K}$.

\subsection{Faraday rotation}\label{sec:frm}
As a measure for the Faraday rotation, the Faraday depth was used \citep{vaneck17}:
\begin{equation}\label{eq:rmdef}
 \phi(d)\,\left[\si{\frac{rad}{cm^2}}\right]= 0.812 \int\limits_d^0 \left(\frac{n_{\mathrm{e}}}{\si{cm^{-3}}}\right) \left(\frac{\vec{B}}{\si{\micro G}}\right) \cdot \left(\frac{\mathrm{d} \vec{l}}{\si{pc}}\right) \ ,
\end{equation}
with $d$ the distance from the source. If $d$ is equal to the distance between observer and source, the Faraday depth is equal to the rotation measure. Instead of using the magnetic induction and the line element as vectors, $\vec{B}$ may be substituted by its component parallel to the LOS $B_{||}$ and $\mathrm{d}\vec{l}$ by its scalar counterpart $\mathrm{d}l$. 

\section{Projection methods}\label{sec:proj}

\begin{figure}
 \centering
 \includegraphics[width=\columnwidth]{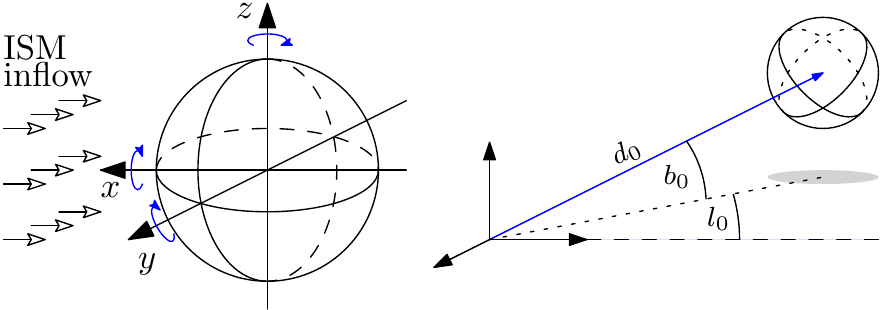}\vspace{0.1cm}
 \includegraphics[width=\columnwidth]{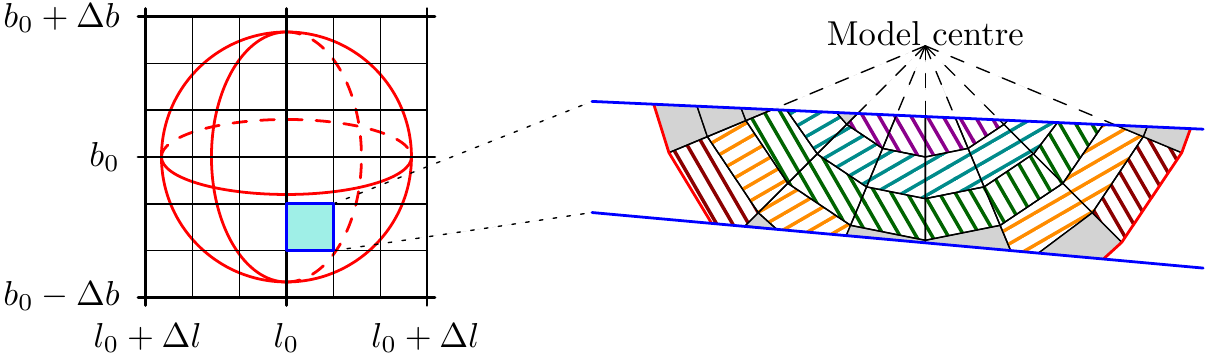}
 \caption{Upper left: Sketch of the model being rotated; the homogeneous ISM inflow comes from the positive $x$-direction. Upper right: Sketch of the model being shifted; in most cases, $b_0=l_0=0$. Bottom: Sketch of the LOS grid covering the projection of a model (red) with the weighting along a single LOS (blue): some cells (grey) are not taken into account because their centres are outside the LOS, while all cells inside the LOS with a common distance from the model centre share the same weight (coloured tiling patterns).}\label{fig:rotshift}
\end{figure}

The model projections were performed in Galactic coordinates $(d,b,l)$, with the modulus $d$, the Galactic latitude $b,$ and the Galactic longitude $l$; $b$ and $l$ cover the ranges $b\in[-90\si{\degree}, 90\si{\degree}]$, $l\in[-180\si{\degree}, 180\si{\degree}]$, with $b=0$ being the Galactic plane (i.e. the $xy$-plane, $z=0$). The model cells $(i,j,k)$ are denominated by the Cartesian coordinates to their centres $\vec{O}(i, j, k)$; henceforth $(\kappa)=(i, j, k)$ is used as a shorthand. The model can be rotated by angles $\alpha, \beta, \gamma$ around the $x$-, $y$-, and $z$-axes via rotation matrices $\mathbf{A}(\alpha)$, $\mathbf{B}(\beta)$, $\mathbf{C}(\gamma)$ and shifted to a new position of the centre of the computational domain  $\vec{D}(d_0, b_0, l_0)$, see Fig.~\ref{fig:rotshift} (upper row). The new position vectors to the model cells are
\begin{equation}
 \vec{N}(\kappa) = \mathbf{A}(\alpha) \cdot \mathbf{B}(\beta) \cdot \mathbf{C}(\gamma) \cdot \vec{O}(\kappa) + \vec{D}(d_0, b_0, l_0) \ ,
\end{equation}
with the rotation matrices 
\begin{align}
 \mathbf{A}(\alpha) &= \left(\begin{array}{ccc}
  1 & 0 & 0 \\
  0 & \phantom{+}\cos\alpha & \sin\alpha \\
  0 & -\sin\alpha & \cos\alpha 
 \end{array}\right) \ , & \\
 \mathbf{B}(\beta) &= \left(\begin{array}{ccc}
  \cos\beta & 0 & -\sin\beta \\
  0 & 1 & 0 \\
  \sin\beta & 0 & \phantom{+}\cos\beta 
 \end{array}\right)\ , & \\
 \mathbf{C}(\gamma) &= \left(\begin{array}{ccc}
  \phantom{+}\cos\gamma & \sin\gamma & 0 \\
  -\sin\gamma & \cos\gamma & 0 \\
  0 & 0 & 1 
 \end{array}\right)\ . & 
\end{align}

The shifting vector $\vec{D}(d_0, b_0, l_0)$ arises as the result of transforming from Galactic to Cartesian coordinates:
\begin{equation}
 \vec{D}(d_0, b_0, l_0)=\left(\begin{array}{l}
  d_0 \cos b_0 \cos l_0 \\
  d_0 \cos b_0 \sin l_0 \\
  d_0 \sin b_0 
 \end{array} \right) \ .
\end{equation}
The scalar values $n(\kappa), T(\kappa)$ of the simulation are given for each cell $(\kappa)$ and need no further modification, whereas the vectorial values must be rotated. Both the fluid velocity and the magnetic induction are available in spherical coordinates $\vec{v}_{\mathrm{s}}(\kappa)$, $\vec{B}_{\mathrm{s}}(\kappa)$, as is a transformation matrix $\mathbf{T}(\kappa)$ from spherical to Cartesian coordinates. The new vectors $\vec{\xi}\in\{\vec{v}, \vec{B}\}$ follow
\begin{equation}
 \vec{\xi} (\kappa) = \mathbf{A}(\alpha) \cdot \mathbf{B}(\beta) \cdot \mathbf{C}(\gamma) \cdot \mathbf{T}(\kappa) \cdot \vec{\xi}_{\mathrm{s}}(\kappa) \ .
\end{equation}
The Galactic coordinates $(d(\kappa), b(\kappa), l(\kappa))$ of the new position of the cell centres $\vec{N}(\kappa)$ were calculated through the usual transformation
\begin{align}
 d(i, j, k) &= \sqrt{N_x^2(\kappa) + N_y^2(\kappa) + N_z^2(\kappa)}\ , \\
 b(i, j, k) &= \arcsin\left(\frac{N_z(\kappa)}{d (\kappa)}\right)\ , \\
 l(i, j, k) &= \arctan\left(\frac{N_y(\kappa)}{N_x(\kappa)}\right) \ .
\end{align}
Because each model is spherical in nature, its radius $R$ is half the difference of the largest and smallest Cartesian $x$-coordinate,
\begin{equation}
 R = \frac{1}{2} \left( \max(N_x(\kappa)) - \min(N_x(\kappa)) \right)
.\end{equation}
The model was then projected onto the surface of a sphere around the origin, where the observer is located. The exact coordinates $(b, l)$ of the projected point of a cell centre on the 2D surface is identical to the angular items of the 3D coordinates of the cell centre. However, the surface of the sphere was divided into a grid that set the resolution of the projection (see below). 

If the distance of the computational domain centre to the origin $d_0$ is shorter than the model radius $R$, that is, if the origin is inside the model, the model projection will occupy the entire sky (the sphere of projection). If $d_0 > R$, the spherical model is projected as a disc with an angular diameter $\diameter < 180\si{\degree}$. The angular radius $\Delta b$ follows from simple geometry as
\begin{equation}
 \Delta b = \frac{1}{2} \diameter = \arcsin \left( \frac{R}{d_0} \right) \ ;
\end{equation}
the symbol $\Delta b$ was chosen because the angular radius is the angular interval of latitude between the disc centre and its edge in every geometric configuration.

If the latitude of the computational domain centre $b_0$ (henceforth called the elevation) is high or low enough for the projection disc to extend to or over the poles of the projection sphere ($|b_0| + \Delta b \geq 90\si{\degree}$), the disc is spread out over all longitudes $l\in[-180\si{\degree}, 180\si{\degree}]$. For an elevation $b_0 \neq 0\si{\degree}$ and a smaller angular radius $|b_0| + \Delta b < 90\si{\degree}$, the maximum interval in the longitude non-trivially depends on $\left\{b_0, \Delta b\right\}$ and is approximated by 
\begin{equation}\label{eq:deltal}
 \Delta l = \arctan\left(\frac{\tan\left(\Delta b\right)}{\cos\left(b_0+\frac{b_0}{90\si{\degree}}\Delta b\right)} \right) \ .
\end{equation}
Because the position of the astrosphere model in the sky (or rather on the projected sphere) has no bearing for most purposes, $b_0=0\si{\degree}$, $l_0=0\si{\degree}$ was used in most cases. In this case, the angular extent in both angular coordinates is identical, $\Delta b = \Delta l$. Even if the actual position of a star was used as $(b_0, l_0)$, most stars appear close enough to the Galactic plane ($b_0\approx 0\si{\degree}$) that $\Delta l \approx \Delta b$. 

Because the outer edges of the 2D grid are now known as $b_0 \pm \Delta b$, $l_0 \pm \Delta l$, its axes were divided into $\dim(b, l)$ intervals. Here $\dim(b, l)$ is the number of model cells in $\vartheta, \varphi$ with additional resolution factors $k_{b, l}$. If the model is centred at the origin ($d_0=0\si{\degree}$) and $k_{b,l}=1$, the 2D grid cells correspond to the columns of 3D model cells in the $\vec{e}_r$-direction; the centres of all cells inside a column line up exactly. If the model is centred outside the origin ($d_0 > R$), resolution factors $k_l = 2 k_b$ must be used to obtain square grid cells because the models usually have twice as many cells in $\varphi$ as in $\vartheta$ to obtain the same angular resolution. 

The integration process was rather simple: when the angular coordinates $(b, l$) of the centre of a model cell appear inside one of the grid cells, its physical parameter $H(\kappa)$ (e.g. the total energy loss rate of bremsstrahlung or the number density) were added to the grid cell value. Depending on the physical parameter $H(\kappa)$, two types of weighting lead to two types of integration: LOS integration weights the cell by its length $L(\kappa)$ along the LOS, whereas volumetric integration weights the cell by its volume $V(\kappa)$ (see Fig.~\ref{fig:rotshift} bottom). 

For most physical parameters, volumetric integration is the appropriate method. For all radiation types, the particles in the entire volume contribute to the total emission. Here, the parameter $H(\kappa)$ is a measure of emission per volume, so the product of $H(\kappa)$ and volume $V(\kappa)$ is the total amount of emission. If $H(\kappa)$ is the number density, $H(\kappa)V(\kappa)$ gives the number of particles in the entire integration volume. Summing $\sum H(\kappa)V(\kappa)$ and dividing by the entire volume ($\sum V(\kappa)$) gives the volume-averaged number density. For the radiation parameters (bremsstrahlung, cyclotron, and H\textalpha), the product $H(\kappa)V(\kappa)$ divided by $4\pi d^2(\kappa)$ gives the radiation flux of the cell, or more formalized,
\begin{equation}\label{eq:volint}
 \mathrm{Res}_{\mathrm{vol}} = \sum\limits_{i=1}^N \frac{H_i V_i}{4\pi d_i^2} \ ,
\end{equation}
with $\mathrm{Res}_{\mathrm{vol}}$ the volumetric integration result, $N$ cells inside the integrated volume, $H_i$, $d_i$, $V_i$ the value of the physical parameter, the distance to the origin, and the volume of cell $i\in[1, N]$. The volume of cell $(\kappa)$ is calculated through
\begin{align}
 V(\kappa) =&\ \int_{r_{\mathrm{in}}(\kappa)}^{r_{\mathrm{out}}(\kappa)} \mathrm{d} r \int_{\varphi(\kappa)-\frac{1}{2}\delta \varphi}^{\varphi(\kappa)+\frac{1}{2}\delta \varphi} \mathrm{d} \varphi \int_{\vartheta-\frac{1}{2}\delta\vartheta}^{\vartheta+\frac{1}{2}\delta\vartheta} \mathrm{d} \vartheta\, r^2 \cos \vartheta \nonumber\\
 =& \frac{1}{3}\left(r_{\mathrm{out}}^3-r_{\mathrm{in}}^3\right)\cdot\delta \varphi\cdot\left(\sin\left(\vartheta+\frac{\delta\vartheta}{2}\right)-\sin\left(\vartheta-\frac{\delta\vartheta}{2}\right)\right) \ ,\label{eq:celvol}
\end{align}
where $r(\kappa)$, $\varphi(\kappa)$, $\vartheta(\kappa)$ are the spherical coordinates of cell $(\kappa)$ in the simulation-centric coordinate system with $\vartheta=0$ the equatorial plane, $\delta r$, $\delta \varphi$, $\delta \vartheta$ are the radial and angular increments of the simulation cells, and the shorthand notation $r_{\mathrm{out,in}}(\kappa)=r(\kappa)\pm\delta r/2$.

For some physical parameters, volumetric integration is not the correct approach. If the observable depends on the length of the LOS instead of the emission volume, as is the case for the column density or the rotation measure, the cells must be weighted by their length $L$ along the LOS instead of their volume $V$. Because $L$ strongly depends on the position and orientation of the cell, an approximation was made to reduce the calculation runtime. Unlike the volumetric integration, where a model cell is treated as a finite volume with uniform physical values such as density or emissivity, a true LOS integration implies an infinitesimal extent in the directions perpendicular to the LOS. For a finite volume there is an infinite number of line segments parallel to the LOS through the volume in question; the length of these line segments along the LOS depends on the problem geometry, but varies widely in general.

In the special case when the centre of a model coincides with the origin ($d_0=0$), all LOS are equal to the radial rows of cells. The length of each cell along the LOS is the radial increment $\delta r$ of the cell, which is equal for all cells. For other geometric arrangements, this is only true for the LOS that goes through the model centre. All other LOS will go through cells at an angle $\alpha(\kappa)$, calculated through 
\begin{equation}
 \alpha(\kappa) = \sqrt{\left(l(\kappa)-l(n)+\frac{1}{2}\delta l\right)^2 + \left(b(\kappa)-b(m)+\frac{1}{2}\delta b\right)^2} \ ,
\end{equation}
with $(b(m), l(n))$ the coordinate of the grid cell $(m,n)$. The corresponding total angular increment is
\begin{equation}
 \delta\alpha(\kappa) = \min \left\{ \left|\frac{\delta b}{\cos\left( x(\kappa)\right)}\right| , \left| \frac{\delta l}{\sin \left( x(\kappa)\right)}\right|\right\} \ ,
\end{equation}
with the shorthand $x=\arctan\left(l(\kappa)/b(\kappa)\right)$, so
\begin{equation}
 \delta\alpha(\kappa) = \min \left\{ \delta b\sqrt{\left(\frac{l(\kappa)}{b(\kappa)}\right)^2+1}\,  ,\ \delta l\frac{\sqrt{l(\kappa)^2+b(\kappa)^2}}{l(\kappa)}\right\} \ .
\end{equation} 
To simplify the problem of a LOS going through a model cell, the latter was approximated by a conical frustum of height $\delta r$ between base and top, an opening angle of $\delta\alpha/2,$ and a central point $r$ (so that the base and top are located at $r\pm\delta r$ along an $r$-axis). The exact model cell takes the shape of a truncated pyramid with a square base and top, where the length of the edges of base and top are equal to the diameters of the base and top of the approximated model cell. The angle $\alpha$ described above is now the angle between the central axis of the cone and the LOS direction. Because the integration method only selects cells with a central point inside the LOS, the length of the LOS through the conical frustrum can be selected as the length of the line segment parallel to the LOS through the cell centre. Thus, the problem is reduced to two dimensions by reducing the conical frustum to an isosceles trapezoid of height $\delta r$ and bottom and top edges $2(r\pm\delta r/2)\sin(\delta\alpha/2)$. The length through the model cell along the LOS now corresponds to the length of a line segment between two of the trapezoid edges through the trapezoid centre. This is further simplified by approximating the trapezoid by a rectangle with edges of the lengths $\delta r$ and $r\delta\alpha,$ which are the lengths of the trapezoid bimedians. In total, the length of a model cell $\kappa$ along the LOS is approximated by 
\begin{equation}\label{eq:loslen}
 L(\kappa) = \min\left\{ \frac{\delta r}{\cos (\alpha(\kappa))} , \frac{r\, \delta\alpha(\kappa)}{\sin(\alpha(\kappa))} \right\} \ .
\end{equation}
The first approximation (square base to circular base of the truncated pyramid) underestimates the trapezoid base length by a factor of $\sqrt{2}$ at most, resulting in an approximation error of $\sqrt{(2(\delta r)^2+(r\delta\alpha)^2)/((\delta r)^2+(r\delta\alpha)^2)}$. The second approximation (trapezoid to rectangle) overestimates the maximum line segment length by less than $\sqrt{(\delta r)^2+(r\delta\alpha)^2}/(r+\delta r/2)$, giving a total maximum approximation error of $\sqrt{2(\delta r)^2+(r\delta\alpha)^2}/(r+\delta r/2)$. The error depends on the distance $r$ of the model cell from its centre and is highest for cells close to the model centre and lowest for those on the outer edge. With the values of Tab.~\ref{tab:boxvals}, this amounts to consistent $10.5\%$ for \textlambda\ Cephei, $5.2\%-8.6\%$ for the heliosphere, $5.2\%-57\%$ for Proxima Centauri, and $3.5\%-17\%$ for V374 Pegasi. To reduce the influence of the approximation error, $L(\kappa)$ was used purely for weighting the cells. The total length of the LOS itself was calculated by adding the radial increment $\delta r$ to the distance between the first and the last cell along the LOS $\Delta s$. Therefore, the result of the LOS integration is
\begin{equation}
 \mathrm{Res}_{\mathrm{LOS}} = \left(\sum\limits_{i=1}^N H_i L_i\right) \left(\sum\limits_{i=1}^N L_i\right)^{-1}(\Delta s + \delta r) \ .
\end{equation}
Both integration methods use the criterion that the angular coordinates $(b_0, l_0)$ of a model cell are inside the respective grid cell. Because the coordinates of a model cell describe the centre of the model cell, the model cell may be positioned so close to the edge of the grid cell that a large part of its volume is outside the grid cell. Likewise, a large part of the simulation cell may be inside the grid cell; as long as the centre is outside the grid cell, the value of its physical parameter is not considered by the integration criterion. It can be assumed, however, that roughly the same number of model cells lie barely outside as inside the cell, so that these effects mostly cancel each other. 

Because we need to transform between spherical and Cartesian coordinates, this effect is exacerbated by numerical uncertainties of the cell coordinates. To compensate for this, a grace interval was introduced: If the angular coordinates of the model cell are only slightly outside the respective grid cell, the model cell was still selected for integration. This grace interval was arbitrarily set to 10\% of the grid cell extent in every direction. However, this led to multiple counting of model cells for different grid cells, resulting in too large volumes, fluxes, etc. In case of the volumetric integration, this error was corrected by calculating the true volume covered by each LOS, changing the results of Eq.~(\ref{eq:volint}) through
\begin{equation}
 \mathrm{Res}_{\mathrm{vol}}^{\mathrm{corr.}}=\frac{\mathrm{Res}_{\mathrm{vol}}}{\sum_{i=1}^N V_i} \cdot V_{\mathrm{LOS}} \ ,
\end{equation}
where $V_{\mathrm{LOS}}$ is the calculated volume of each LOS, approximated by an octahedron. The vertices of this octahedron are the points where the LOS intersects a sphere of the simulation radius ($R_{\mathrm{ISM}}$ in Tab.~\ref{tab:modvals}). When the this grace interval was set to zero, then $V_{\mathrm{LOS}}\approx \sum_{i=1}^N V_i$ with little error, validating this approach. When LOS integration was used instead of volumetric integration, the result was divided by the total number of integrated-over cells and multiplied with the total number of cells, thereby correcting for cells that were selected for multiple lines of sight.

In addition to the integration criterion, other parameters can be used to exclude specific cells: the number density $n(\kappa)$, the temperature $T(\kappa),$ and the distance to the origin $d(\kappa)$. This is useful for selecting or disregarding certain regions for integrations, for instance, the outer astrosheath.

The results of the integration are displayed in an orthogonal grid by default, with the displayed grid cells corresponding to the integration grid cells, and the value of the physical parameter represented by colour with a logarithmic colour scale. When the angular extent of the projection is large, the grid can be projected stereographically or in a Hammer projection. In practice, the stereographic projection is not necessary for most cases, while the Hammer projection is a much better depiction of all-sky projections ($d_0=0$) than the orthogonal grid.

Because the simulations only extend over a finite volume, the modelled astrospheres are cut off at the outer border of the simulated domain. Parts of the astrosphere structure were therefore not simulated and not taken into account for the projections. However, for all four examined models, all expected structures lie at least partially inside the simulated volume. Thus, a larger simulated domain is unlikely to yield any new fundamental results but would only quantitatively alter the projections.

\section{Results}\label{sec:res}
The projection methods are applied to the models of \textlambda~Cephei in Sec.~\ref{sec:lcb} and of the heliosphere in Sec.~\ref{sec:h3d}. Because the different types of radiation are the most interesting observables that are currently included in the projection methods, we focus on their fluxes. The rotation measure as an observable is also examined. A more limited investigation of the models of Proxima Centauri and V374 Pegasi is presented in Sec.~\ref{sec:pxbpeg}. 

\subsection{\textlambda~Cephei}\label{sec:lcb}

\begin{figure}
 \centering
 \includegraphics[width=0.5\columnwidth]{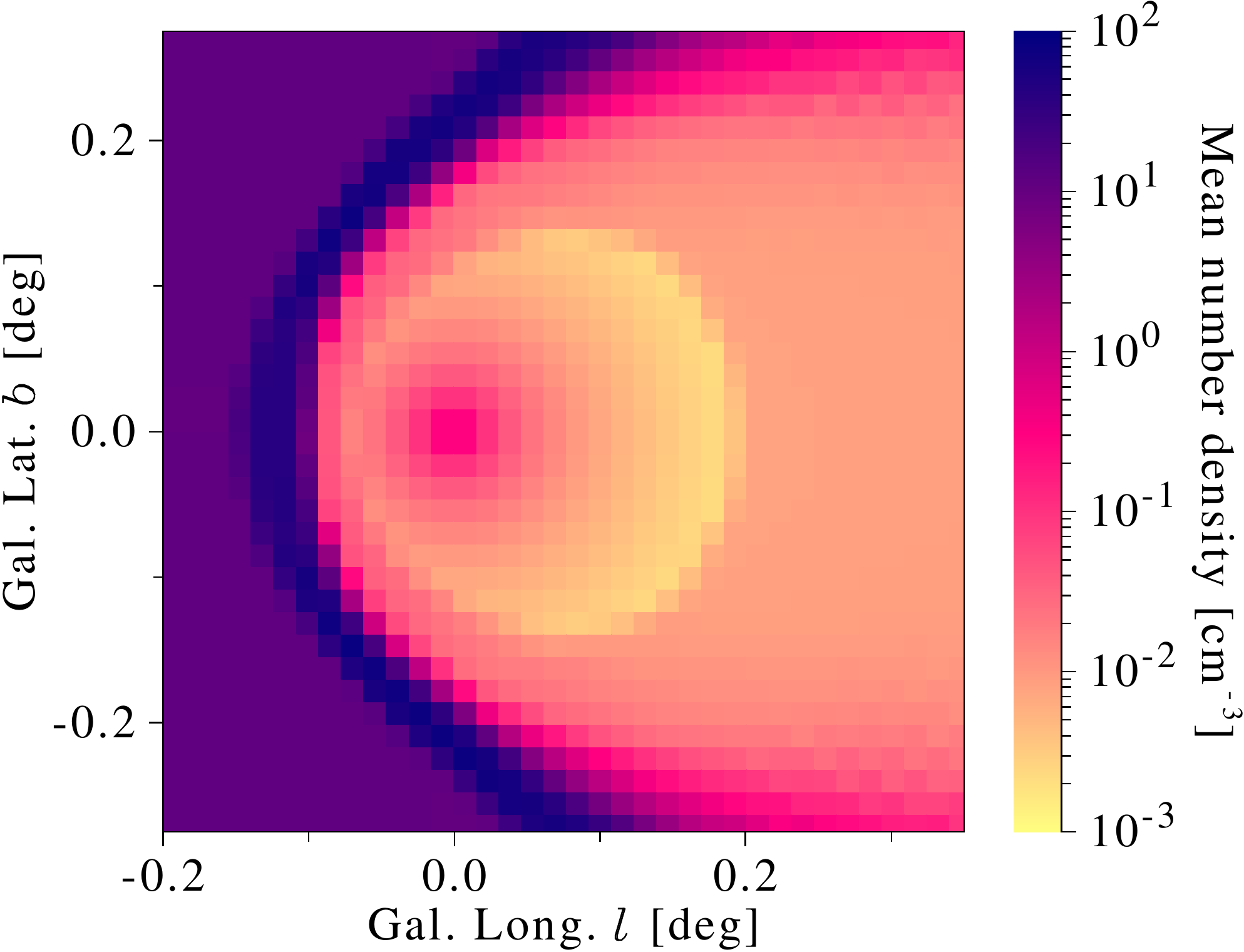}\hfill
 \includegraphics[width=0.5\columnwidth]{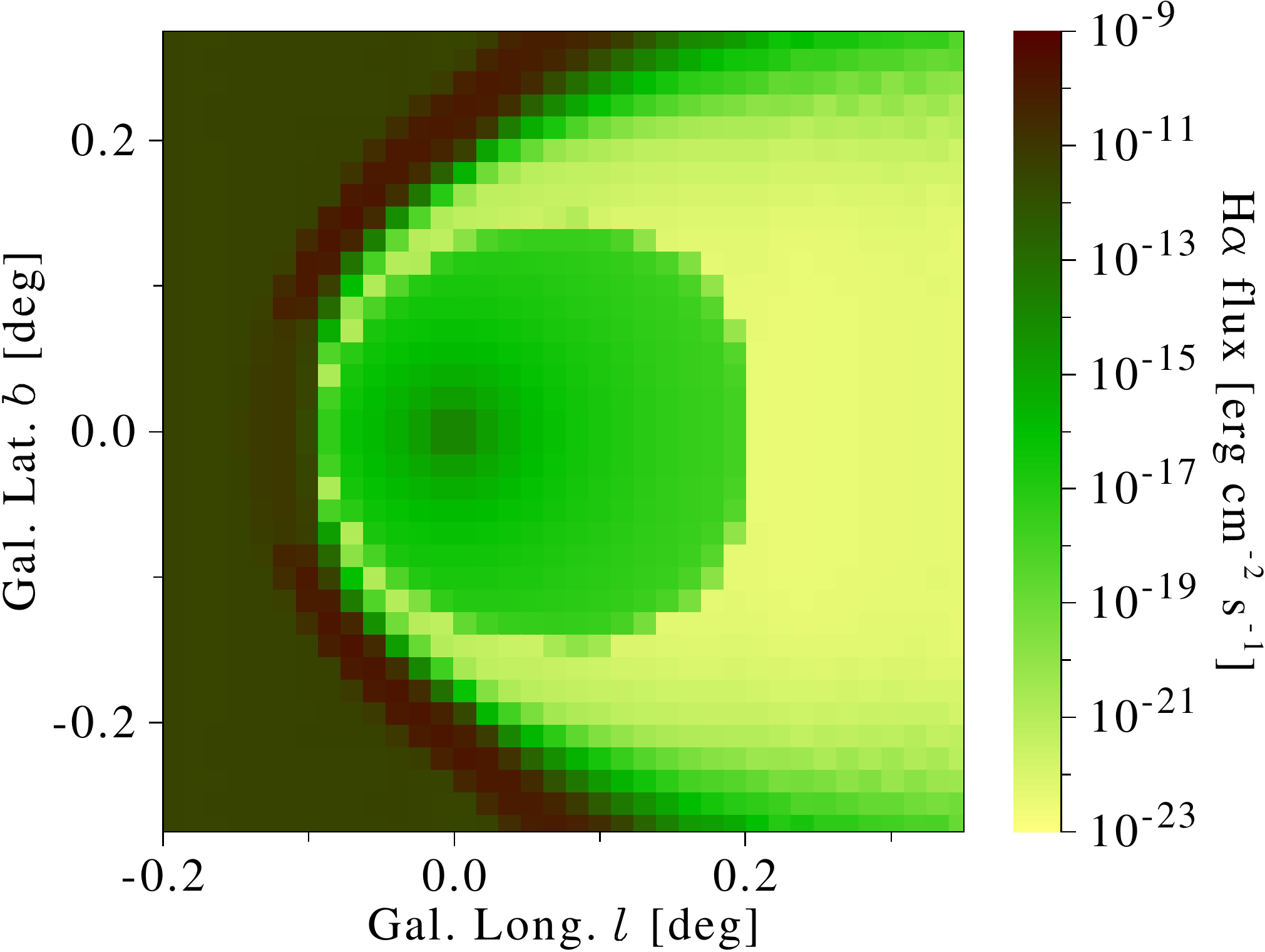}
 \caption{Mean number density (left) and H\textalpha\ flux (right) of a flat box at the \textlambda~Cephei model central $xz$-plane with a thickness of $0.6\,\si{pc}$.}\label{fig:eclip}
\end{figure}

\begin{figure*}
 \centering
 \includegraphics[scale=0.31]{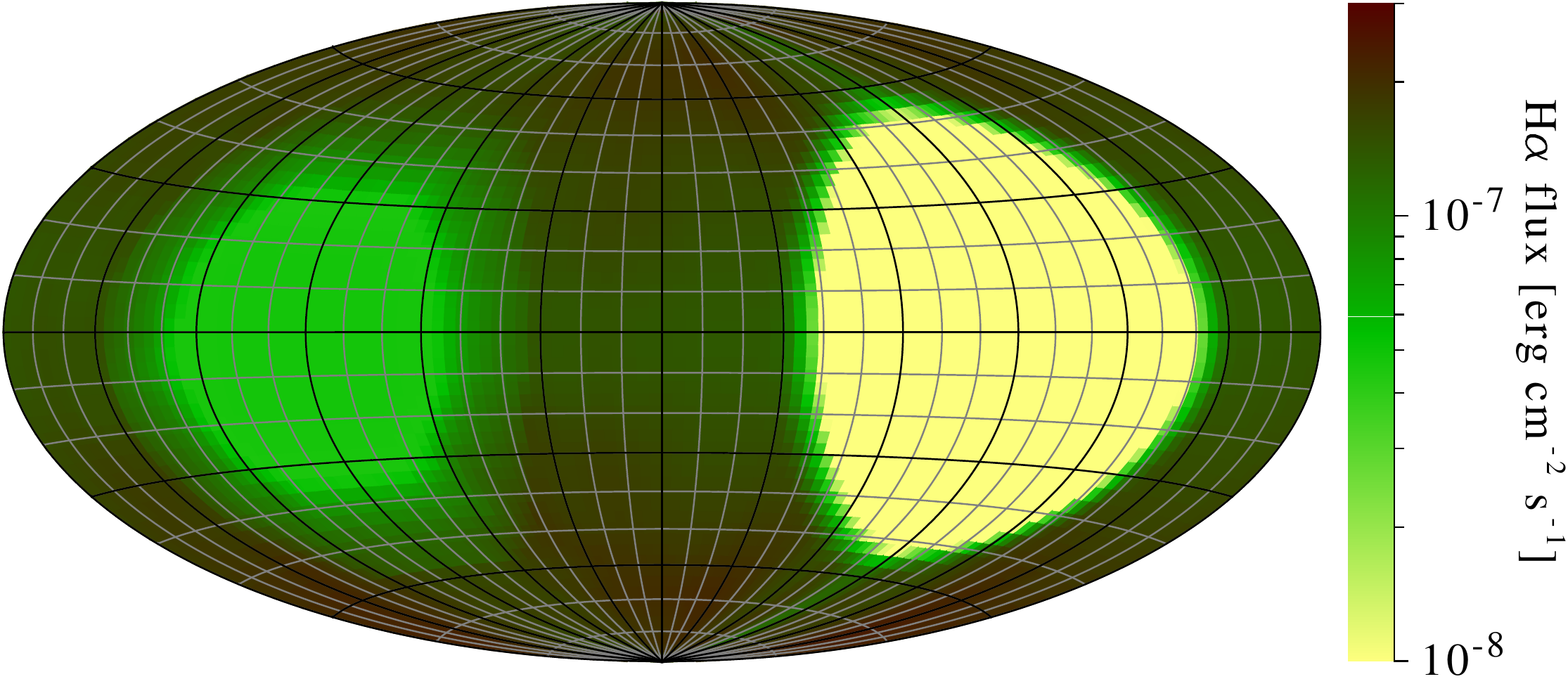}\hfill
 \includegraphics[scale=0.22]{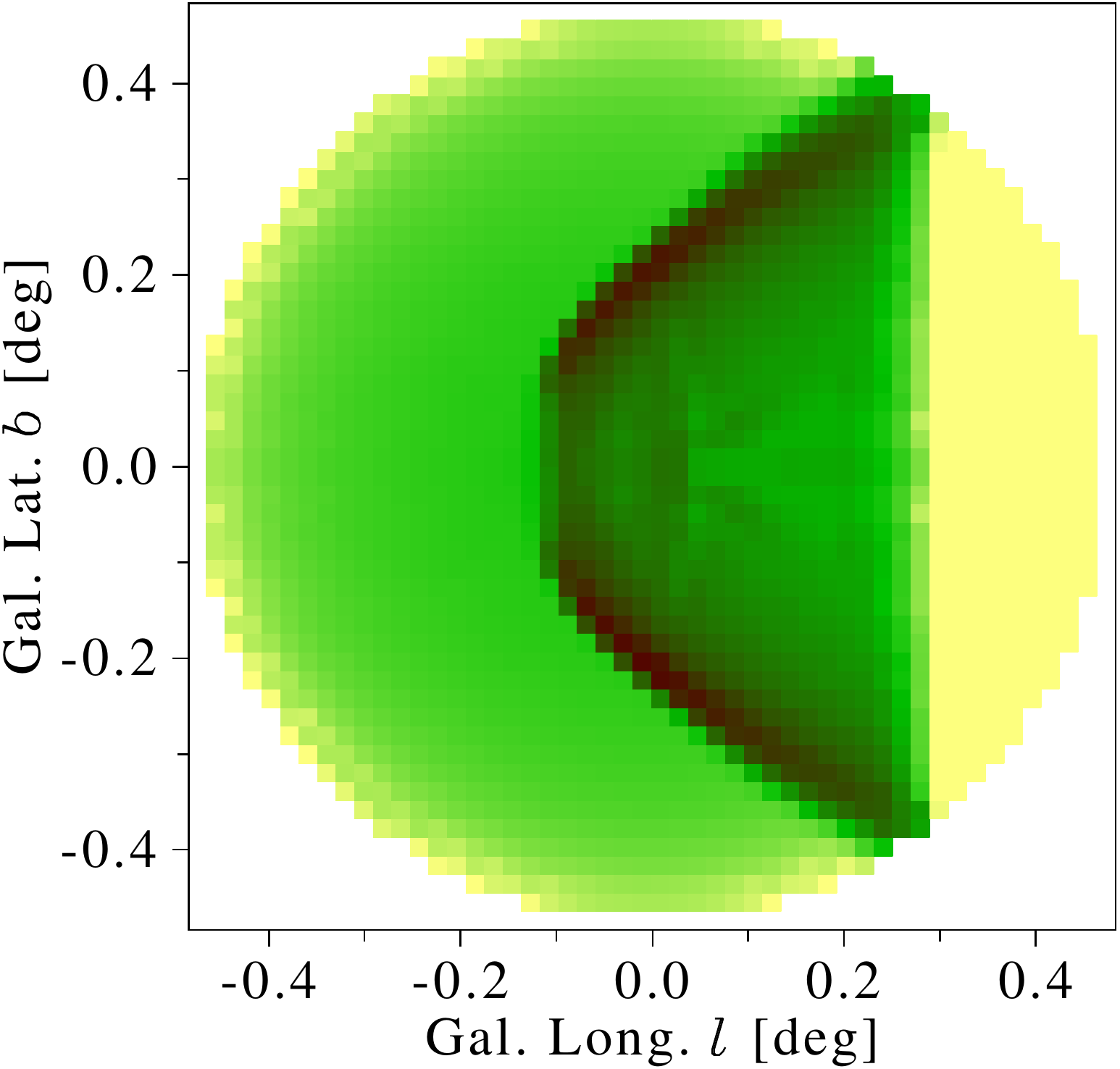}\hfill
 \includegraphics[scale=0.22]{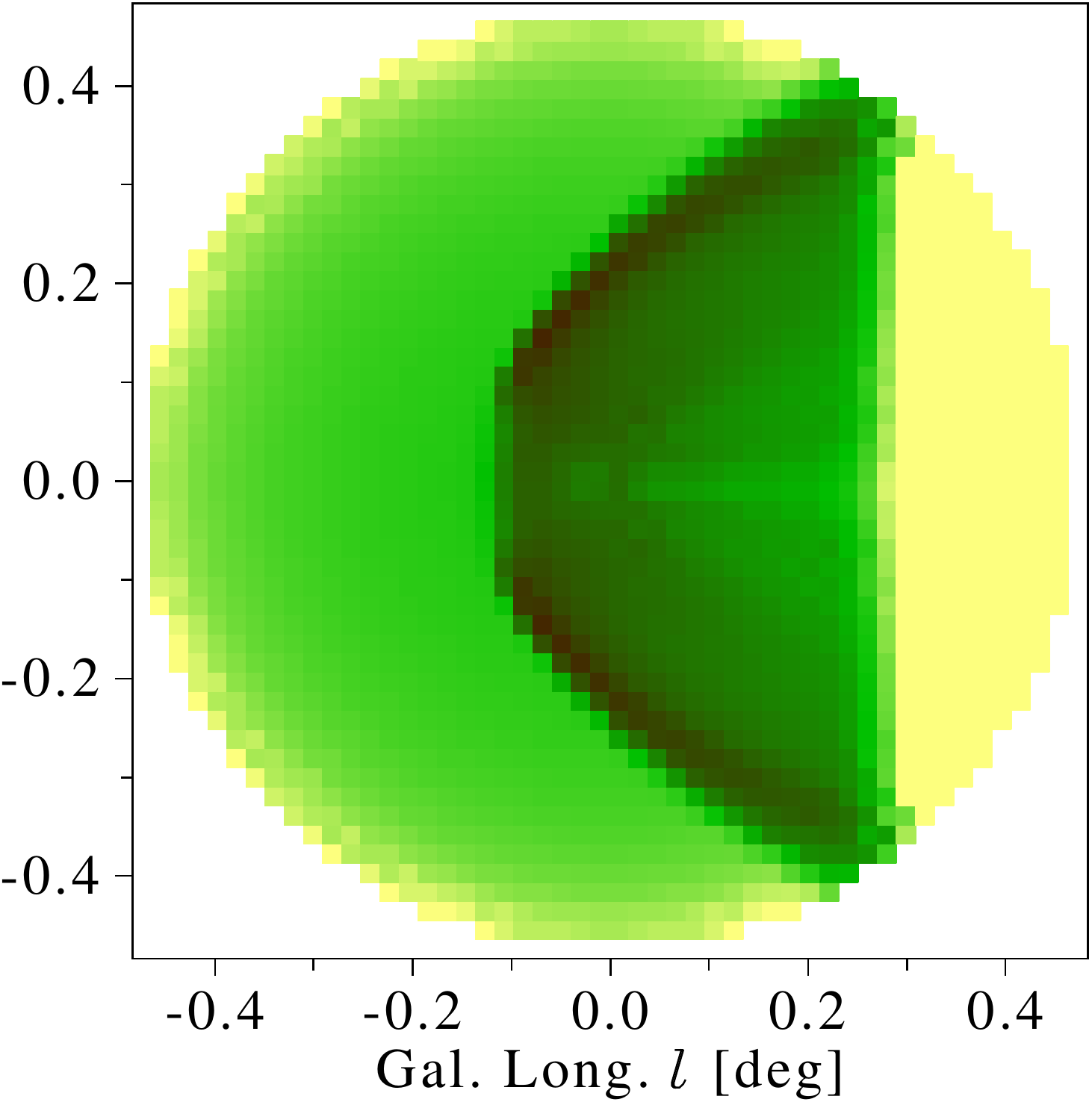}\hfill
 \includegraphics[scale=0.22]{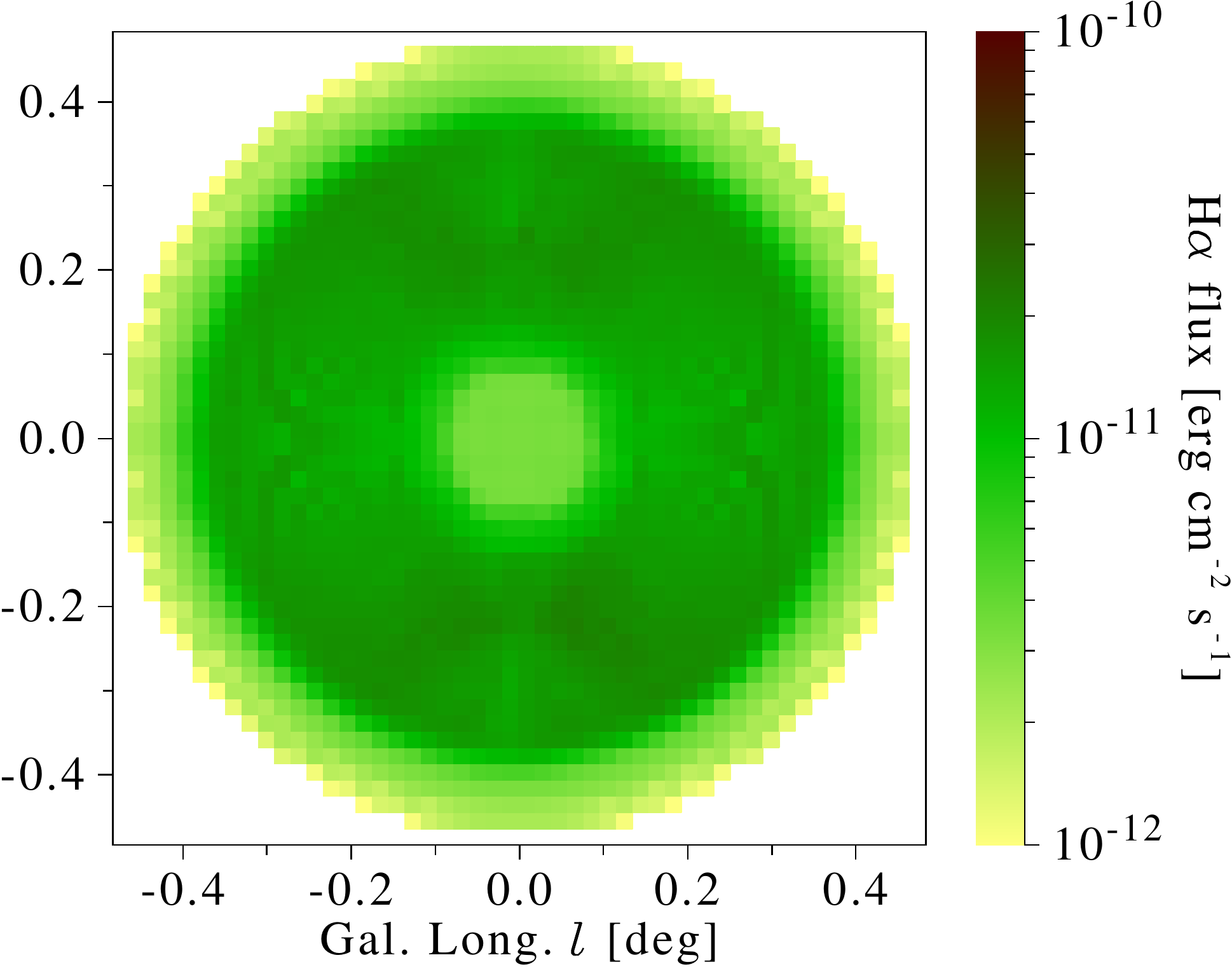}\\
 \includegraphics[scale=0.31]{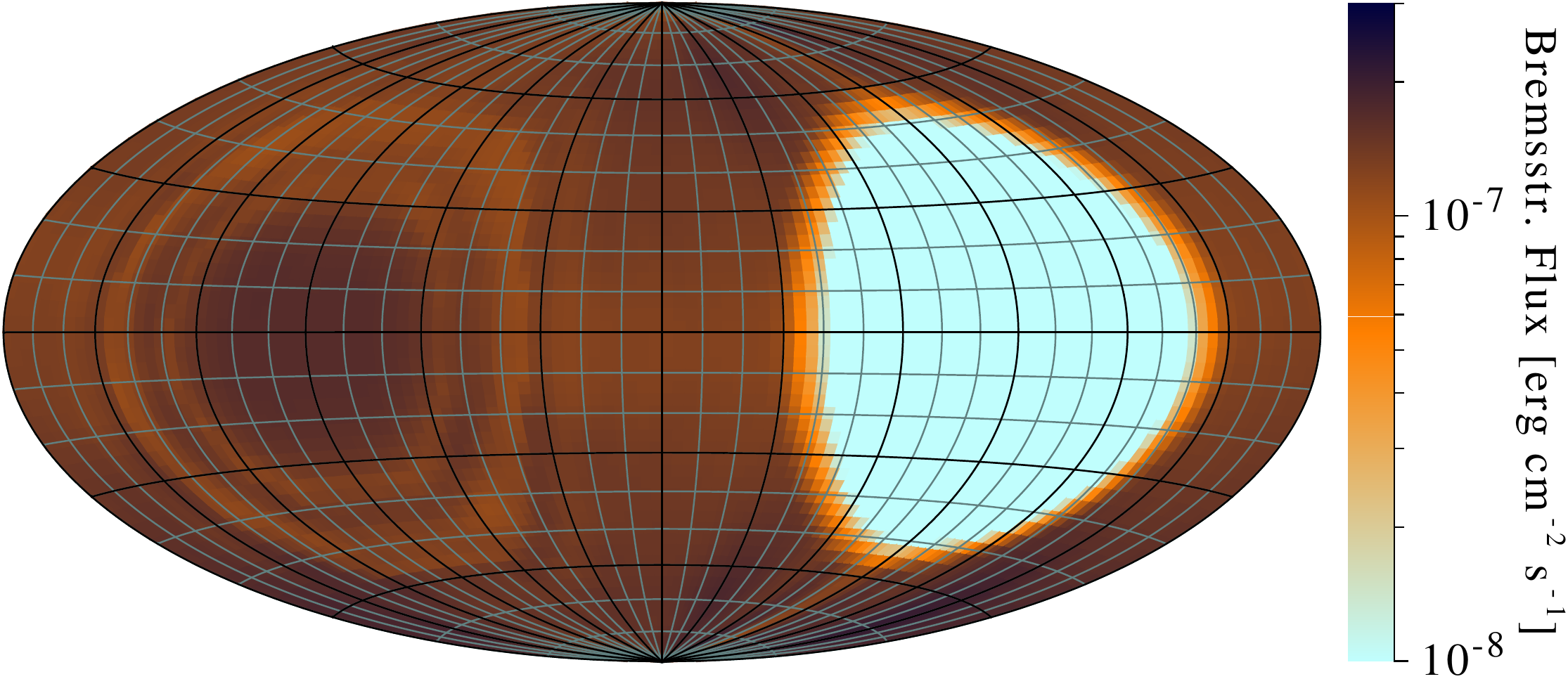}\hfill
 \includegraphics[scale=0.22]{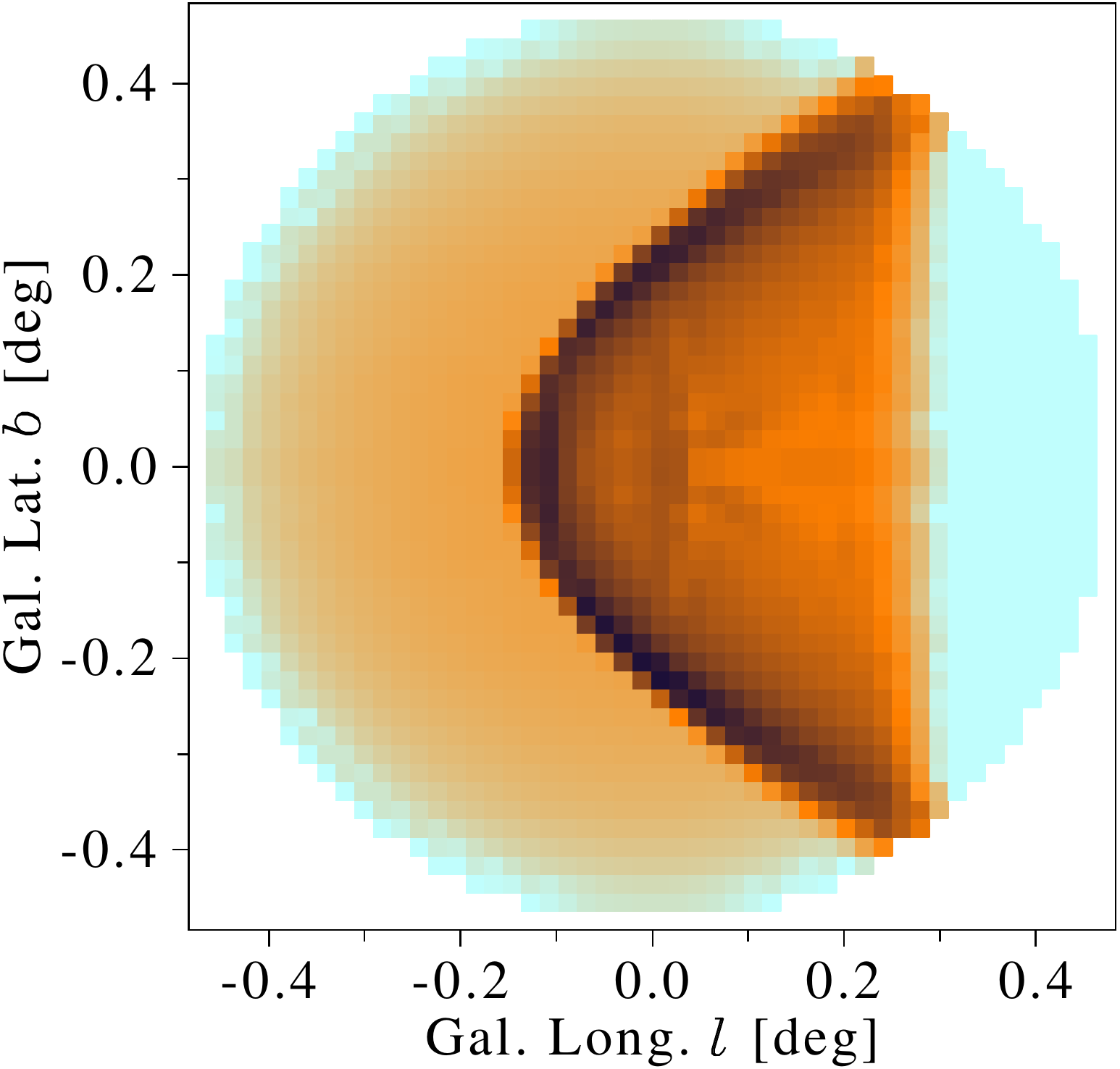}\hfill
 \includegraphics[scale=0.22]{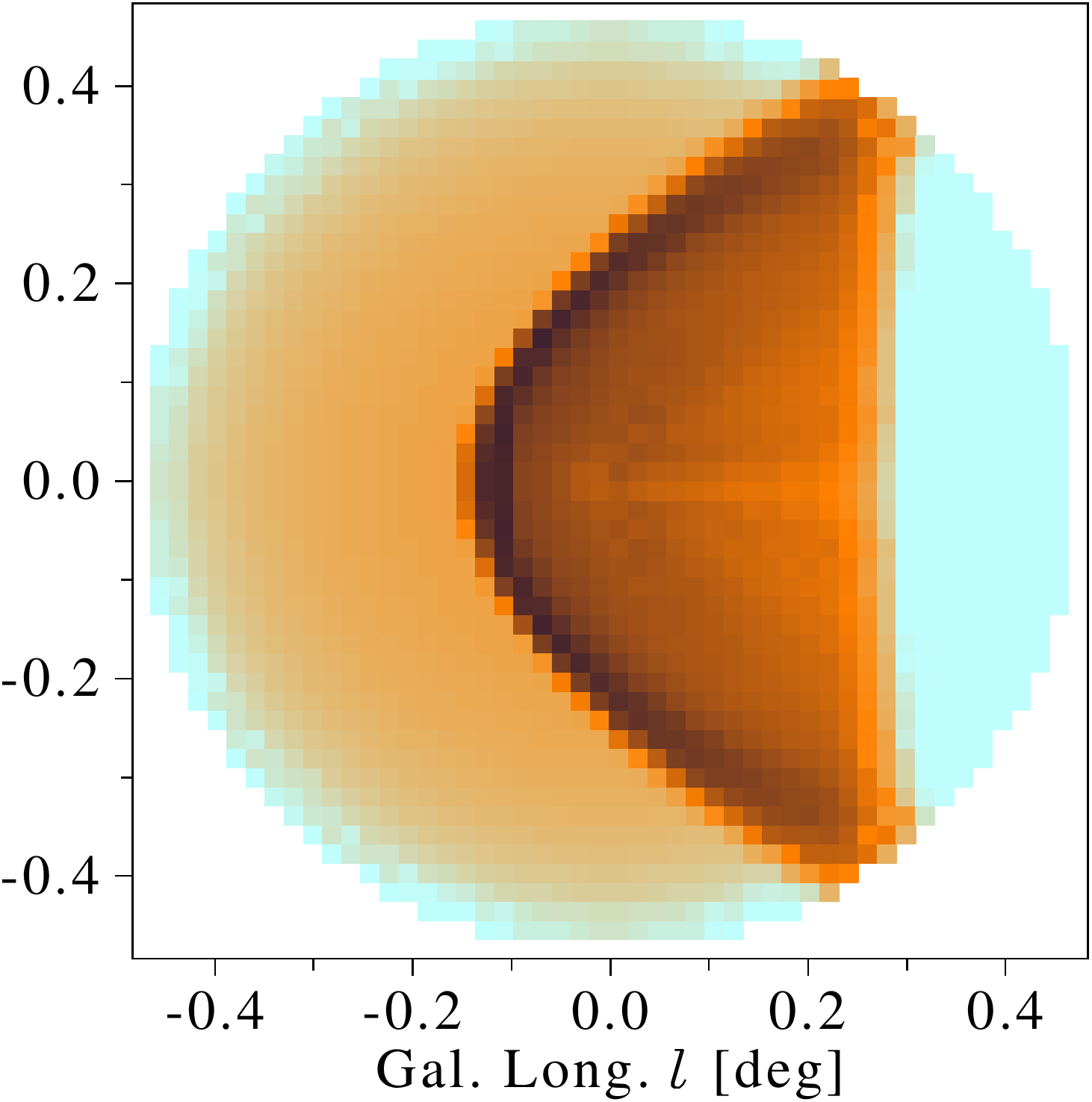}\hfill
 \includegraphics[scale=0.22]{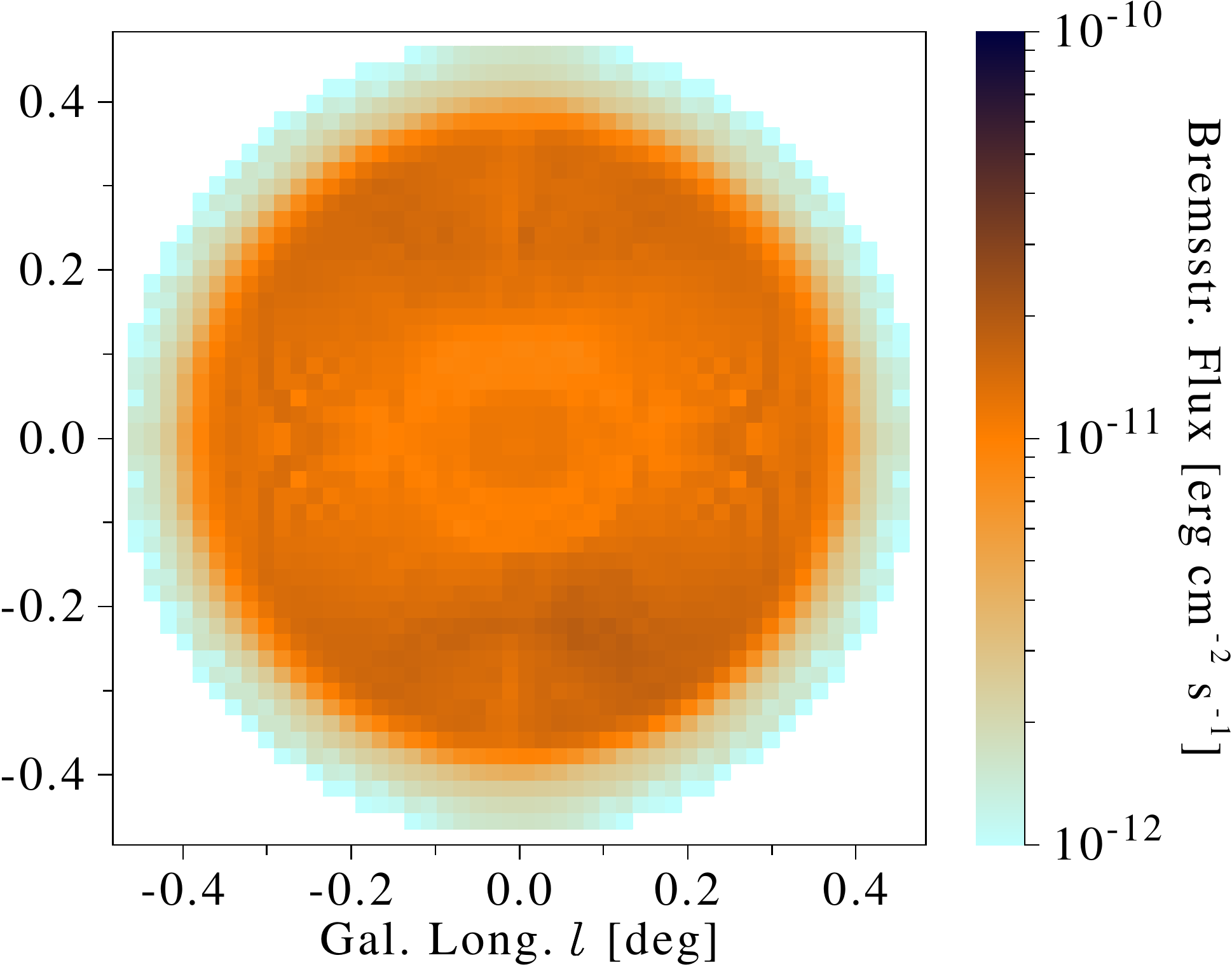}\\
 \includegraphics[scale=0.31]{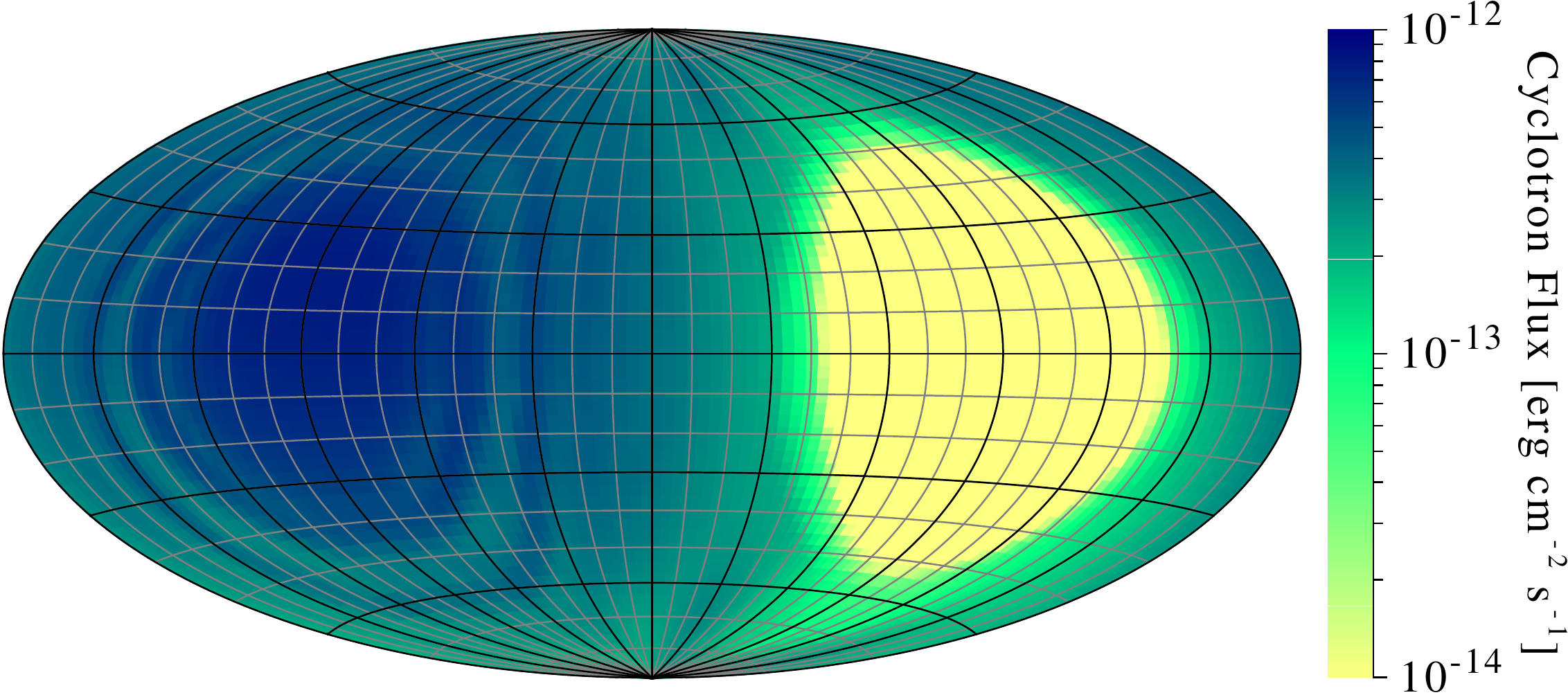}\hfill
 \includegraphics[scale=0.22]{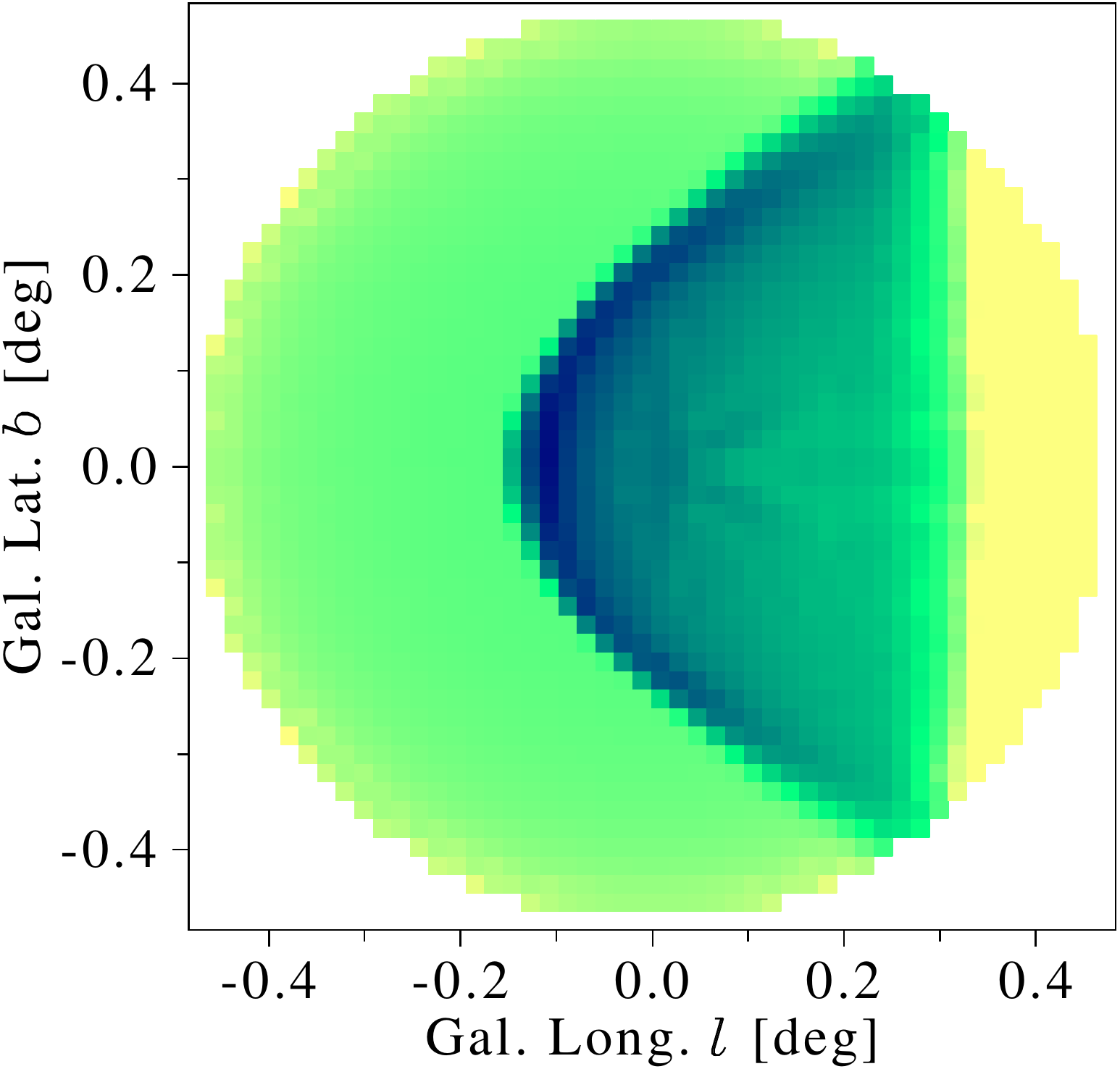}\hfill
 \includegraphics[scale=0.22]{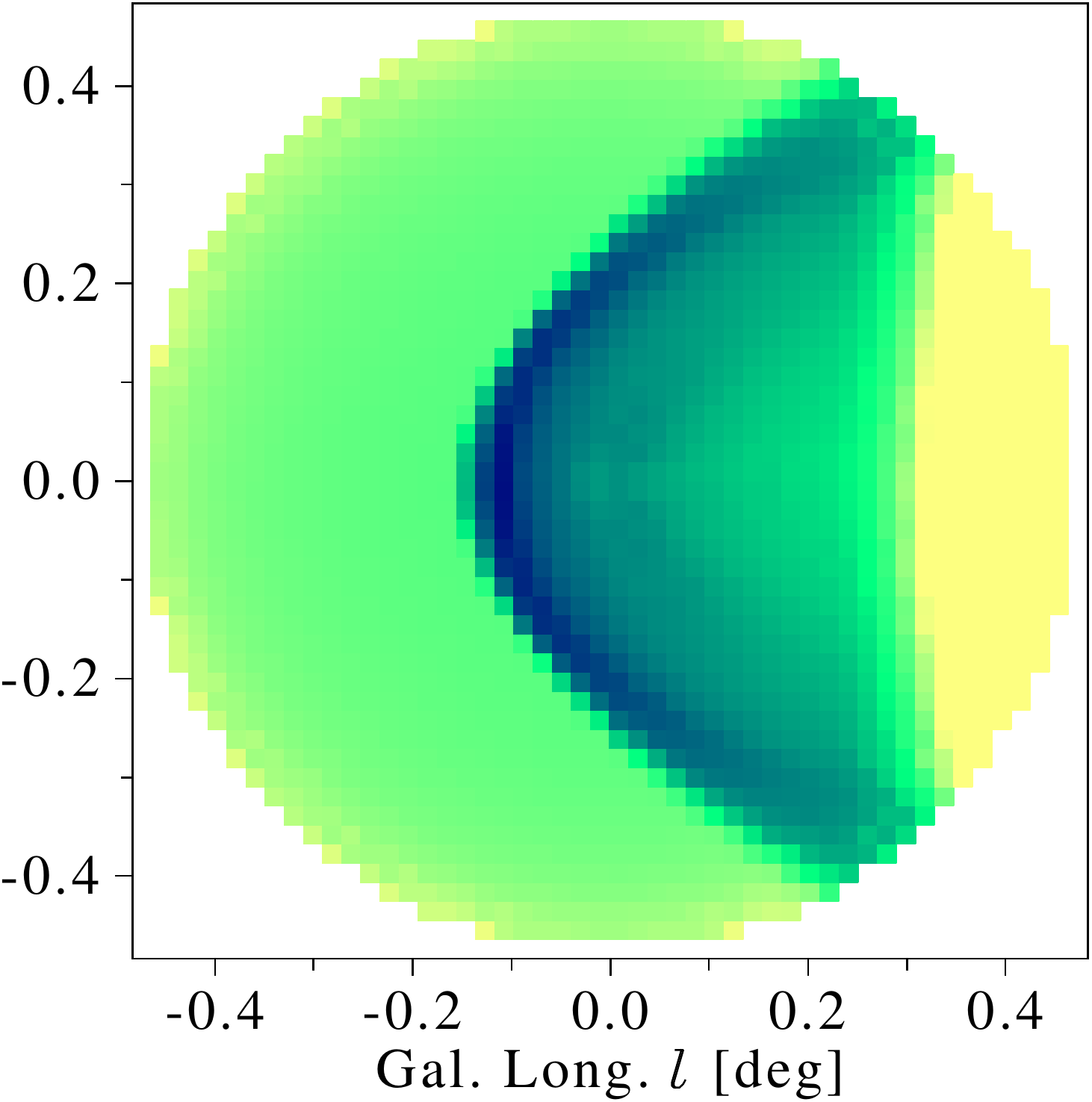}\hfill
 \includegraphics[scale=0.22]{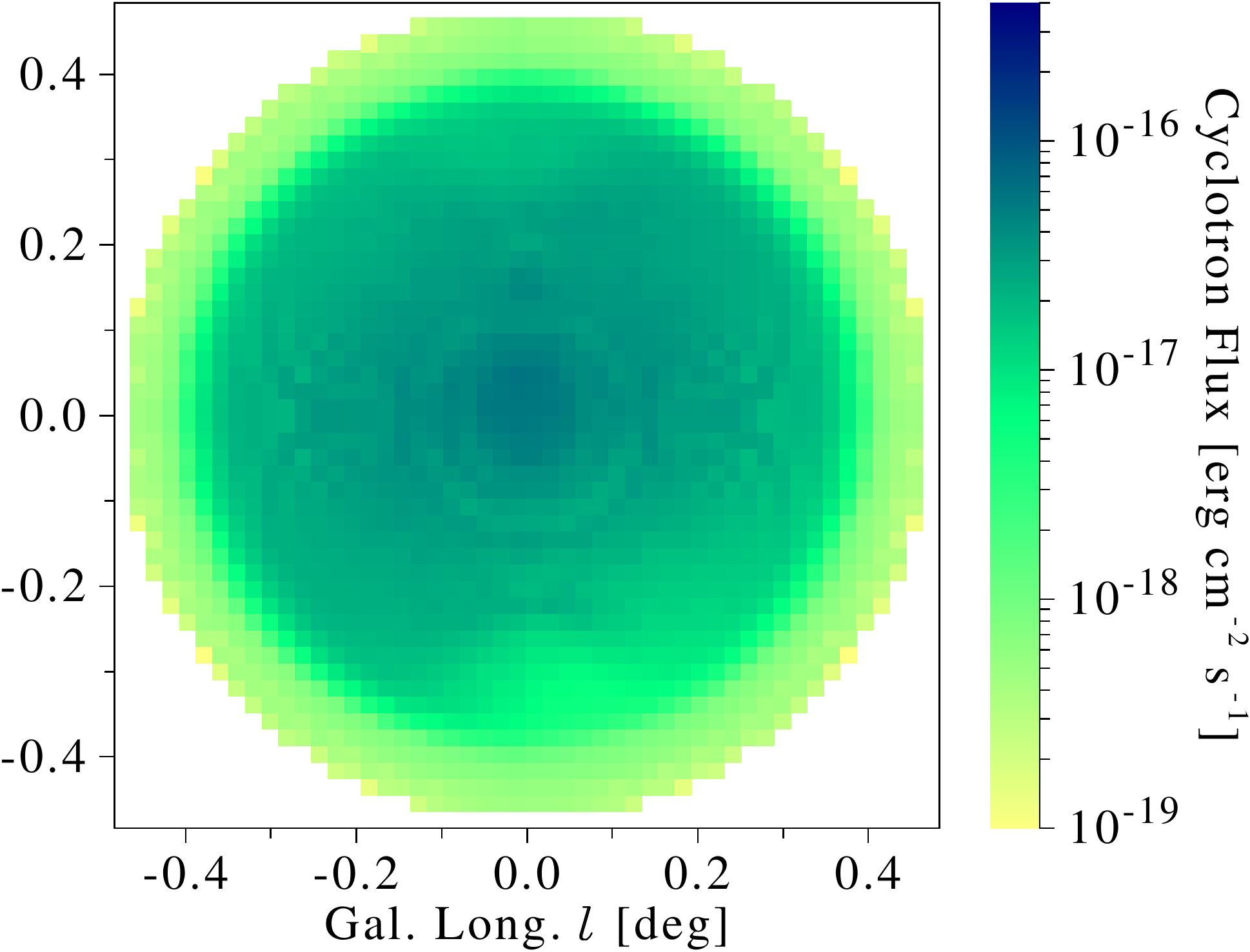}\\
 \includegraphics[scale=0.31]{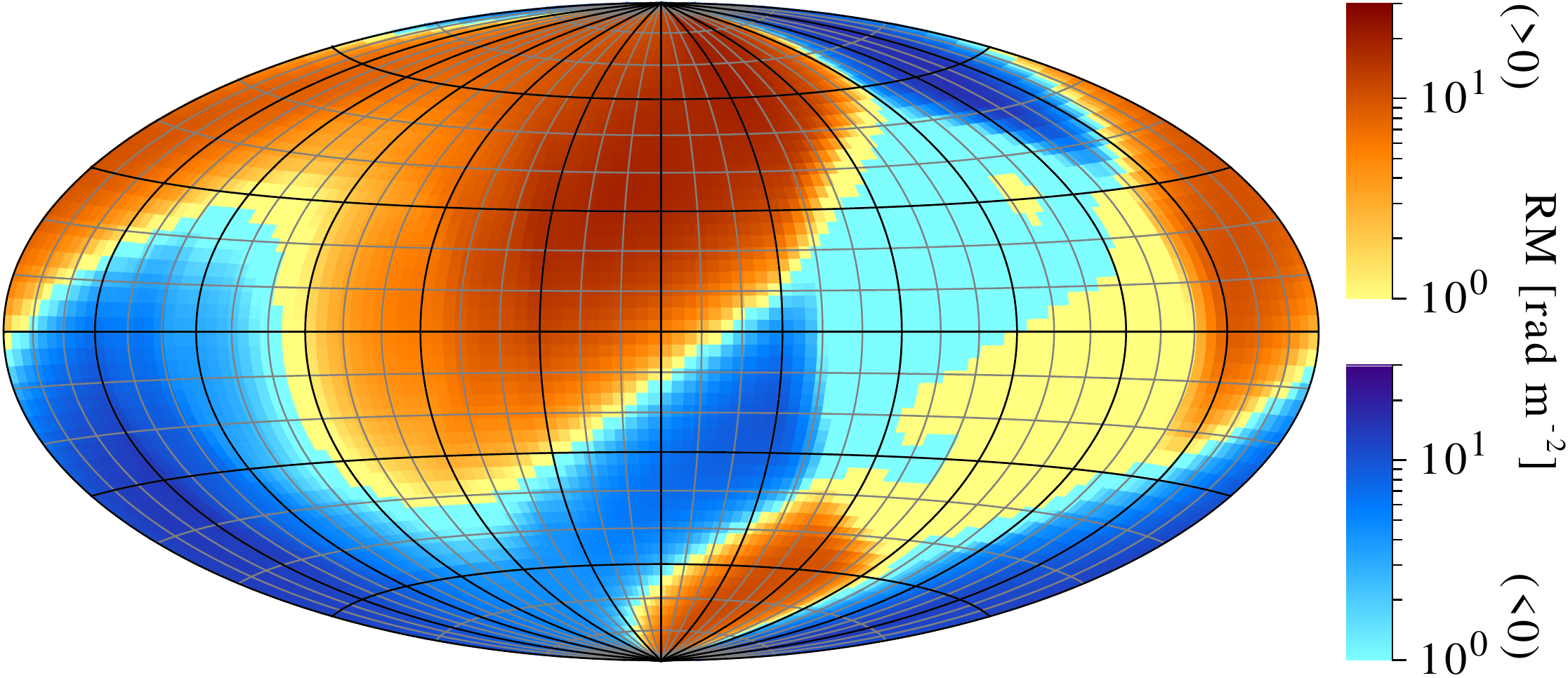}\hfill
 \includegraphics[scale=0.22]{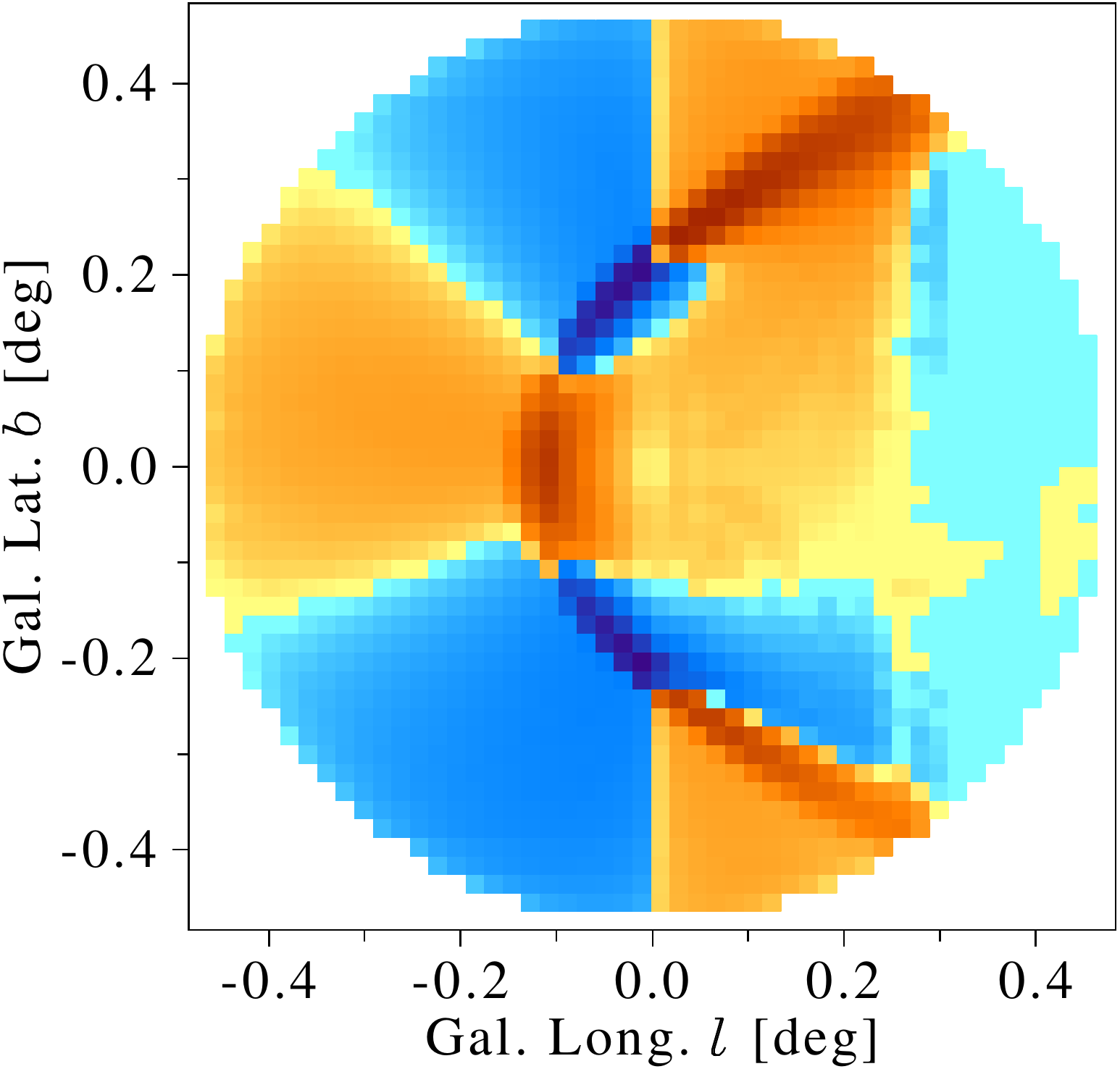}\hfill
 \includegraphics[scale=0.22]{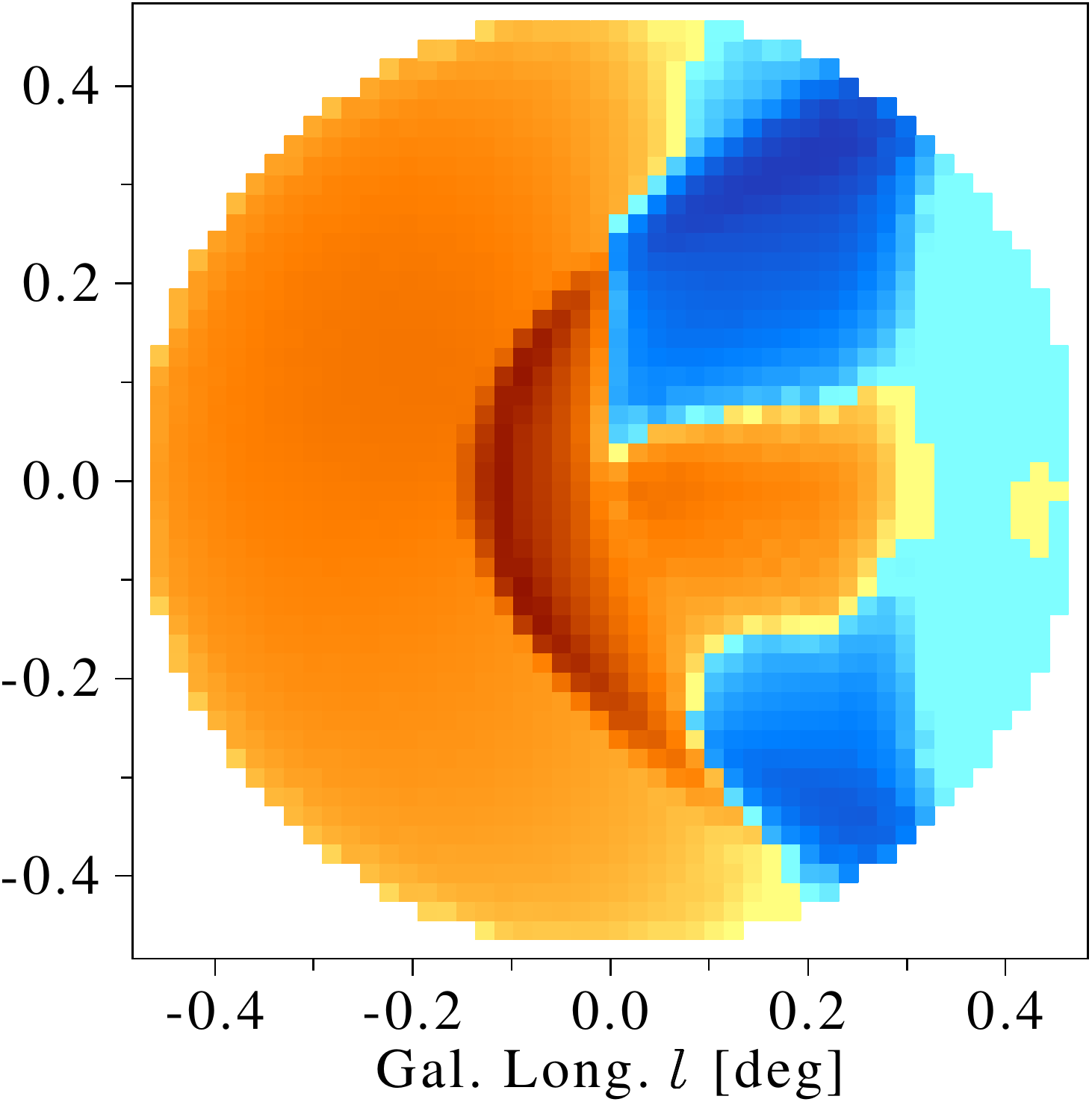}\hfill
 \includegraphics[scale=0.22]{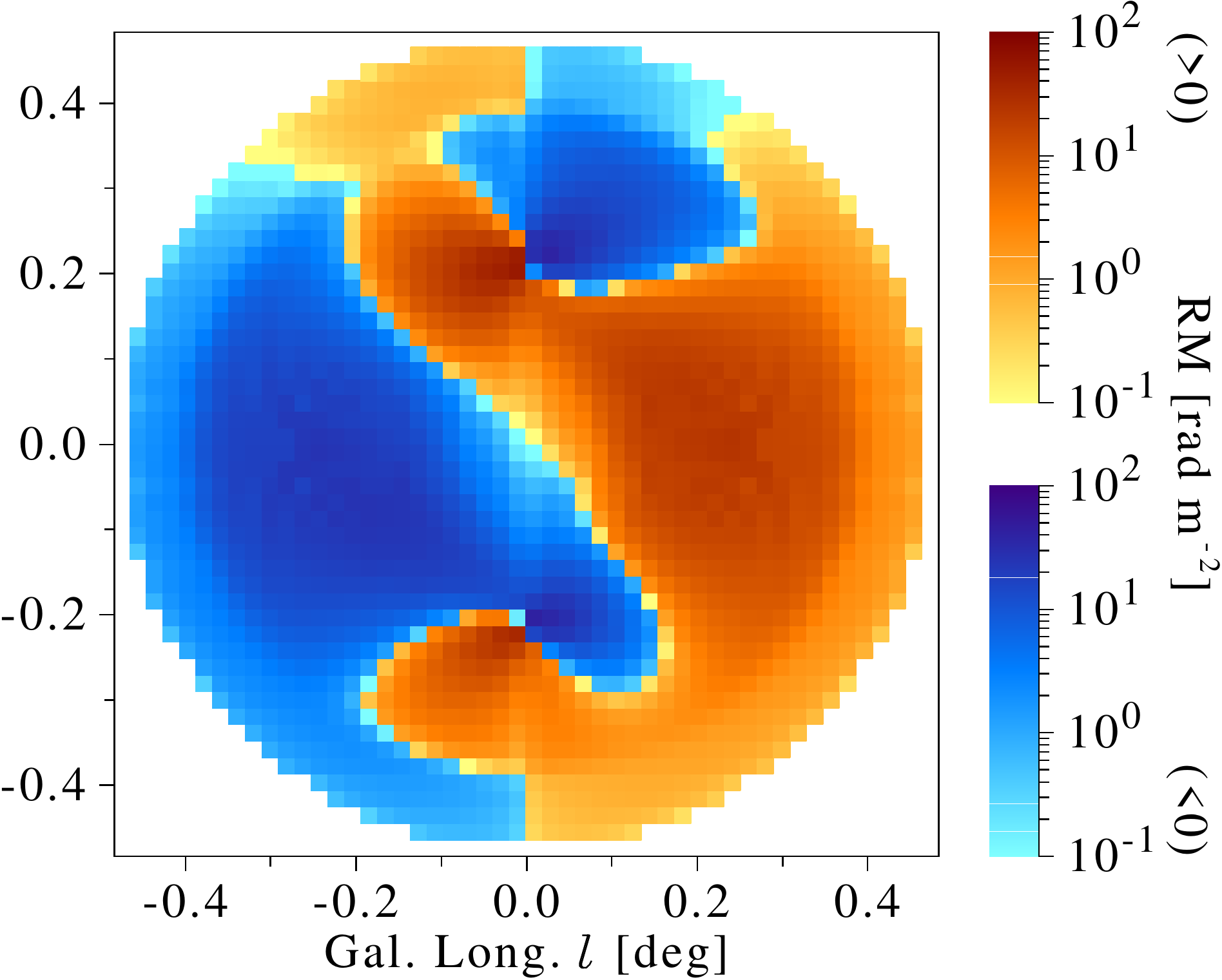}
 \caption{Projections of the \textlambda~Cephei model in (from top to bottom) H\textalpha\ flux, bremsstrahlung flux, cyclotron flux, and rotation measure at distances $d_0=0$ (left column) and $d_0=617\,\si{pc}$ (right columns), projected onto (from left to right) the sky and the $xz$-, $xy$-, and $yz$-planes. Values lower than the minimum value of the colour scales are plotted in the colour of that minimum value.}
 \label{fig:lcbproj}
\end{figure*}

\begin{figure*}
 \centering
 \includegraphics[scale=0.22]{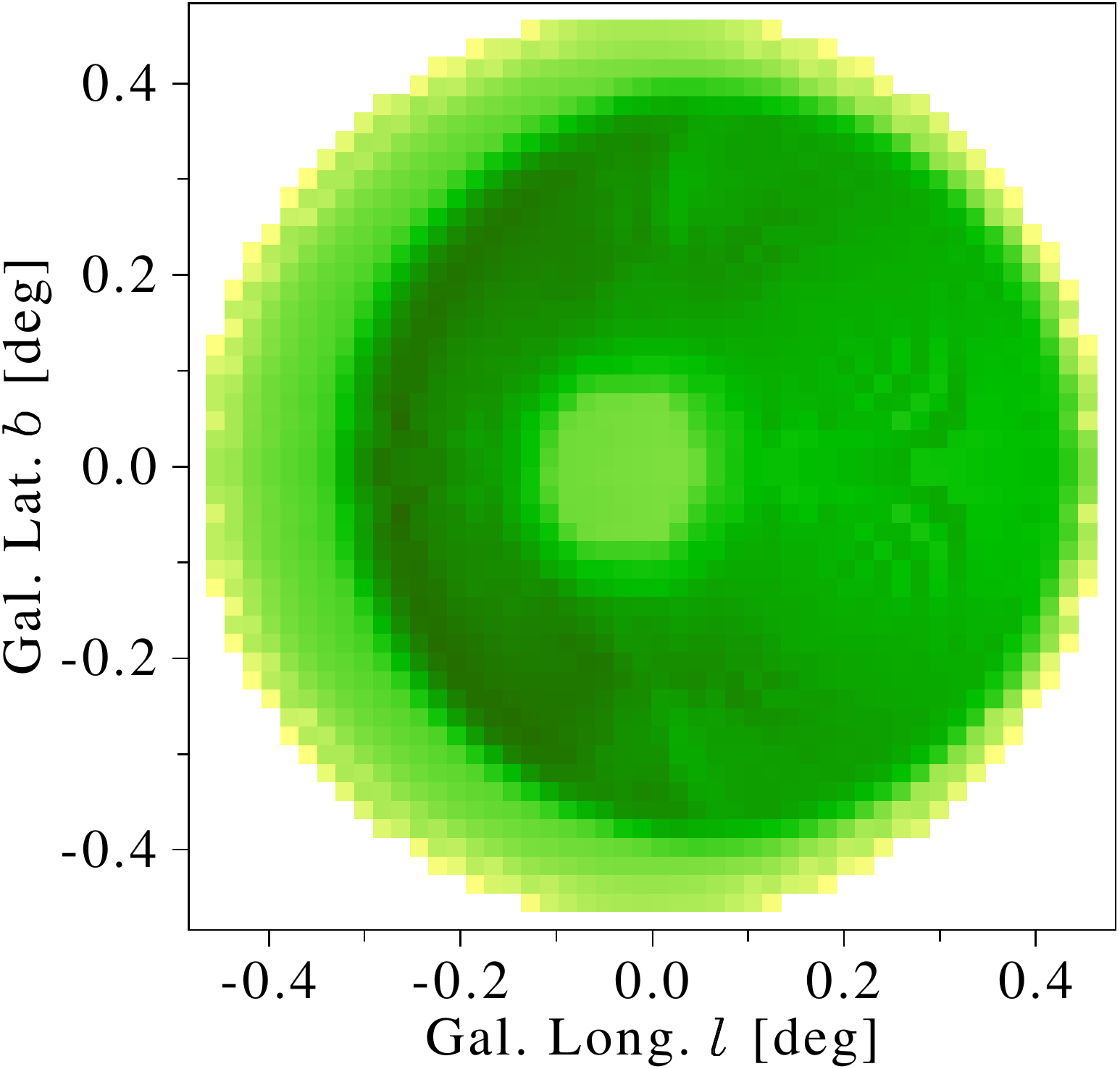}\hfill
 \includegraphics[scale=0.22]{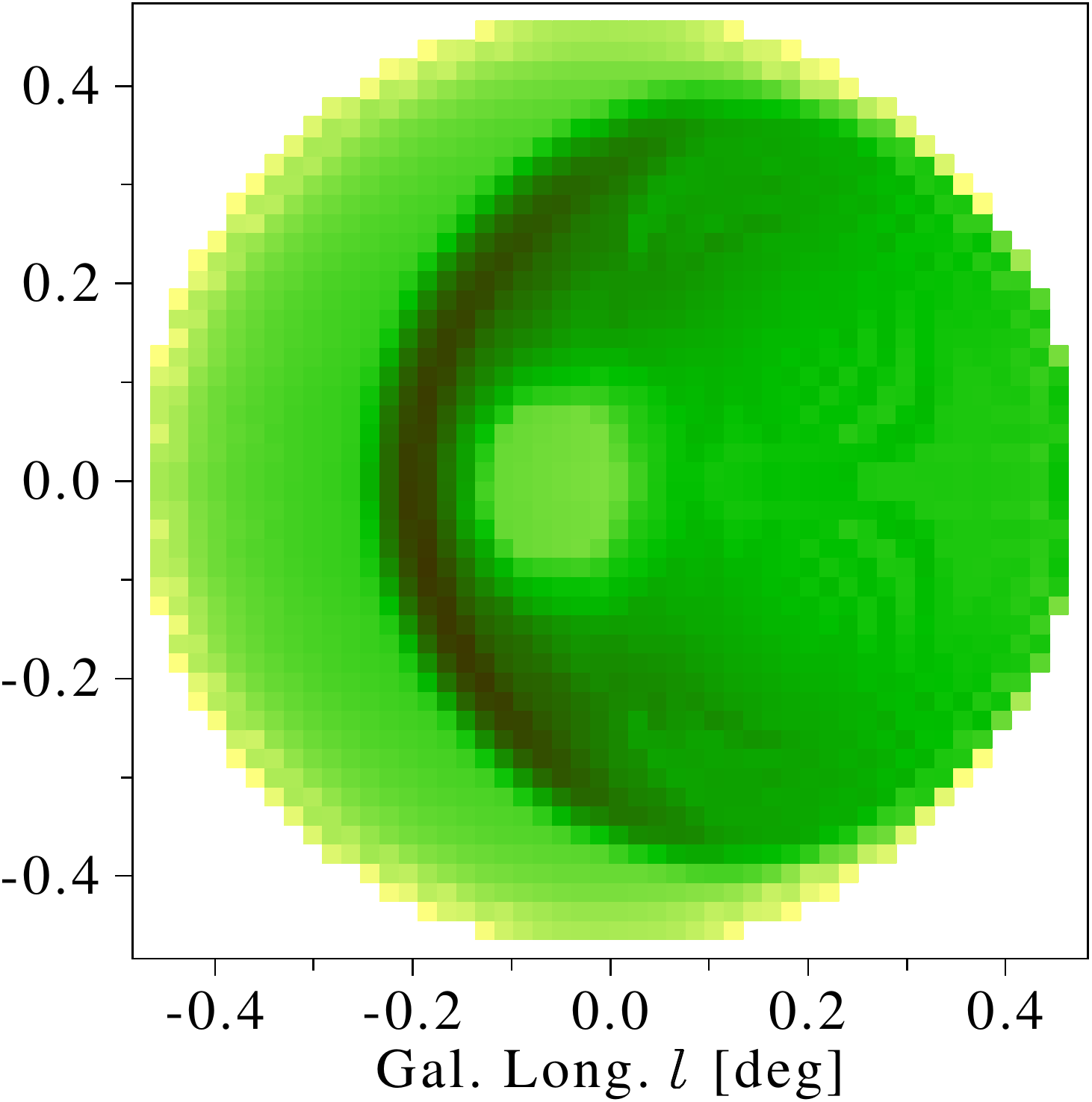}\hfill
 \includegraphics[scale=0.22]{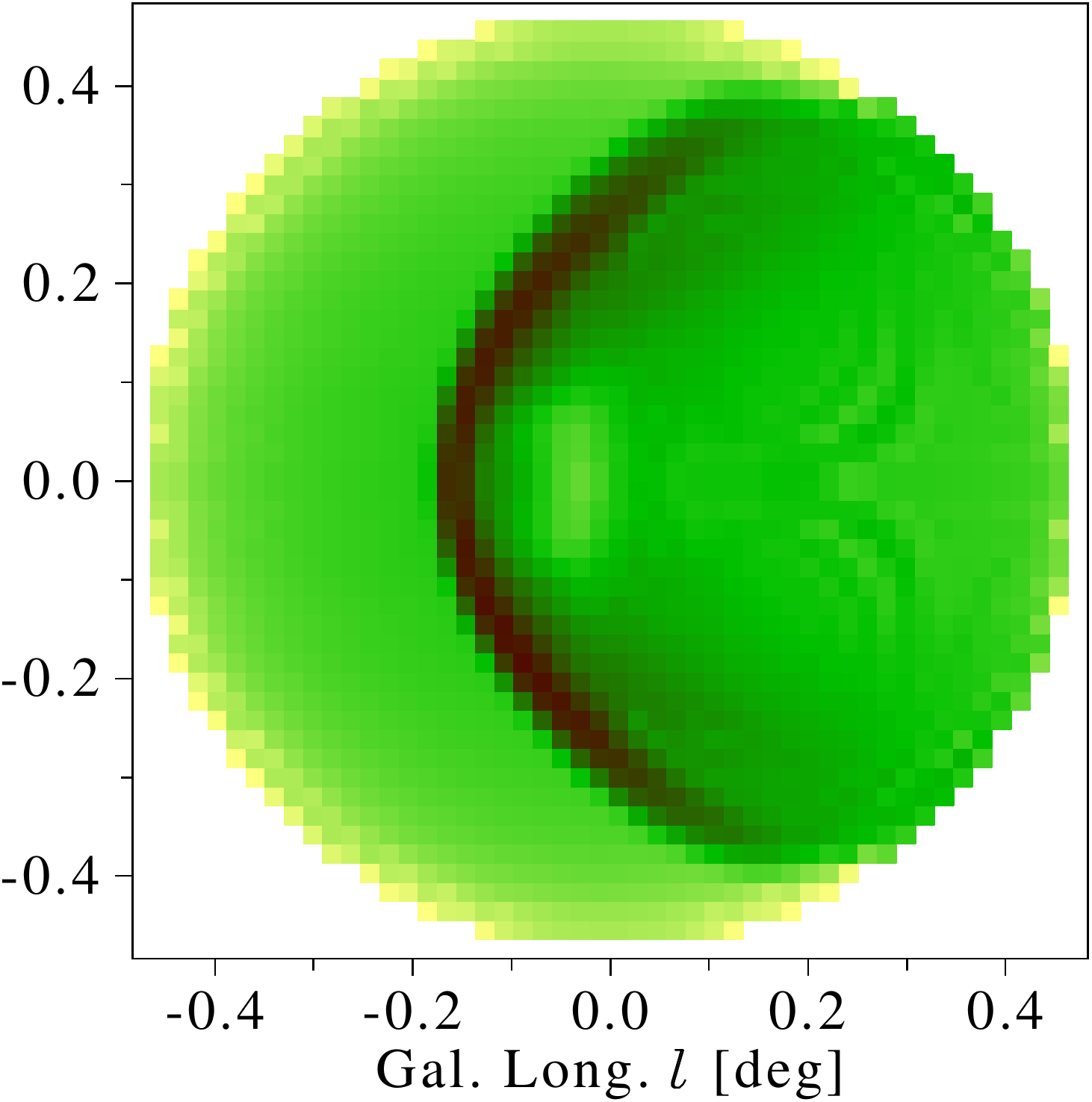}\hfill
 \includegraphics[scale=0.22]{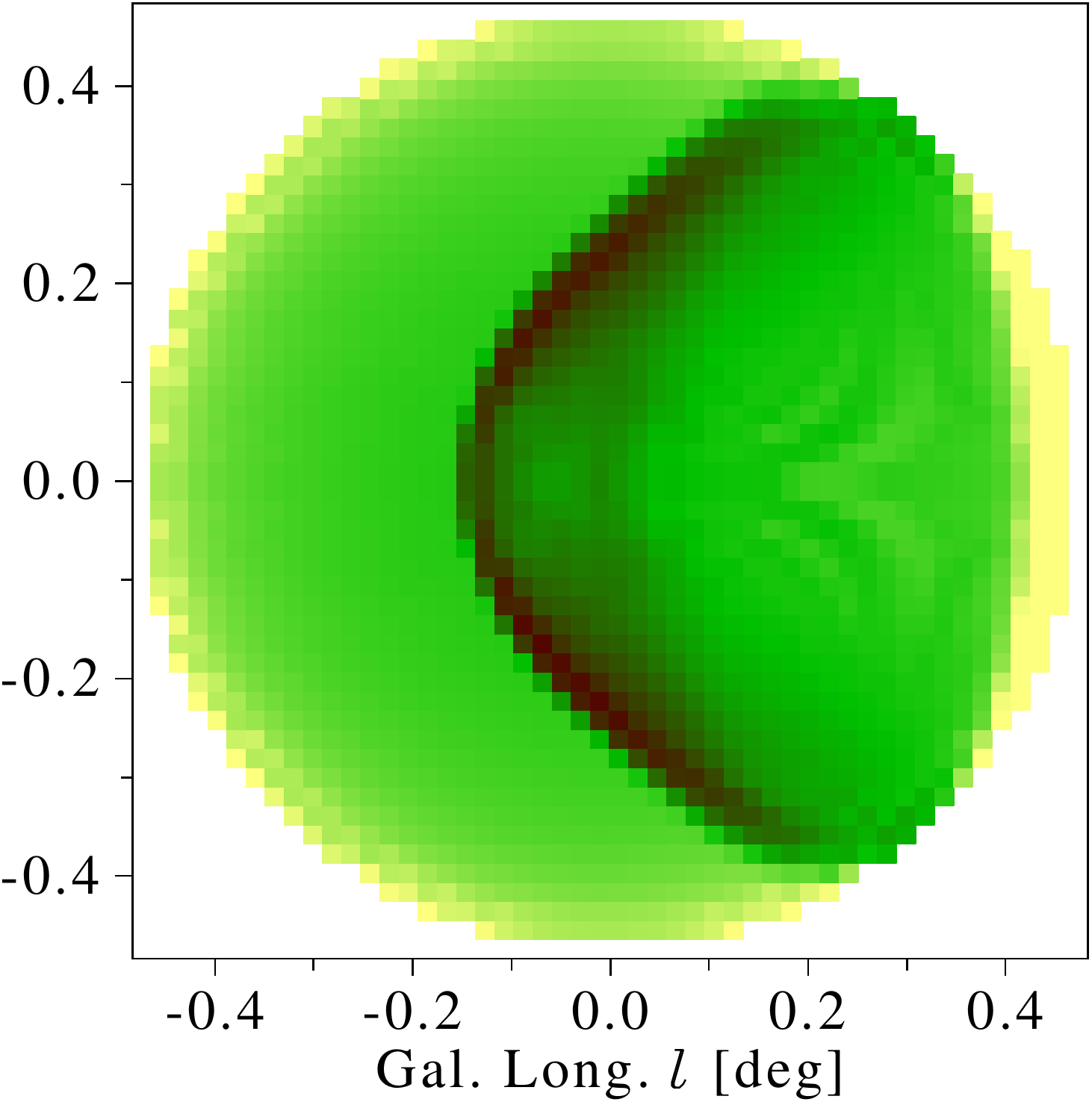}\hfill
 \includegraphics[scale=0.22]{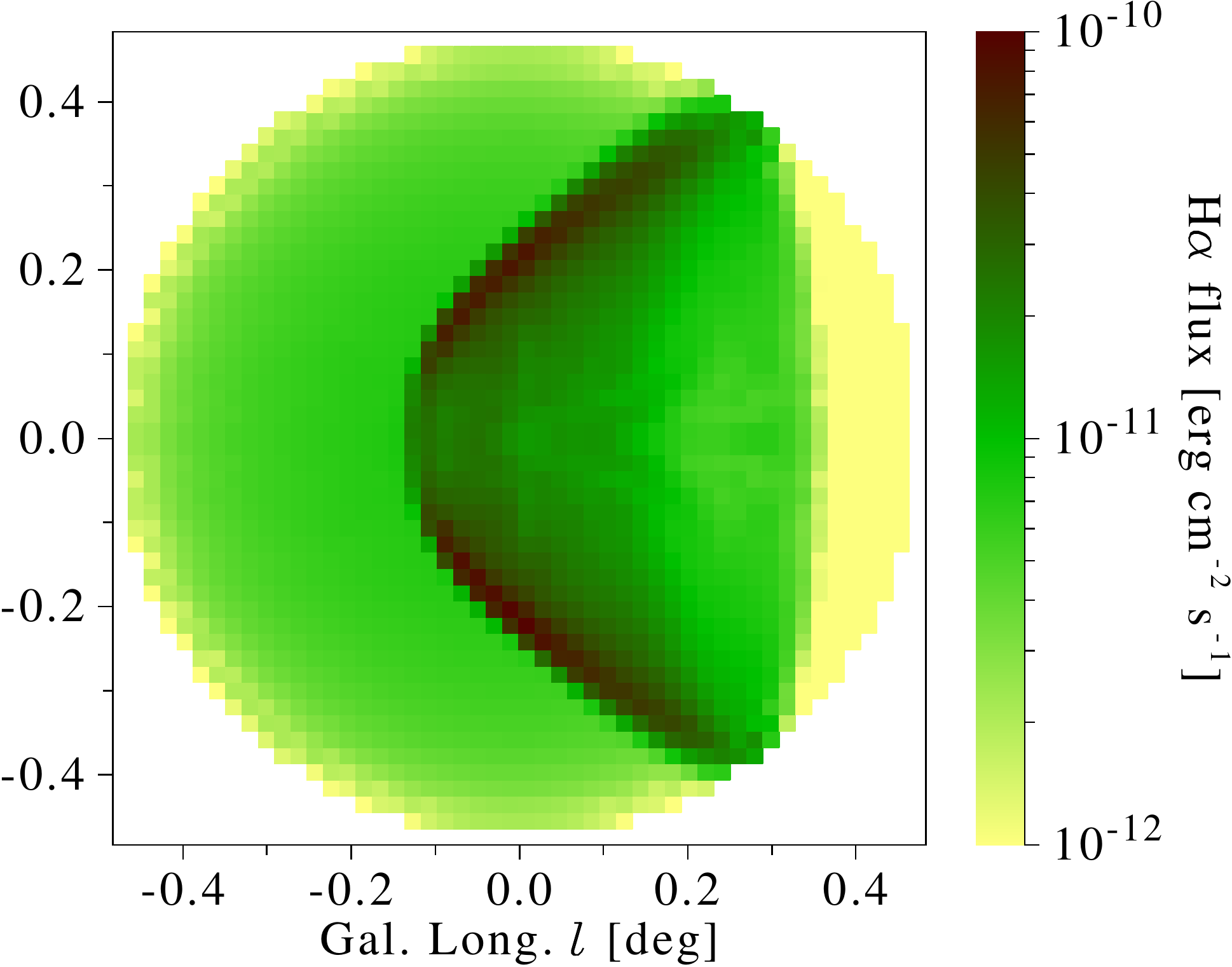}
 \caption{Further projections of the \textlambda~Cephei model in the H\textalpha\ flux, rotated about the $x$-axis by $0\si{\degree}$, and about the $z$-axis by (from left to right) $15\si{\degree}$, $30\si{\degree}$, $45\si{\degree}$, $60\si{\degree}$ , and $75\si{\degree}$. A rotation of $90\si{\degree}$ ($180\si{\degree}$) results in the same projection on the $xz$- ($yz$-) plane as in Fig.~\ref{fig:lcbproj}. Scales are the same as in Fig.~\ref{fig:lcbproj} (top row).}
 \label{fig:proj}
\end{figure*}

\begin{figure*}
 \centering
 \includegraphics[scale=0.31]{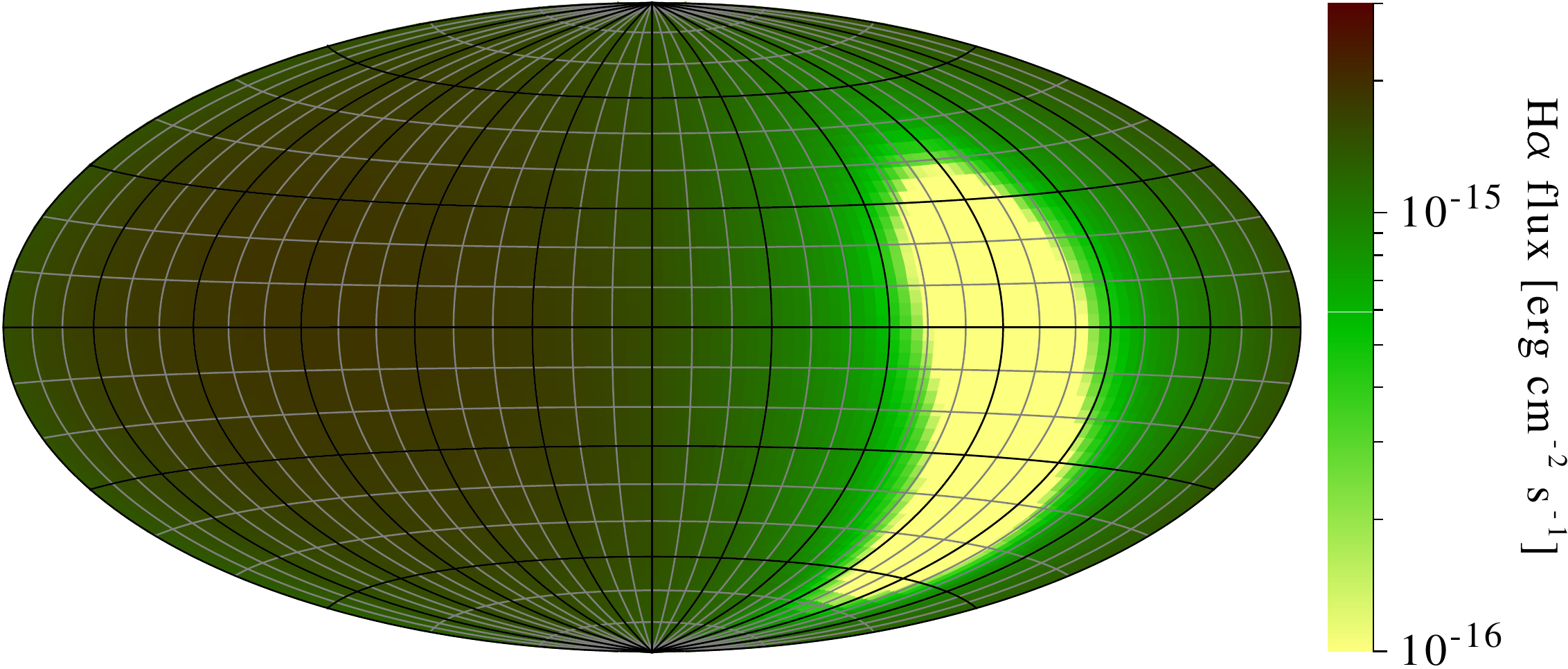}\hfill
 \includegraphics[scale=0.22]{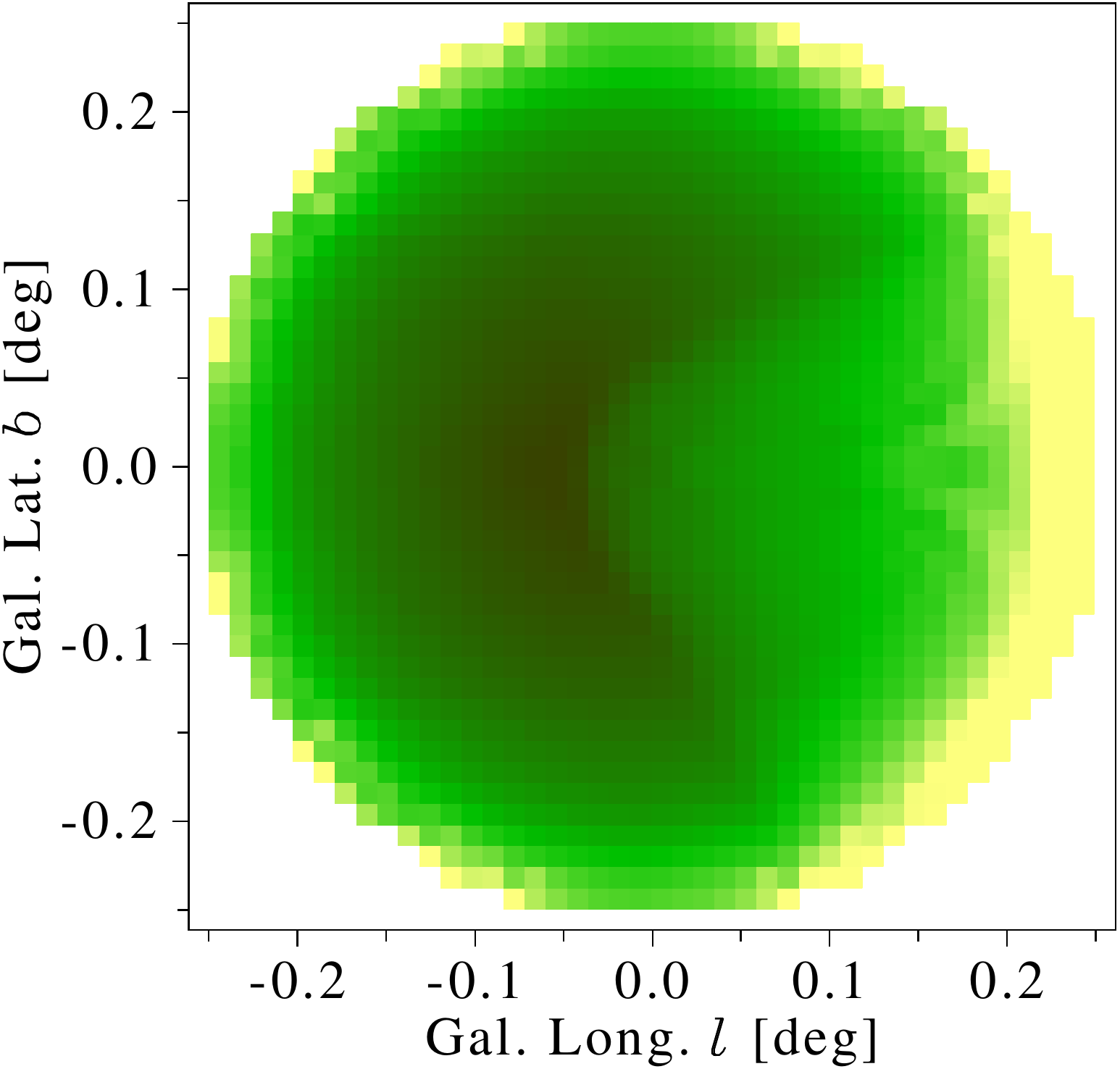}\hfill
 \includegraphics[scale=0.22]{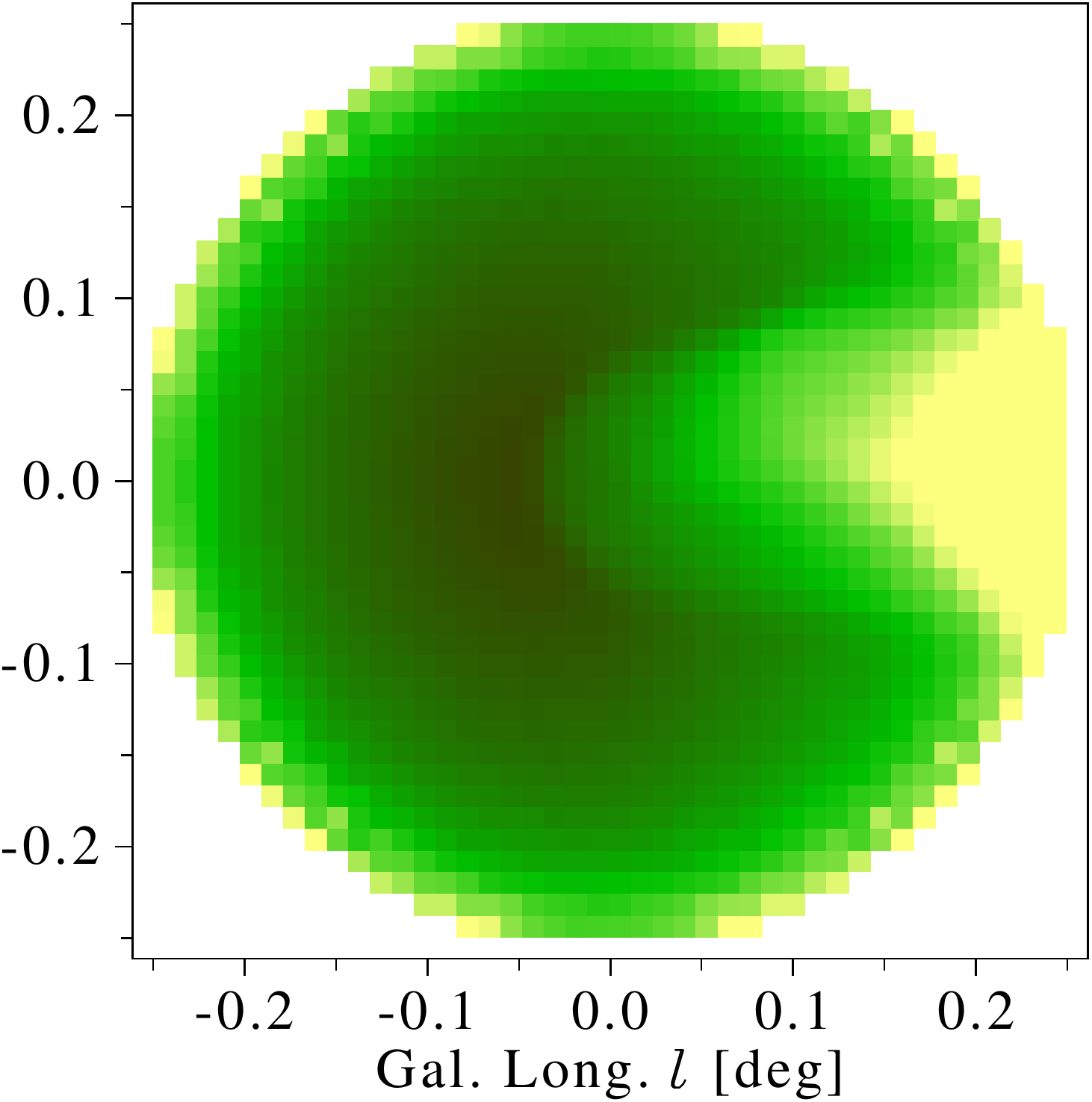}\hfill
 \includegraphics[scale=0.22]{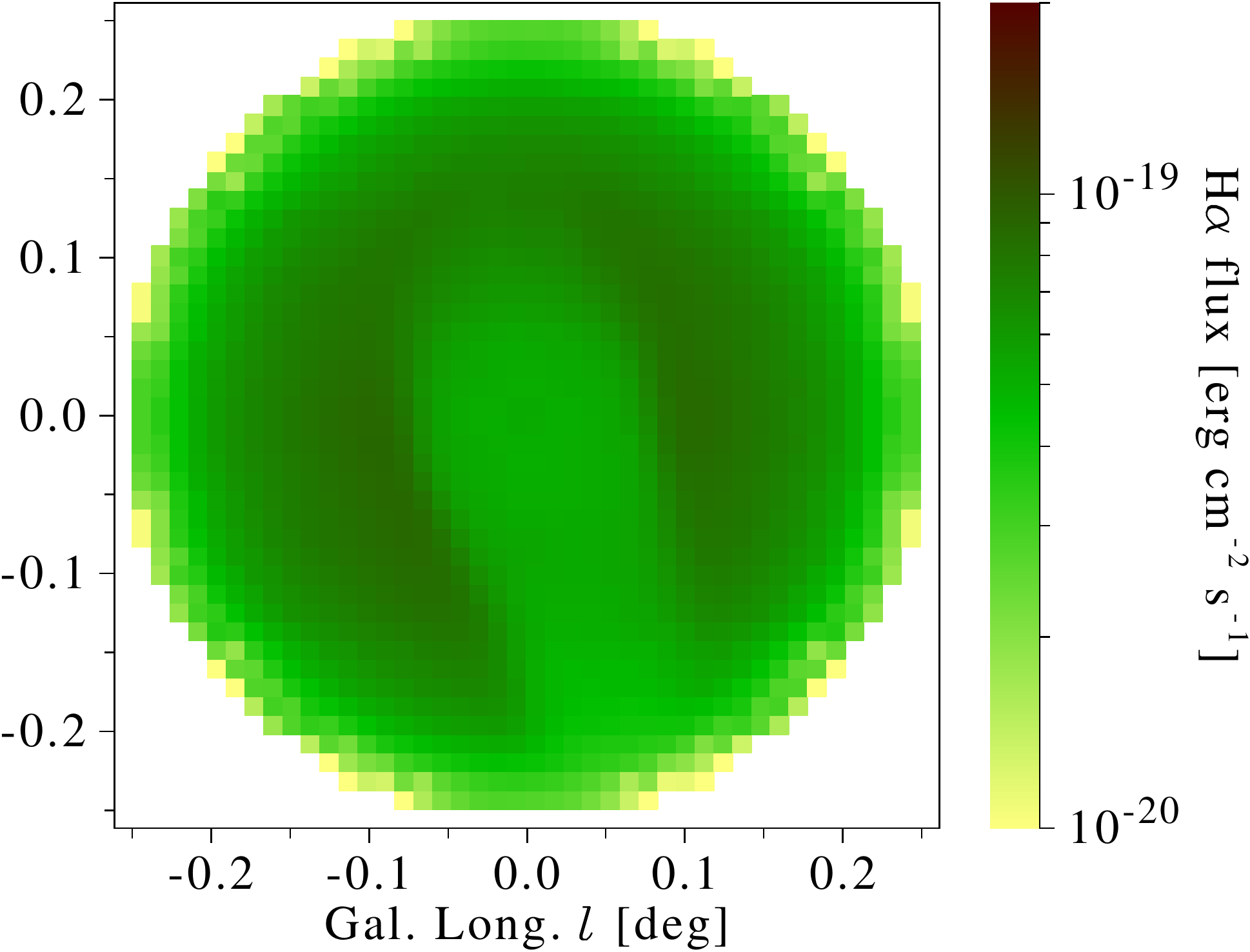}
 \caption{H\textalpha\ flux of the heliosphere model at distances $d_0=0$ (left column) and $d_0=1\,\si{pc}$ (right columns), projected onto (from left to right) the sky and the $xz$-, $xy$-, and $yz$-planes. Values lower than the minimum value of the colour scales are plotted in the colour of that minimum value.}
 \label{fig:h3dproj}
\end{figure*}

In Fig.~\ref{fig:eclip} the mean number density (left panel) and H\textalpha\ flux (right panel) of an integration area close to the central $xz$-plane of the  \textlambda~Cephei model  are displayed. The integration area has a thickness along the LOS of $0.6\,\si{pc}$ ($6\%$ of the total LOS length). Not the entire computed domain is shown. This configuration was chosen to show the structure in a comparably thin box similar to an infinitesimally thin $xz$-plane. The ISM inflow comes from the left; the undisturbed ISM is the homogeneous area of high number density ($10\,\si{cm}^{-3}$) and high H\textalpha\ flux ($10^{-12}\,\si{erg\,cm^{-2}\,s^{-1}}$) left of the parabolic shape, which is the BS. The domain of low number density ($10^{-3}-10^{-2}\,\si{cm}^{-3}$) and medium H\textalpha\ flux ($10^{-16}-10^{-14}\,\si{erg\,cm^{-2}\,s^{-1}}$) around $(l,b)=(0\si{\degree},0\si{\degree})$ is the undisturbed SW. The TS and the MD are visible as the outside border of this domain, where the number density jumps from roughly $10^{-3}\,\si{cm}^{-3}$ to $10^{-2}\,\si{cm}^{-3}$ and the H\textalpha\ flux jumps from roughly $10^{-17}\,\si{erg\,cm^{-2}\,s^{-1}}$ to $10^{-23}\,\si{erg\,cm^{-2}\,s^{-1}}$. The AP separates the inner astrosheath (number density $\sim10^{-2}\,\si{cm}^{-3}$, H\textalpha\ flux $\leq 10^{-22}\,\si{erg\,cm^{-2}\,s^{-1}}$) from the outer astrosheath (number density $(0.1-1)\,\si{cm}^{-3}$, H\textalpha\ flux $(10^{-18}-10^{-15})\,\si{erg\,cm^{-2}\,s^{-1}}$), appearing as the colour transition between the TS and the BS. The domain in the inner astrosheath extending to the right from the MD ($l>0.2\si{\degree}$) is also called the astrotail.

We display the calculated fluxes of H\textalpha, bremsstrahlung, and cyclotron radiation for the \textlambda~Cephei model in Fig.~\ref{fig:lcbproj}. In the left panels, the model centre was placed at the origin ($d_0=0$) and rotated about the $z$-axis by $90\si{\degree}$ so that the ISM inflow comes from $l=-90\si{\degree}$. Different values of the respective physical parameter are mapped onto colour scales, with values lower than the minimum value of the colour scale plotted in that colour. The scaling is bounded to give the best possible representation of all observable features at the cost of cutting off multiple orders of magnitude of the minima, which all occur in the astrotail alone. In all three radiation types, this global minimum is clearly visible at around $(b,l)=(0\si{\degree}, 90\si{\degree})$, forming a near-perfect circle that appears distorted due to the projection. This corresponds to the astrotail, that is, the direction opposite to the stellar motion, or in the stellar rest frame the direction opposite to the ISM inflow. The astrotail is a region of low density (roughly $10^{-2}\,\si{cm^{-3}}$); therefore all radiation processes are rather weak in this region (cf. Eqs. (\ref{eq:brs}, \ref{eq:harb}), Sec.~\ref{sec:cyc}). The global maximum is expected in the opposite direction, where the number density jumps by a factor of 4 at the BS. This is true for the cyclotron radiation (third panel from the top), which is proportional not only to $n$ but also to $B^2$ (cf. Sec.~\ref{sec:cyc}), which also jumps to higher values at the BS. In a roughly circular region with a radius of $\sim 30\si{\degree}$ around the inflow direction $(b,l)=(0\si{\degree}, -90\si{\degree})$, the cyclotron flux reaches its almost constant maximum value of $7\times 10^{-13}\,\si{erg\,cm^{-2}\,s^{-1}}$. This angular extent coincides with the size of the HOA, an additional layer of high-temperature cells ($T\approx 7\times 10^4\,\si{K}$) in the radial direction that is confined to the central $30\si{\degree}$ around the inflow axis (the axis through the star parallel to the homogeneous ISM inflow). The high temperature results in high thermal velocities and therefore also in high cyclotron emission. Outside this maximum, two concentric rings of high emission ($5\times 10^{-13}\,\si{erg\,cm^{-2}\,s^{-1}}$) surround the central maximum, stemming from several regions with high density, high temperature, and/or strong magnetic fields close to the BS. Outside these rings, the cyclotron flux has values of $(1-3)\times 10^{-13}\,\si{erg\,cm^{-2}\,s^{-1}}$  until it sharply declines due to the astrotail. Unlike the other emission types, the cyclotron flux shows an asymmetry in the astrotail minimum, being elongated to an elliptical shape whose semi-major axis lies at an angle of about $30\si{\degree}$ to the vertical axis. This is due to the orientation of the magnetic field (cf. Tab.~\ref{tab:modvals}).

While the bremsstrahlung flux (second panel from the top) also shows a central maximum ($\sim2\times 10^{-7}\,\si{erg\,cm^{-2}\,s^{-1}}$) around the inflow direction, this maximum is local instead of global. A ring of low fluxes ($\sim1.5\times 10^{-7}\,\si{erg\,cm^{-2}\,s^{-1}}$) with a radius of $30\si{\degree}$ around the inflow direction separates the central maximum from another high-emission ring ($(1-2)\times 10^{-7}\,\si{erg\,cm^{-2}\,s^{-1}}$) that itself is surrounded by a second low-emission ring. Outside these rings, the bremsstrahlung flux remains at values higher than $1\times 10^{-7}\,\si{erg\,cm^{-2}\,s^{-1}}$, reaching its maximum values of $(2-3)\times 10^{-7}\,\si{erg\,cm^{-2}\,s^{-1}}$ close to the astrotail minimum at high latitudes ($b\approx\pm 80\si{\degree}$). These maxima come from long LOSs through the high-temperature outer astrosheath, while the inner rings correspond to those of the cyclotron flux. The bremsstrahlung flux is not symmetric to the equatorial plane of the model ($b=0\si{\degree}$) but shows lower values close the central maximum in the northern hemisphere ($(0.8-1.0)\times 10^{-7}\,\si{erg\,cm^{-2}\,s^{-1}}$). The high-emission rings described above likewise have lower values in the northern hemisphere than in the southern one. 

The global maxima of the bremsstrahlung flux are also global for the H\textalpha\ flux (top panel, $2\times 10^{-7}\,\si{erg\,cm^{-2}\,s^{-1}}$). Unlike the other emission types, there is no central maximum at the inflow direction $(b,l)=(0\si{\degree}, -90\si{\degree})$ but a local minimum ($4\times 10^{-8}\,\si{erg\,cm^{-2}\,s^{-1}}$). This minimum has a homogeneous flux inside a circular region of $30\si{\degree}$ around the inflow direction and a gradient ring of $(30-60)\si{\degree}$ with fluxes up to $1.3\times 10^{-7}\,\si{erg\,cm^{-2}\,s^{-1}}$  at the outer edge. This edge is well defined, separating the inner minimum from the higher-emission domain ($(1.5-2)\times 10^{-7}\,\si{erg\,cm^{-2}\,s^{-1}}$) in which the maxima lie. The minimum is caused by the decrease with higher temperatures  of the effective radiation rate coefficient $\alpha_{\mathrm{eff},3\rightarrow 2}$ (cf. Sec.~\ref{sec:rec}), which is therefore much lower in the HOA than in the COA. This feature can also be observed in Fig.~\ref{fig:eclip} (right panel), disrupting the arc of the BS at its vertex.

The various fluxes of the \textlambda~Cephei model at distance $d_0=617\,\si{pc}$ (the right three panels of Figs.~\ref{fig:lcbproj}) have a rather similar appearance for each geometric arrangement. In both the $xz$- (left) and the $xy$-plane (centre), corresponding to the side and top view of the astrosphere, the parabolic shape of the BS is clearly visible. The projected image is a result of the overlying parabolas of the outer astrosheath from different slices of the computational domain: in the central plane, the BS extends the farthest to the left (smaller $l$), while the BSs of the slices further from the central plane lie closer to the right (larger $l$). Because the shifting distance of the parabolas to the right decreases the closer the slice is to the central plane, the positions of the BSs in these slices are close to the position of the leftmost BS in the central slice. Therefore, the LOSs going through the leftmost edge of the parabolic shape encounter multiple BSs from different slices, giving the edge of the parabolic shape its high flux.  Because the left and central images look similar, the structure of the astrosphere must be strongly symmetric around the inflow axis. Because the structure of a purely HD model is completely symmetric, it can be concluded that the influence of the magnetic field compared to that of thermal effects must be rather weak, or in other words, that the plasma $\beta$ must be rather high. This has also been concluded by \citet{scherer18} with different arguments. For $l>0.3\si{\degree}$, the LOSs of higher longitudes encounter only the astrotail, resulting in fluxes that are orders of magnitude below those from the outer astrosheath or the ISM, therefore causing the global minima. This unphysical effect is due to the finite extent of the model; in a larger model, the outer astrosheath would extend farther outward and would therefore appear in front of and behind the astrotail among these LOS, causing fluxes similar to the region inside the arc structure with $0.0\si{\degree}<l<0.3\si{\degree}$.

The difference between the fluxes from the outer astrosheath (the arc) and the undisturbed ISM (left from the parabolic structure) is strongest for cyclotron radiation (note the different scales for cyclotron radiation and bremsstrahlung/H\textalpha). This is because cyclotron emission is proportional to $n B^2$ (cf. Sec.~\ref{sec:cyc}), with both $n$ and $B$ jumping to higher values at the BS. Bremsstrahlung and H\textalpha\ emission both follow $n^2$ (cf. Eqs.~(\ref{eq:brs}, \ref{eq:harb})) and are therefore similar to each other. The H\textalpha\ flux differs from the other two emission types by the behaviour of the BS at the parabola vertex: While the cyclotron and bremsstrahlung fluxes have their maxima here (taking the abovementioned geometric effect into account), the H\textalpha\ flux is comparatively low. This is due to the high temperature of the HOA, analogous to the minimum in the H\textalpha\ projection at $d_0=0$.

The right panels of Fig.~\ref{fig:lcbproj} show the model projected on the $yz$-plane, the plane perpendicular to the inflow direction, corresponding to the front view of the astrosphere. In all three fluxes a distinct outer ring of low emission can be observed. This is created by the LOSs that do not encounter any cells from the outer astrosheath, corresponding to the top and bottom lines of the projections on the $xy$- and $xz$-planes (central columns of Fig.~\ref{fig:lcbproj}). For H\textalpha, a central minimum is created by the HOA, in analogy to the central minima of the other H\textalpha\ projections. For the bremsstrahlung and the cyclotron fluxes, faint ring-like structures can be observed, corresponding to the ring-like structures of the projections for $d_0=0$ (left column).

Projections at more angles are displayed in Fig.~\ref{fig:proj}. H\textalpha\ was chosen as the projected observable because the visible structures are similar for all three radiation types, and also because this observable is best suited for comparison with astronomical data. The rotational angle about the $z$-axis, henceforth labelled $\alpha$, was varied for these projections in steps of $15\si{\degree}$ from $15\si{\degree}$ to $75\si{\degree}$. The projection for $\alpha=90\si{\degree}$ corresponds to the projection on the $xz$- plane and is displayed in Fig.~\ref{fig:lcbproj} (first row, second column); a projection for $\alpha=0\si{\degree}$ results in the projection on the $yz$-plane (Fig.~\ref{fig:lcbproj}, first row, fourth column), mirrored about the $z$-axis. Projections for $90\si{\degree}<\alpha<180\si{\degree}$ result in the same images as for $180\si{\degree}-\alpha$ and for $180\si{\degree}<\alpha<360\si{\degree}$ in the mirror images of projections for $360\si{\degree}-\alpha$. Fig.~\ref{fig:proj} shows that a crescent of high emission ($\geq 3\times 10^{-11}\,\si{erg\,cm^{-2}\,s^{-1}}$) arises to the left of the projection centre with increasing $\alpha$, which becomes more narrow in width and extends into the right half of the projection until it obtains the distinctive parabolic shape of the bow shock at $\alpha=90\si{\degree}$. The area to the left of this crescent is free from any influence of the astrosphere and only consists of the undisturbed homogeneous ISM. The central minimum from the HOA moves to the left with increasing $\alpha$ until it touches the bow shock crescent, interrupting the arc of high emission at its vertex. The astrotail effect decrease the flux in the area to the right of the crescent, reducing the otherwise mostly homogeneous high emission from the outer astrosheath to lower values at the inflow axis ($b=0$) and resulting in the global minimum of the astrotail for $\alpha\geq 60\si{\degree}$. Except for the HOA-caused central minimum, the same holds true for the other radiation types.

The rotation measure (RM) of the \textlambda~Cephei model is displayed in the bottom row of Fig.~\ref{fig:lcbproj}; the red part (blue part) of the colour scale denominates positive (negative) RMs. The astrotail is clearly visible as the structure of low RMs in the all-sky projection (left panel, $RM\leq 1\,\si{rad\,m^{-2}}$) and also on the far right in the projections on the $xz$- and $xy$-plane (central panels, $RM\leq 0.1\,\si{rad\,m^{-2}}$). The alternating structure of positive and negative RM bands in the all-sky projection mostly comes from the homogeneous magnetic field of the undisturbed ISM. The BS can clearly be recognized in the $xz$- and $xy$-planes. In the $yz$-plane projection, only the radial gradient can be recognized as a structure.

\subsection{Heliosphere}\label{sec:h3d}

The heliosphere model was evaluated by the same method. The projections of all three radiation types look very similar, therefore we only display the H\textalpha\ flux in Fig.~\ref{fig:h3dproj}. Unlike the \textlambda~Cephei model, the ISM fluid speed is too low for a BS to emerge, at least in single-fluid modelling \citep[see, however,][]{scherer14}. Instead of a sudden jump, the ISM parameters increase or decrease gradually in the entire region outside the heliopause (HP), forming a so-called bow wave \citep{mccomas12,zank13}. 

The left panel shows the all-sky projection with the model centre at origin ($d_0=0$). As with the \textlambda~Cephei model, the heliotail is clearly visible as the global minimum. However, unlike the \textlambda~Cephei model, the tail of the heliosphere model is elongated in the $b$-direction. This stems from an asymmetric shape of the heliosphere: the average plasma $\beta$ is much lower for the heliosphere model than for the \textlambda~Cephei model, therefore the effect of the magnetic field is much stronger. Because there is no BS, there also is no (hot) outer heliosheath, and the flux accordingly follows a simple gradient outside the heliotail instead of showing a ring-like structure around the bow direction. The lack of a BS is much more apparent in the projections at larger distances ($d_0=1\,\si{pc}$, right panels), where a radial gradient corresponding to the bow wave is apparent in the $xz$-plane until the inflow hits the heliopause. The projection on the $xy$-plane shows an indentation on the right side, caused by the low emission of the heliotail. As in the \textlambda~Cephei model, the heliotail causes a ring-like structure in the projection on the $yz$-plane, although the ring is elongated in the $b$-direction due to the asymmetry of the heliotail. 

The RM in the all-sky projection shows a similar structure of alternating bands of positive and negative RMs as the \textlambda~Cephei model, only with much lower values ($\mathrm{RM}<10^{-7}\,\si{rad\, m^{-2}}$). Like the astrotail of the \textlambda~Cephei model, the heliotail causes a structure of lower RMs; no further structures can be seen. It is unlikely that such low values of the RM could be detected because fluctuations of the density and the magnetic field in the ISM outside the model range are assumed to cause RM values of similar magnitude.

\subsection{Proxima Centauri and V374 Pegasi}\label{sec:pxbpeg}

\begin{figure}
 \centering
 \includegraphics[height=0.37\columnwidth]{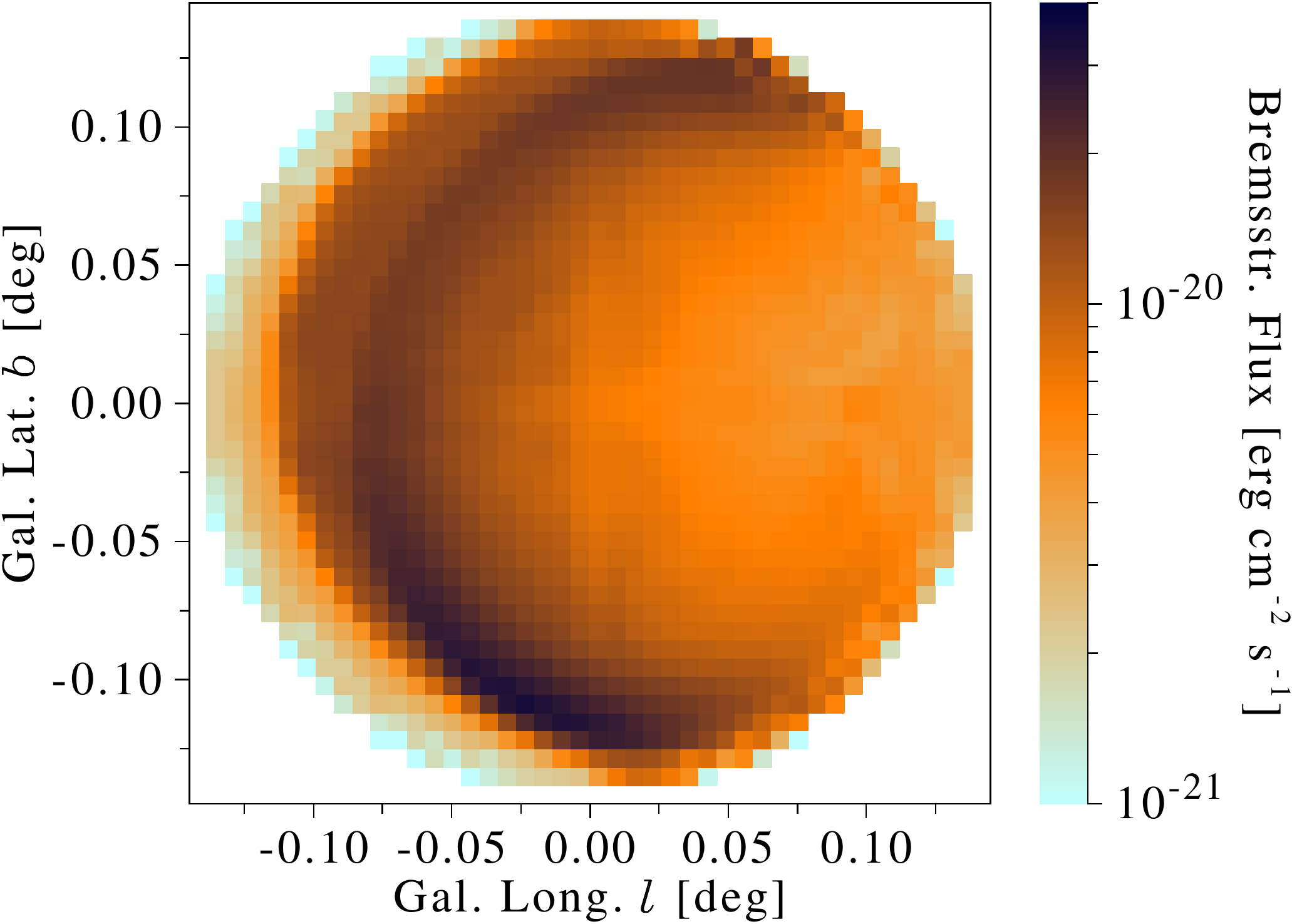}\hfill
 \includegraphics[height=0.37\columnwidth]{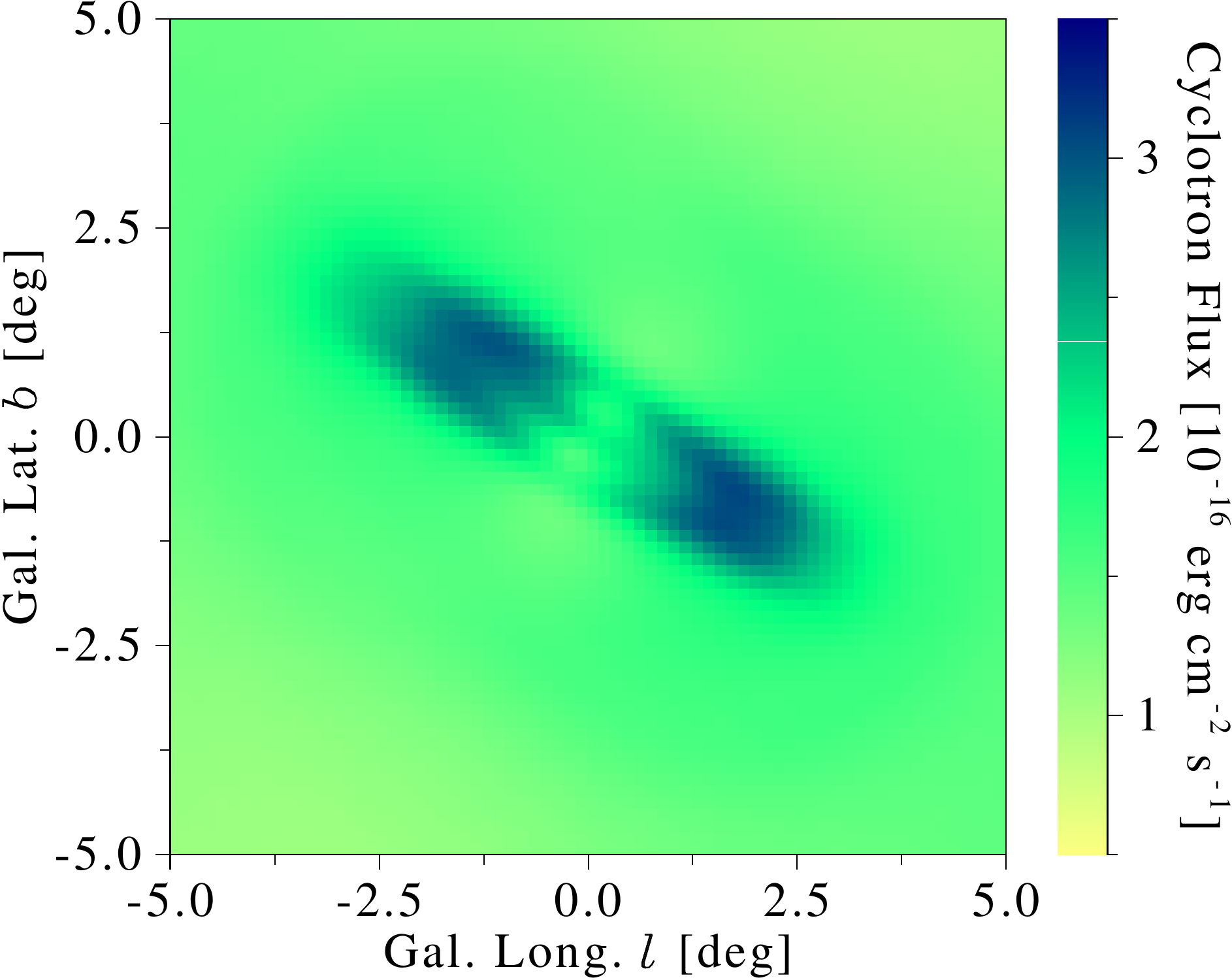}
 \caption{Left: Bremsstrahlung flux of the Proxima Centauri model at a distance of $1\,\si{pc}$, projected at the $xz$-plane. Right: Cyclotron flux of the central region of the V374 Pegasi model at a distance of $9\,\si{pc}$, projected at the $xy$-plane. The cyclotron flux is scaled linearly.}\label{fig:peg}
\end{figure}

The speed of Proxima Centauri relative to the ISM is high enough to produce a BS; the entire astrotail is contained in the model. The radiation fluxes are of the same order of magnitude as in the heliosphere model. Like in the \textlambda~Cephei model, the origin of the highest fluxes lies in the outer astrosheath, where the density and the magnetic induction are comparably high ($n\lesssim 10^{-2}\,\si{cm^{-3}}$, $B\lesssim 10^{-5}\,\si{G}$) and the temperature is moderate ($T\lesssim 10^5\,\si{K}$). The RM is about one magnitude lower than the RM of the heliosphere model and shows a similar large-scale structure in the all-sky projection as the previous two models. Because the astrotail is completely contained in the model, there is only a faint minimum in the direction opposite to the inflow. As an example, the bremsstrahlung flux in the $xz$-plane is shown in Fig.~\ref{fig:peg} (left); the other fluxes are similar. 

No BS appears to be produced in the V374 Pegasi model, although the structure outside the TS is complicated and requires further investigation. The projections show an elongated structure in the cyclotron flux of the $xy$-plane (see Fig.~\ref{fig:peg} right) that is also faintly visible in the other projection planes and observables. The fluxes are higher than those of the heliosphere model, but still far lower than the detection limits. It is noticeable, however, that the bremsstrahlung flux is higher by three orders of magnitude while the H\textalpha\ flux is only higher by one order magnitude than that of the heliosphere model; this is due to the high temperatures of the V374 Pegasi model (see Tab.~\ref{tab:modvals}), which lead to low effective recombination rate coefficients. Because of the high magnetic induction ($3\times 10^{-5}\,\si{G}$ outside the TS), the cyclotron flux is higher than that of the heliosphere model by six orders of magnitude. The critical frequency of bremsstrahlung is $\nu_{\mathrm{c}}=10^{18}\,\si{Hz}$ for the V374 Pegasi model, which is higher by two orders of magnitude than $\nu_{\mathrm{c}}$ of the other three models; this is also caused by the high temperature ($T\gtrsim 10^6\,\si{K}$ outside the TS) of the model. 

\section{Observability}\label{sec:obs}

To evaluate the model projections, the computed fluxes are first compared to current observational limits in Sec.~\ref{sec:obslims}. Subsequently, in Sec.~\ref{sec:obsdat} observational images of \textlambda\ Cephei are compared to the corresponding projection. 

\subsection{Observational limits}\label{sec:obslims}

The detectable minimum flux of H\textalpha\ emission was assumed to be $10^{-19}\,\si{erg\, cm^{-2}\,s^{-1}\,arcsec^{-2}}$ \citep{donahue95}. The angular area of the projection grid cells for the \textlambda\ Cephei model at $d_0=0$ (left column) is $1.14\times 10^8\,\si{arcsec^2}=2.68\times 10^{-3}\,\si{sr}$; at $d_0=617\,\si{pc}$ (right columns) it is $5.05\times 10^4\,\si{arcsec^2}$. Multiplying the detection limit with the respective angular area for $d_0=0$ gives a minimum flux of $1.14\times 10^{-11}\,\si{erg\, cm^{-2}\,s^{-1}}$; for the $d_0=617\,\si{pc}$ projections, the minimum flux amounts to $5.05\times 10^{-15}\,\si{erg\, cm^{-2}\,s^{-1}}$. The entire projections of the \textlambda\ Cephei model would be clearly detectable.

The emission spectrum of bremsstrahlung was assumed to be flat up to the critical frequency $\nu_{\max}=10^{16}\,\si{Hz}$, so that the spectral flux was calculated from the total flux by dividing by $\nu_{\max}$. Two estimates for the detectable minimum flux were made, one of which is given by the NLAO VLA Sky Survey (NVSS) with a rms brightness fluctuation of $\sigma=0.45\,\si{mJy}/(4\si{\degree}\times 4\si{\degree})\approx 2.2\times 10^{-35}\,\si{erg\,s^{-1}\,cm^{-2}\,Hz^{-1}\,arcsec^{-2}}$ at a frequency of $1.4\,\si{GHz}$ \citep{condon98}. In combination with the solid angles described above, this amounts to detection limits of $2.5\times 10^{-27}\,\si{erg\,s^{-1}\,cm^{-2}\,Hz^{-1}}$ at $d_0=0$ and $1.1\times 10^{-30}\,\si{erg\,s^{-1}\,cm^{-2}\,Hz^{-1}}$ at $d_0=617\,\si{pc}$. These values are below the calculated bremsstrahlung fluxes by some orders of magnitude. Spectrum-modifying effects such as self-absorption are most effective for low frequencies, therefore we assumed that the observed bremsstrahlung flux may be lower than the emitted flux by many orders of magnitude. Therefore, an estimate of the detectable minimum flux at higher frequencies is of value. A detection limit of $10^{-15}\,\si{erg\,s^{-1}\,cm^{-2}}$ is given for a spectral range $[0.5, 2]\,\si{keV}\hat{=}[1.2, 4.8]\times 10^{17}\,\si{Hz}$ by \citet{hasinger04}. The corresponding spectral flux is $2.8\times 10^{-33}\,\si{erg\,s^{-1}\,cm^{-2}\,Hz^{-1}}$ and lower than the calculated values by several orders of magnitude. While the detection limit is given for a spectral range one order of magnitude higher than the calculated critical frequency, we assumed that the detection limit of the correct spectral range is in a similar order of magnitude. This means that the entire projection of the \textlambda\ Cephei model would be detectable.

Because cyclotron radiation is emitted at extremely low frequencies, the minimum detectable flux of the LOw Fequency ARray \citep[LOFAR,][]{lofar13} was used as a detection limit. LOFAR is sensitive in a spectral range $[10, 240]\,\si{MHz}$; a sensitivity of $\sigma_{60\,\si{MHz}}=20\,\si{\micro Jy}$ is given for frequency $\nu=60\,\si{MHz}$. The sensitivity is frequency-dependent with $\sigma_{\nu} \propto \nu^{-0.7}$. Extrapolated to a frequency of $300\,\si{Hz}$, the detection limit would be $\sigma_{300\,\si{Hz}}=103\,\si{mJy}=1.03\times 10^{-24}\,\si{erg\,s^{-1}\,cm^{-2}\,Hz^{-1}}$. While neither LOFAR nor any other astronomical instrument is sensitive in this spectral range, we still used $\sigma_{300\,\si{Hz}}$   as a minimum detectable flux for the sake of comparison. Compared to the calculated cyclotron fluxes, $\sigma_{300\,\si{Hz}}$ is lower by a few orders of magnitude. If an instrument with detection limits comparable to LOFAR were sensitive for the \textlambda~Cephei model gyration frequencies ($\nu_{\mathrm{g}}<500\,\si{Hz}$), it would be able to detect such astrospheres with ease. However, cyclotron (or synchrotron) radiation can be safely discarded as an origin to the detected emission of astrospheres because the frequencies are far too low.

For the heliosphere model, the fluxes are all much lower than those of the \textlambda~Cephei model, about eight to nine orders of magnitude for all three types of radiation. The gyration frequencies, at which cyclotron radiation is emitted, are lower than $20\,\si{Hz}$, while the critical bremsstrahlung frequency remains at $\nu=10^{16}\,\si{Hz}$. The angular area of the projection cells is $1.14\times 10^8\,\si{arcsec^2}$ for $d_0=0$ and $1.90\times 10^3\,\si{arcsec^2}$ for $d_0=1\,\si{pc}$. The resulting detection limits of H\textalpha\ are $1.14\times 10^{-11}\,\si{arcsec^2}$ ($d_0=0$) and $1.90\times 10^{-16}\,\si{arcsec^2}$ ($d_0=1\,\si{pc}$), which is far too high for any chance of detection. Similarly, the detection limits of bremsstrahlung and cyclotron fluxes are higher than the calculated fluxes by many orders of magnitude. 

Because the fluxes for the model of Proxima Centauri are on the same order of magnitude as those of the heliosphere model, the model astrosphere would not be detectable with the assumed detection limits. While the fluxes for the model of V374 Pegasi are higher than those of the heliosphere model, they still remain far lower than the detection limits.

\subsection{Observational data}\label{sec:obsdat}

\begin{figure*}
 \centering
 \includegraphics[width=0.32\textwidth]{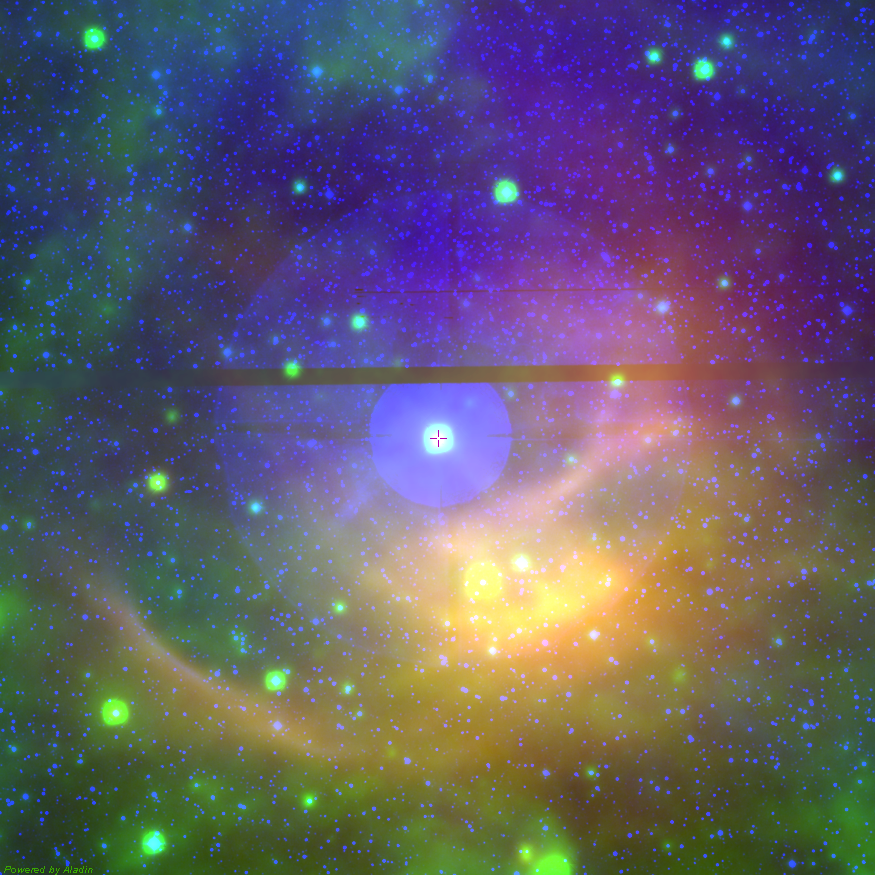}
 \includegraphics[width=0.32\textwidth]{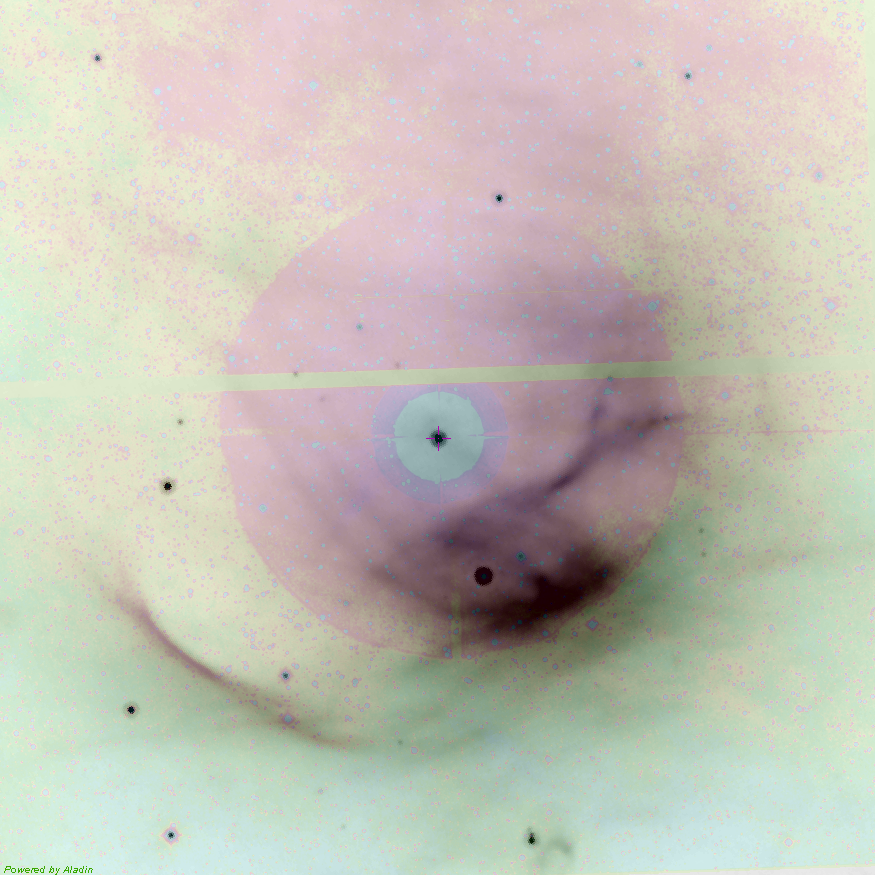}
 \includegraphics[width=0.32\textwidth]{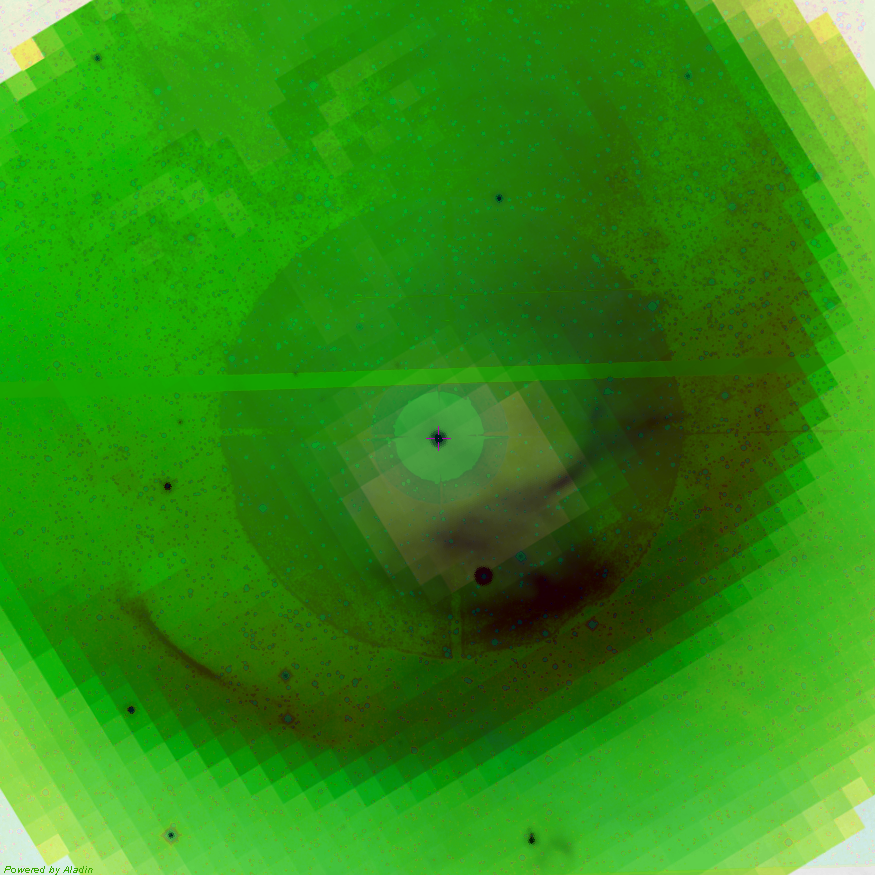}
 \caption{Composite images of the shock structure around \textlambda~Cephei. The star is at the central position, marked by a magenta cross; north is up, east is left. The size of both plotted fields is $20'\times 20'$. Left: WISE W3 band ($12\,\si{\micro m}$, red), WISE W4 band ($22\,\si{\micro m}$, green), IPHAS H\textalpha\ (blue); centre: MIPS $24\,\si{\micro m}$ (greyscale), IPHAS H\textalpha\ (magenta). Right: Composite image of the simulated shock structure, rotated about the $z$-axis by $25\si{\degree}$, and the image from the centre panel. The simulation image was further rotated by $90\si{\degree}+31.373\si{\degree}$ anticlockwise to align the BS directions.}
 \label{fig:obs}
\end{figure*}

Composite images of the area around \textlambda~Cephei are displayed in Fig.~\ref{fig:obs} by combining observational data from three infrared bands, observed by the Wide-field Infrared Survey Explorer (WISE, left panel, \citet{wise2010}) and the Multiband Imaging Photometer for Spitzer (MIPS, centre panel, \citet{spitzer2004,mips2004}), and H\textalpha, obtained as part of the the INT Photometric H\textalpha\ Survey of the Northern Galactic Plane (IPHAS, left and centre panels, \citet{iphas2005,iphas2014}). The star moves in the south-southwest direction (the bottom right direction in the image plane; east is left and west is right), so that the bow shock is faintly visible as a wide arc southeast to west of the star in the long-wave infrared. A small part of the bow shock is clearly visible in both H\textalpha\ and the infrared southeast of the star. While this structure is very apparent in H\textalpha\ and the long-wave infrared, it is faint at $12\,\si{\micro m}$. The large disc-like structure in H\textalpha\ centred on \textlambda~Cephei stems from ghosting effects and is purely instrumental. A much larger structure closer to the star in the southwest, visible at the edge of the ghosting disc, can be seen in the infrared bands but not in H\textalpha .

To compare observed and simulated astrospheres, the model was projected with observed geometry of \textlambda~Cephei: Gaia measures a parallax angle of $(1.6199\pm 0.1265)\,\si{mas}$ \citep{gaia18}, resulting in a distance of roughly $d=617\,\si{pc}$; \citet{bouret12} calculated a distance of $950\,\si{pc}$. Gaia measures a proper motion of $\mu_{\alpha}\cos(\delta)=-6.328\,\si{mas/yr}$, $\mu_{\delta}=-10.378\,\si{mas/yr}$ at right ascension $\alpha=103.83\si{\degree}$ and declination $\delta=2.61\si{\degree}$. The radial velocity is $v_{\mathrm{rad}}=-75.10\,\si{km/s}$ \citep{gontcharov06}, pointing towards the observer. Assuming the distance derived from Gaia measurements, the angle of motion, that is, the angle between proper and radial motion, which in the simulation is the rotation about the $z$-axis, is 
\begin{equation}\label{eq:angle}
\gamma=\arctan\left(\frac{d}{v_{\mathrm{rad}}}\sqrt{\mu_{\alpha}^2\cos^2(\delta)+\mu_{\delta}^2}\right) \approx 25.35\si{\degree} \ .
\end{equation}
With an assumed distance of $950\,\si{pc}$, the angle of motion would be $36.11\,\si{\degree}$. The proper motion angle, that is, the angle of the proper motion to the image coordinate system, is $\phi=\arctan\left(\mu_{\alpha}\cos(\delta)/\mu_{\delta}\right)\approx 31.373\si{\degree}$. 

The projection of the simulated astrosphere has a distance between BS and star of about $0.2\si{\degree}= 12'$, whereas the observed distance between BS and star is about $8'$. Because the size of the simulated astrosphere scales with the boundary conditions, especially the inner number density and fluid velocity, the difference in scales is most likely caused by incorrect estimates of these values. The overall shape of the simulated astrosphere does not depend on the boundary conditions as strongly; scaling the images so that the bow shocks align is therefore a valid approach. 

This was done for the right panel of Fig.~\ref{fig:obs}: the MIPS composite image (central panel) was combined with a projection of the simulated astrosphere. This projection was generated with an angle about the $z$-axis of $\gamma=25\si{\degree}$ (see Eq.~(\ref{eq:angle})); the image was further rotated by $90\si{\degree}+\phi=121.373\si{\degree}$ anticlockwise to align the inflow direction of the simulated astrosphere with the stellar proper motion. The projection image was scaled so that the BSs appear at the same distance to the star. The shapes of the BSs align well: the filament southeast of the star (bottom left quadrant) can be identified as part of the BS. The unknown infrared emission structure inside the BSs roughly coincides with the astropause, but the inner structure of the astrosphere unfortunately disappears inside the H\textalpha\ ghosting disc.

\section{Conclusions}\label{sec:conc}
We presented and evaluated projections of 3D single-fluid MHD simulations of the astrospheres of \textlambda~Cephei and the heliosphere. The projections of two further astrosphere models, those of Proxima Centauri and V374 Pegasi, were discussed. It was shown that the region between the astropause and the bow shock, the outer astrosheath, causes high fluxes in all examined radiation processes, resulting in a variety of shapes following the apparent shape of the outer astrosheath. This shape depends on the inclination angle between the LOS and the direction of the stellar motion, resulting in a ring with a central maximum when the angle is close to $0\si{\degree}$, a filled parabola when the angle is close to $90\si{\degree}$ , and a crescent with a barely detached or attached central maximum when the angle is in between. Likewise, high values of the rotation measure follow the surface of the bow shock in the central plane, leading to parabolic shapes. This is consistent with the results of different approaches, such as the 3D HD models by \citet{mohamed12} or the 2D MHD models by \citet{meyer17}, the latter of which also gives similar maps of their axisymmetric model at different inclination angles.

In astrospheres without a bow shock, for instance, a single-fluid model of the heliosphere, this region does not exist. No distinguishable shapes are visible in the fluxes or the rotation measure. It is unlikely that detailed observations of such astrospheres can be made. While the astrospheres of low-mass stars with supersonic motion relative to their ISM (such as Proxima Centauri) produce a bow shock and display comparably high fluxes in their outer astrosheaths, these fluxes are still much lower than those of the astrosheaths of runaway high-mass stars such as \textlambda~Cephei. Even though many more of these low-mass stars are far closer to us than high-mass stars, their fluxes are most likely too low to be detectable.

Unlike with other approaches \citep[e.g.][]{delvalle18}, no synchrotron radiation could be obtained with these models. This is due to the non-relativistic thermal speed of the electrons, which do not exceed temperatures of $10^8\,\si{K}$. For relativistic electrons, which can be acquired by higher initial temperatures and (less likely by) different heating terms in the governing MHD equations, synchrotron radiation must be expected. While cyclotron radiation is emitted at very low and ultra-low frequencies, far below the spectral range of observational instruments, the bremsstrahlung spectrum reaches frequencies in the extreme ultraviolet and soft X-ray regime. Higher thermal speeds of electrons would result in even higher frequencies, as were observed by \citet{ayaso2018}, for example.

\begin{acknowledgements}
KS is grateful to the \textit{Deutsche Forschungsgemeinschaft} (DFG), funding the
project SCHE334/9-2. JK acknowledges financial support through the \textit{Ruhr Astroparticle and Plasma Physics (RAPP) Center}, funded as MERCUR project St-2014-040.

This publication makes use of data products from the Wide-field Infrared Survey Explorer, which is a joint project of the University of California, Los Angeles, and the Jet Propulsion Laboratory/California Institute of Technology, funded by the National Aeronautics and Space Administration. 

This work is based in part on observations made with the Spitzer Space Telescope, which is operated by the Jet Propulsion Laboratory, California Institute of Technology under a contract with NASA. 

This paper makes use of data obtained as part of the INT Photometric H\textalpha\ Survey of the Northern Galactic Plane (IPHAS, www.iphas.org) carried out at the Isaac Newton Telescope (INT). The INT is operated on the island of La Palma by the Isaac Newton Group in the Spanish Observatorio del Roque de los Muchachos of the Instituto de Astrofisica de Canarias. All IPHAS data are processed by the Cambridge Astronomical Survey Unit, at the Institute of Astronomy in Cambridge.  The bandmerged DR2 catalogue was assembled at the Centre for Astrophysics Research, University of Hertfordshire, supported by STFC grant ST/J001333/1.
\end{acknowledgements}

\bibliography{2018paper.bib}

\begin{thebibliography}{56}
\expandafter\ifx\csname natexlab\endcsname\relax\def\natexlab#1{#1}\fi

\bibitem[{{Arthur}(2012)}]{arthur12}
{Arthur}, S.~J. 2012, \mnras, 421, 1283

\bibitem[{{Barentsen} {et~al.}(2014){Barentsen}, {Farnhill}, {Drew},
  {Gonz{\'a}lez-Solares}, {Greimel}, {Irwin}, {Miszalski}, {Ruhland}, {Groot},
  {Mampaso}, {Sale}, {Henden}, {Aungwerojwit}, {Barlow}, {Carter}, {Corradi},
  {Drake}, {Eisl{\"o}ffel}, {Fabregat}, {G{\"a}nsicke}, {Gentile Fusillo},
  {Greiss}, {Hales}, {Hodgkin}, {Huckvale}, {Irwin}, {King}, {Knigge},
  {Kupfer}, {Lagadec}, {Lennon}, {Lewis}, {Mohr-Smith}, {Morris}, {Naylor},
  {Parker}, {Phillipps}, {Pyrzas}, {Raddi}, {Roelofs}, {Rodr{\'{\i}}guez-Gil},
  {Sabin}, {Scaringi}, {Steeghs}, {Suso}, {Tata}, {Unruh}, {van Roestel},
  {Viironen}, {Vink}, {Walton}, {Wright}, \& {Zijlstra}}]{iphas2014}
{Barentsen}, G., {Farnhill}, H.~J., {Drew}, J.~E., {et~al.} 2014, \mnras, 444,
  3230

\bibitem[{{Blumenthal} \& {Gould}(1970)}]{blumenthal70}
{Blumenthal}, G.~R. \& {Gould}, R.~J. 1970, Rev. Mod. Phys., 42, 237

\bibitem[{{Bouret} {et~al.}(2012){Bouret}, {Hillier}, {Lanz}, \&
  {Fullerton}}]{bouret12}
{Bouret}, J.~C., {Hillier}, D.~J., {Lanz}, T., \& {Fullerton}, A.~W. 2012,
  \aap, 544, A67

\bibitem[{Condon \& Odishaw(1958)}]{condon58}
Condon, E. \& Odishaw, H. 1958, Handbook of Physics (McGraw-Hill)

\bibitem[{{Condon} {et~al.}(1998){Condon}, {Cotton}, {Greisen}, {Yin},
  {Perley}, {Taylor}, \& {Broderick}}]{condon98}
{Condon}, J.~J., {Cotton}, W.~D., {Greisen}, E.~W., {et~al.} 1998, \aj, 115,
  1693

\bibitem[{{del Valle} \& {Pohl}(2018)}]{delvalle18}
{del Valle}, M.~V. \& {Pohl}, M. 2018, \apj, 864, 19

\bibitem[{{Dialynas} {et~al.}(2017){Dialynas}, {Krimigis}, {Mitchell},
  {Decker}, \& {Roelof}}]{dialynas2017}
{Dialynas}, K., {Krimigis}, S.~M., {Mitchell}, D.~G., {Decker}, R.~B., \&
  {Roelof}, E.~C. 2017, Nature Astronomy, 1, 0115

\bibitem[{{Donahue} {et~al.}(1995){Donahue}, {Aldering}, \&
  {Stocke}}]{donahue95}
{Donahue}, M., {Aldering}, G., \& {Stocke}, J.~T. 1995, \apj, 450, L45

\bibitem[{{Draine}(2011)}]{draine}
{Draine}, B.~T. 2011, {Physics of the Interstellar and Intergalactic Medium}
  (Princeton University Press)

\bibitem[{{Drew} {et~al.}(2005){Drew}, {Greimel}, {Irwin}, {Aungwerojwit},
  {Barlow}, {Corradi}, {Drake}, {G{\"a}nsicke}, {Groot}, {Hales}, {Hopewell},
  {Irwin}, {Knigge}, {Leisy}, {Lennon}, {Mampaso}, {Masheder}, {Matsuura},
  {Morales-Rueda}, {Morris}, {Parker}, {Phillipps}, {Rodriguez-Gil}, {Roelofs},
  {Skillen}, {Sokoloski}, {Steeghs}, {Unruh}, {Viironen}, {Vink}, {Walton},
  {Witham}, {Wright}, {Zijlstra}, \& {Zurita}}]{iphas2005}
{Drew}, J.~E., {Greimel}, R., {Irwin}, M.~J., {et~al.} 2005, \mnras, 362, 753

\bibitem[{{Gaia Collaboration}(2018)}]{gaia18}
{Gaia Collaboration}. 2018, VizieR Online Data Catalog, I/345

\bibitem[{{Garraffo} {et~al.}(2016){Garraffo}, {Drake}, \&
  {Cohen}}]{garraffo2016}
{Garraffo}, C., {Drake}, J.~J., \& {Cohen}, O. 2016, \apjl, 833, L4

\bibitem[{{Gontcharov}(2006)}]{gontcharov06}
{Gontcharov}, G.~A. 2006, Astronomy Letters, 32, 759

\bibitem[{{Guo} {et~al.}(2019){Guo}, {Florinski}, \& {Wang}}]{guo2019}
{Guo}, X., {Florinski}, V., \& {Wang}, C. 2019, \apj, 879, 87

\bibitem[{{Gvaramadze} {et~al.}(2018){Gvaramadze}, {Alexashov}, {Katushkina},
  \& {Kniazev}}]{gvaramadze18}
{Gvaramadze}, V.~V., {Alexashov}, D.~B., {Katushkina}, O.~A., \& {Kniazev},
  A.~Y. 2018, \mnras, 474, 4421

\bibitem[{{Gvaramadze} {et~al.}(2012){Gvaramadze}, {Langer}, \&
  {Mackey}}]{gvaramadze12}
{Gvaramadze}, V.~V., {Langer}, N., \& {Mackey}, J. 2012, \mnras, 427, L50

\bibitem[{{Hasinger}(2004)}]{hasinger04}
{Hasinger}, G. 2004, Nuclear Physics B Proceedings Supplements, 132, 86

\bibitem[{{Kissmann} {et~al.}(2018){Kissmann}, {Kleimann}, {Krebl}, \&
  {Wiengarten}}]{kissmann18}
{Kissmann}, R., {Kleimann}, J., {Krebl}, B., \& {Wiengarten}, T. 2018, The
  Astrophysical Journal Supplement Series, 236, 53

\bibitem[{{Kobulnicky} {et~al.}(2017){Kobulnicky}, {Schurhammer}, {Baldwin},
  {Chick}, {Dixon}, {Lee}, \& {Povich}}]{kobulnicky17}
{Kobulnicky}, H.~A., {Schurhammer}, D.~P., {Baldwin}, D.~J., {et~al.} 2017,
  \aj, 154, 201

\bibitem[{{Kornbleuth} {et~al.}(2018){Kornbleuth}, {Opher}, {Michael}, \&
  {Drake}}]{kornbleuth2018}
{Kornbleuth}, M., {Opher}, M., {Michael}, A.~T., \& {Drake}, J.~F. 2018, \apj,
  865, 84

\bibitem[{{Kosi{\'n}ski} \& {Hanasz}(2006)}]{kosinski2006}
{Kosi{\'n}ski}, R. \& {Hanasz}, M. 2006, \mnras, 368, 759

\bibitem[{Longair(1992)}]{longair92}
Longair, M. 1992, {High Energy Astrophysics: Volume 1, Particles, Photons and
  Their Detection}, High Energy Astrophysics (Cambridge University Press)

\bibitem[{{Mao} \& {Kaastra}(2016)}]{mao16}
{Mao}, J. \& {Kaastra}, J. 2016, \aap, 587, A84

\bibitem[{{McComas} {et~al.}(2012){McComas}, {Alexashov}, {Bzowski}, {Fahr},
  {Heerikhuisen}, {Izmodenov}, {Lee}, {M{\"o}bius}, {Pogorelov}, {Schwadron},
  \& {Zank}}]{mccomas12}
{McComas}, D.~J., {Alexashov}, D., {Bzowski}, M., {et~al.} 2012, Science, 336,
  1291

\bibitem[{{McKee} \& {Ostriker}(1977)}]{mckee1977}
{McKee}, C.~F. \& {Ostriker}, J.~P. 1977, \apj, 218, 148

\bibitem[{{Meyer} {et~al.}(2017){Meyer}, {Mignone}, {Kuiper}, {Raga}, \&
  {Kley}}]{meyer17}
{Meyer}, D.~M.~A., {Mignone}, A., {Kuiper}, R., {Raga}, A.~C., \& {Kley}, W.
  2017, \mnras, 464, 3229

\bibitem[{{Mohamed} {et~al.}(2012){Mohamed}, {Mackey}, \& {Langer}}]{mohamed12}
{Mohamed}, S., {Mackey}, J., \& {Langer}, N. 2012, \aap, 541, A1

\bibitem[{{Opher}(2016)}]{opher16}
{Opher}, M. 2016, \ssr, 200, 475

\bibitem[{{Opher} {et~al.}(2016){Opher}, {Drake}, {Zieger}, {Swisdak}, \&
  {Toth}}]{opher2016}
{Opher}, M., {Drake}, J.~F., {Zieger}, B., {Swisdak}, M., \& {Toth}, G. 2016,
  Physics of Plasmas, 23, 056501

\bibitem[{{Parker}(1958)}]{parker1958}
{Parker}, E.~N. 1958, \apj, 128, 664

\bibitem[{{Pogorelov} {et~al.}(2017){Pogorelov}, {Fichtner}, {Czechowski},
  {Lazarian}, {Lembege}, {le Roux}, {Potgieter}, {Scherer}, {Stone}, {Strauss},
  {Wiengarten}, {Wurz}, {Zank}, \& {Zhang}}]{pogorelov17}
{Pogorelov}, N.~V., {Fichtner}, H., {Czechowski}, A., {et~al.} 2017, \ssr, 212,
  193

\bibitem[{{Reiners} \& {Basri}(2008)}]{reiners2008}
{Reiners}, A. \& {Basri}, G. 2008, \aap, 489, L45

\bibitem[{{Reynolds}(1984)}]{reynolds84}
{Reynolds}, R.~J. 1984, \apj, 282, 191

\bibitem[{{Reynolds} {et~al.}(1999){Reynolds}, {Haffner}, \&
  {Tufte}}]{reynolds1999}
{Reynolds}, R.~J., {Haffner}, L.~M., \& {Tufte}, S.~L. 1999, \apjl, 525, L21

\bibitem[{{Rieke} {et~al.}(2004){Rieke}, {Young}, {Engelbracht}, {Kelly},
  {Low}, {Haller}, {Beeman}, {Gordon}, {Stansberry}, {Misselt}, {Cadien},
  {Morrison}, {Rivlis}, {Latter}, {Noriega-Crespo}, {Padgett}, {Stapelfeldt},
  {Hines}, {Egami}, {Muzerolle}, {Alonso-Herrero}, {Blaylock}, {Dole}, {Hinz},
  {Le Floc'h}, {Papovich}, {P{\'e}rez-Gonz{\'a}lez}, {Smith}, {Su}, {Bennett},
  {Frayer}, {Henderson}, {Lu}, {Masci}, {Pesenson}, {Rebull}, {Rho}, {Keene},
  {Stolovy}, {Wachter}, {Wheaton}, {Werner}, \& {Richards}}]{mips2004}
{Rieke}, G.~H., {Young}, E.~T., {Engelbracht}, C.~W., {et~al.} 2004, \apjs,
  154, 25

\bibitem[{{S{\'a}nchez-Ayaso} {et~al.}(2018){S{\'a}nchez-Ayaso}, {del Valle},
  {Mart{\'\i}}, {Romero}, \& {Luque-Escamilla}}]{ayaso2018}
{S{\'a}nchez-Ayaso}, E., {del Valle}, M.~V., {Mart{\'\i}}, J., {Romero}, G.~E.,
  \& {Luque-Escamilla}, P.~L. 2018, \apj, 861, 32

\bibitem[{{Scherer} {et~al.}(2019){Scherer}, {Baalmann}, {Fichtner},
  {Kleimann}, {Bomans}, {Weis}, {Ferreira}, \& {Herbst}}]{scherer18}
{Scherer}, K., {Baalmann}, L.~R., {Fichtner}, H., {et~al.} 2019, {MNRAS,
  submitted}

\bibitem[{{Scherer} {et~al.}(2016{\natexlab{a}}){Scherer}, {Bomans},
  {Ferreira}, {Fichtner}, {Kleimann}, {Strauss}, \& {Weis}}]{scherer16b}
{Scherer}, K., {Bomans}, D.~J., {Ferreira}, S.~E.~S., {et~al.}
  2016{\natexlab{a}}, in Journal of Physics Conference Series, Vol. 767, 012024

\bibitem[{{Scherer} \& {Fichtner}(2014)}]{scherer14}
{Scherer}, K. \& {Fichtner}, H. 2014, \apj, 782, 25

\bibitem[{{Scherer} {et~al.}(2016{\natexlab{b}}){Scherer}, {Fichtner},
  {Kleimann}, {Wiengarten}, {Bomans}, \& {Weis}}]{scherer16}
{Scherer}, K., {Fichtner}, H., {Kleimann}, J., {et~al.} 2016{\natexlab{b}},
  \aap, 586, A111

\bibitem[{{Scherer} {et~al.}(2016{\natexlab{c}}){Scherer}, {Strauss},
  {Ferreira}, \& {Fichtner}}]{scherer16c}
{Scherer}, K., {Strauss}, R.~D., {Ferreira}, S.~E.~S., \& {Fichtner}, H.
  2016{\natexlab{c}}, Astroparticle Physics, 82, 93

\bibitem[{{Scherer} {et~al.}(2015){Scherer}, {van der Schyff}, {Bomans},
  {Ferreira}, {Fichtner}, {Kleimann}, {Strauss}, {Weis}, {Wiengarten}, \&
  {Wodzinski}}]{scherer15}
{Scherer}, K., {van der Schyff}, A., {Bomans}, D.~J., {et~al.} 2015, \aap, 576,
  A97

\bibitem[{{Schure} {et~al.}(2009){Schure}, {Kosenko}, {Kaastra}, {Keppens}, \&
  {Vink}}]{schure09}
{Schure}, K.~M., {Kosenko}, D., {Kaastra}, J.~S., {Keppens}, R., \& {Vink}, J.
  2009, \aap, 508, 751

\bibitem[{{Tarango-Yong} \& {Henney}(2018)}]{tarangoyong18}
{Tarango-Yong}, J.~A. \& {Henney}, W.~J. 2018, \mnras, 477, 2431

\bibitem[{{Van Eck} {et~al.}(2017){Van Eck}, {Haverkorn}, {Alves, M. I. R.},
  {Beck, R.}, {de Bruyn, A. G.}, {Enßlin, T.}, {Farnes, J. S.}, {Ferrière,
  K.}, {Heald, G.}, {Horellou, C.}, {Horneffer, A.}, {Iacobelli, M.}, {Jelić,
  V.}, {Martí-Vidal, I.}, {Mulcahy, D. D.}, {Reich, W.}, {Röttgering, H. J.
  A.}, {Scaife, A. M. M.}, {Schnitzeler, D. H. F. M.}, {Sobey, C.}, \&
  {Sridhar, S. S.}}]{vaneck17}
{Van Eck}, C.~L., {Haverkorn}, M., {Alves, M. I. R.}, {et~al.} 2017, \aap, 597,
  A98

\bibitem[{{van Haarlem} {et~al.}(2013){van Haarlem}, {Wise}, {Gunst}, {Heald},
  {McKean}, {Hessels}, {de Bruyn}, {Nijboer}, {Swinbank}, {Fallows},
  {Brentjens}, {Nelles}, {Beck}, {Falcke}, {Fender}, {H{\"o}randel},
  {Koopmans}, {Mann}, {Miley}, {R{\"o}ttgering}, {Stappers}, {Wijers},
  {Zaroubi}, {van den Akker}, {Alexov}, {Anderson}, {Anderson}, {van Ardenne},
  {Arts}, {Asgekar}, {Avruch}, {Batejat}, {B{\"a}hren}, {Bell}, {Bell}, {van
  Bemmel}, {Bennema}, {Bentum}, {Bernardi}, {Best}, {B{\^\i}rzan}, {Bonafede},
  {Boonstra}, {Braun}, {Bregman}, {Breitling}, {van de Brink}, {Broderick},
  {Broekema}, {Brouw}, {Br{\"u}ggen}, {Butcher}, {van Cappellen}, {Ciardi},
  {Coenen}, {Conway}, {Coolen}, {Corstanje}, {Damstra}, {Davies}, {Deller},
  {Dettmar}, {van Diepen}, {Dijkstra}, {Donker}, {Doorduin}, {Dromer}, {Drost},
  {van Duin}, {Eisl{\"o}ffel}, {van Enst}, {Ferrari}, {Frieswijk}, {Gankema},
  {Garrett}, {de Gasperin}, {Gerbers}, {de Geus}, {Grie{\ss}meier}, {Grit},
  {Gruppen}, {Hamaker}, {Hassall}, {Hoeft}, {Holties}, {Horneffer}, {van der
  Horst}, {van Houwelingen}, {Huijgen}, {Iacobelli}, {Intema}, {Jackson},
  {Jelic}, {de Jong}, {Juette}, {Kant}, {Karastergiou}, {Koers}, {Kollen},
  {Kondratiev}, {Kooistra}, {Koopman}, {Koster}, {Kuniyoshi}, {Kramer},
  {Kuper}, {Lambropoulos}, {Law}, {van Leeuwen}, {Lemaitre}, {Loose}, {Maat},
  {Macario}, {Markoff}, {Masters}, {McFadden}, {McKay-Bukowski}, {Meijering},
  {Meulman}, {Mevius}, {Middelberg}, {Millenaar}, {Miller-Jones}, {Mohan},
  {Mol}, {Morawietz}, {Morganti}, {Mulcahy}, {Mulder}, {Munk}, {Nieuwenhuis},
  {van Nieuwpoort}, {Noordam}, {Norden}, {Noutsos}, {Offringa}, {Olofsson},
  {Omar}, {Orr{\'u}}, {Overeem}, {Paas}, {Pandey-Pommier}, {Pandey}, {Pizzo},
  {Polatidis}, {Rafferty}, {Rawlings}, {Reich}, {de Reijer}, {Reitsma},
  {Renting}, {Riemers}, {Rol}, {Romein}, {Roosjen}, {Ruiter}, {Scaife}, {van
  der Schaaf}, {Scheers}, {Schellart}, {Schoenmakers}, {Schoonderbeek},
  {Serylak}, {Shulevski}, {Sluman}, {Smirnov}, {Sobey}, {Spreeuw}, {Steinmetz},
  {Sterks}, {Stiepel}, {Stuurwold}, {Tagger}, {Tang}, {Tasse}, {Thomas},
  {Thoudam}, {Toribio}, {van der Tol}, {Usov}, {van Veelen}, {van der Veen},
  {ter Veen}, {Verbiest}, {Vermeulen}, {Vermaas}, {Vocks}, {Vogt}, {de Vos},
  {van der Wal}, {van Weeren}, {Weggemans}, {Weltevrede}, {White}, {Wijnholds},
  {Wilhelmsson}, {Wucknitz}, {Yatawatta}, {Zarka}, {Zensus}, \& {van
  Zwieten}}]{lofar13}
{van Haarlem}, M.~P., {Wise}, M.~W., {Gunst}, A.~W., {et~al.} 2013, \aap, 556,
  A2

\bibitem[{{van Hoof} {et~al.}(2014){van Hoof}, {Williams}, {Volk}, {Chatzikos},
  {Ferland}, {Lykins}, {Porter}, \& {Wang}}]{vanhoof14}
{van Hoof}, P.~A.~M., {Williams}, R.~J.~R., {Volk}, K., {et~al.} 2014, \mnras,
  444, 420

\bibitem[{{van Marle} {et~al.}(2014){van Marle}, {Decin}, \&
  {Meliani}}]{vanmarle14}
{van Marle}, A.~J., {Decin}, L., \& {Meliani}, Z. 2014, \aap, 561, A152

\bibitem[{{Vidotto} {et~al.}(2011){Vidotto}, {Jardine}, {Opher}, {Donati}, \&
  {Gombosi}}]{vidotto2011}
{Vidotto}, A.~A., {Jardine}, M., {Opher}, M., {Donati}, J.~F., \& {Gombosi},
  T.~I. 2011, \mnras, 412, 351

\bibitem[{{Weaver} {et~al.}(1977){Weaver}, {McCray}, {Castor}, {Shapiro}, \&
  {Moore}}]{weaver77}
{Weaver}, R., {McCray}, R., {Castor}, J., {Shapiro}, P., \& {Moore}, R. 1977,
  \apj, 218, 377

\bibitem[{{Werner} {et~al.}(2004){Werner}, {Roellig}, {Low}, {Rieke}, {Rieke},
  {Hoffmann}, {Young}, {Houck}, {Brandl}, {Fazio}, {Hora}, {Gehrz}, {Helou},
  {Soifer}, {Stauffer}, {Keene}, {Eisenhardt}, {Gallagher}, {Gautier}, {Irace},
  {Lawrence}, {Simmons}, {Van Cleve}, {Jura}, {Wright}, \&
  {Cruikshank}}]{spitzer2004}
{Werner}, M.~W., {Roellig}, T.~L., {Low}, F.~J., {et~al.} 2004, \apjs, 154, 1

\bibitem[{{Wiese} \& {Fuhr}(2009)}]{wiese2009}
{Wiese}, W.~L. \& {Fuhr}, J.~R. 2009, Journal of Physical and Chemical
  Reference Data, 38, 565

\bibitem[{{Wright} {et~al.}(2010){Wright}, {Eisenhardt}, {Mainzer}, {Ressler},
  {Cutri}, {Jarrett}, {Kirkpatrick}, {Padgett}, {McMillan}, {Skrutskie},
  {Stanford}, {Cohen}, {Walker}, {Mather}, {Leisawitz}, {Gautier}, {McLean},
  {Benford}, {Lonsdale}, {Blain}, {Mendez}, {Irace}, {Duval}, {Liu}, {Royer},
  {Heinrichsen}, {Howard}, {Shannon}, {Kendall}, {Walsh}, {Larsen}, {Cardon},
  {Schick}, {Schwalm}, {Abid}, {Fabinsky}, {Naes}, \& {Tsai}}]{wise2010}
{Wright}, E.~L., {Eisenhardt}, P.~R.~M., {Mainzer}, A.~K., {et~al.} 2010, \aj,
  140, 1868

\bibitem[{{Zank}(1999)}]{zank1999}
{Zank}, G.~P. 1999, \ssr, 89, 413

\bibitem[{{Zank} {et~al.}(2013){Zank}, {Heerikhuisen}, {Wood}, {Pogorelov},
  {Zirnstein}, \& {McComas}}]{zank13}
{Zank}, G.~P., {Heerikhuisen}, J., {Wood}, B.~E., {et~al.} 2013, \apj, 763, 20

\end{thebibliography}
\end{document}